\documentclass{elsarticle}

\usepackage[british]{babel}
\usepackage{graphicx}
\usepackage{amsmath}
\usepackage{amssymb}
\usepackage{amsfonts}
\usepackage{amscd}
\usepackage{bbold}
\usepackage{tikz}

\begin{document}

\begin{frontmatter}

\title{Electronic optics in graphene in the semiclassical approximation}

\author[RUN]{K. J. A. Reijnders}
\ead{K.Reijnders@science.ru.nl}
\author[IPM]{D. S. Minenkov}
\ead{minenkov.ds@gmail.com}
\author[RUN]{M. I. Katsnelson}
\ead{M.Katsnelson@science.ru.nl}
\author[IPM,MIPT]{S. Yu. Dobrokhotov}
\ead{dobr@ipmnet.ru}
\address[RUN]{Radboud University, Institute for Molecules and Materials, \\Heyendaalseweg 135, 6525 AJ Nijmegen, The Netherlands}
\address[IPM]{Ishlinsky Institute for Problems in Mechanics of Russian Academy of Sciences, \\Moscow, Russia}
\address[MIPT]{Moscow Institute of Physics and Technology, Dolgoprudny, Moscow Region, Russia}

\begin{abstract}

We study above-barrier scattering of Dirac electrons by a smooth electrostatic potential combined with a coordinate-dependent mass in graphene. We assume that the potential and mass are sufficiently smooth, so that we can define a small dimensionless semiclassical parameter $h \ll 1$. This electronic optics setup naturally leads to focusing and the formation of caustics, which are singularities in the density of trajectories. We construct a semiclassical approximation for the wavefunction in all points, placing particular emphasis on the region near the caustic, where the maximum of the intensity lies.
Because of the matrix character of the Dirac equation, this wavefunction contains a nontrivial semiclassical phase, which is absent for a scalar wave equation and which influences the focusing.
We carefully discuss the three steps in our semiclassical approach: the adiabatic reduction of the matrix equation to an effective scalar equation, the construction of the wavefunction using the Maslov canonical operator and the application of the uniform approximation to the integral expression for the wavefunction in the vicinity of a caustic.
We consider several numerical examples and show that our semiclassical results are in very good agreement with the results of tight-binding calculations. In particular, we show that the semiclassical phase can have a pronounced effect on the position of the focus and its intensity.

\end{abstract}


\end{frontmatter}

Graphene is a two-dimensional allotrope of carbon, in which the carbon atoms are arranged in a honeycomb lattice~\cite{Katsnelson13}. 
Considering its electronic structure, one observes that the valence and conduction bands touch at two nonequivalent corners of the Brillouin zone, known as $K$ and $K'$. In the vicinity of these points, that is, for energies near the Fermi energy, the dispersion relation can be approximated by a cone.
In particular, the behavior of the effective low-energy charge carriers can be described by the two-dimensional Dirac equation~\cite{Wallace47,McClure57,Slonczewski58,Semenoff84,Novoselov05,Zhang05,CastroNeto09,Katsnelson13}.
We need a matrix Hamiltonian to describe the system, since the honeycomb lattice is made up out of two sublattices. Therefore, the wavefunction is given by a two-dimensional spinor, whose components represent the contribution of each of the two sublattices.
When we consider standard graphene, in which these sublattices are equivalent, the mass term in the Dirac Hamiltonian vanishes.
However, when we consider graphene on a substrate, e.g. hexagonal boron nitride, this equivalence is generally broken and a mass term naturally arises~\cite{Sachs11,Bokdam14,Slotman15,Wallbank13,Diez14,Chizhova14,Yankowitz14}.

In contrast to scalar Hamiltonians, matrix Hamiltonians such as the Dirac Hamiltonian can give rise to nontrivial adiabatic phases in the wavefunction, even in a time-independent scattering problem.
The most prominent example of such a phase is the Berry phase~\cite{Berry84,Shapere89}.
In graphene, it acquires a value of $\pm \pi$ upon a full rotation around the Dirac point, depending on whether one considers the $K$ or $K'$ valley~\cite{Berry84,Ando98,Katsnelson13}.
By studying the massless Dirac equation, one can show that this Berry phase affects the Fabry-P\'erot condition for resonant scattering in \emph{n-p-n} junctions in a magnetic field~\cite{Shytov08,Young09}.
In the absence of a mass term, the Berry phase also enters the semiclassical quantization condition for electrons in a strong magnetic field, which determines the positions of the Landau levels.
The Berry phase therefore strongly affects the quantum Hall effect in graphene~\cite{Novoselov05,Zhang05,Mikitik99,Zheng02,Gusynin05,Peres05}.
In the presence of a mass term, the situation is however more complicated. 
Instead of the Berry phase, the more general semiclassical phase now enters the wavefunction~\cite{Carmier08,Littlejohn91}. However, one can show that in this case the Landau levels are still determined by the winding number in momentum space~\cite{Fuchs10}.

However, the influence of the semiclassical phase is not limited to setups with a magnetic field.
It is important in all problems in which interference plays a role, such as focusing in electronic optics. 
In this type of problems, one focuses electrons using electrostatic potentials in the same way as one focuses light using an optical lens.
In graphene, these potentials can be experimentally realized by gating. The mass, which typically arises from the substrate, also plays a role in focusing.
Basic understanding of focusing can be obtained by considering the classical trajectories, which are analogous to the rays in geometrical optics. However, to understand how interference affects the focusing, one needs to perform a quantum mechanical analysis, for instance using the semiclassical approximation.

Focusing in graphene has mostly been considered in the context of \emph{n-p} junctions, for both straight~\cite{Cheianov07,Reijnders17a,Reijnders17b,Milovanovic15,Lee15,Chen16} and circular~\cite{Cserti07,Peterfalvi10,Wu14,Matulis11} interfaces. 
At such an interface, charge carriers are refracted with a negative refractive index, creating an electronic lens.
This type of lens is known as a Veselago lens~\cite{Veselago68} and can be effectively realized in graphene because of the high tunneling probability across the \emph{n-p} interface. A normally incident electron is even transmitted with unit probability, a process known as Klein tunneling~\cite{Klein29,Katsnelson06,Cheianov06,Shytov08,Tudorovskiy12,Reijnders13,Young09,Stander09,Sutar12,Chen16}.
Recently, it has been suggested that graphene Veselago lenses could be used to create a two-dimensional analog of a Scanning Tunneling Microscope: the Dirac fermion microscope~\cite{Boggild17}.

Previous studies have shown that the matrix structure of the Dirac equation influences focusing in graphene Veselago lenses~\cite{Reijnders17a,Reijnders17b}, in particular through the initial sublattice polarization.
However, in almost all of the theoretical studies of Veselago lensing in graphene, the authors considered a sharp junction interface with a very special shape, either straight of circular.
This is mainly due to the fact that smooth junctions are much more complicated to treat analytically, and the methods that exist are limited to (effectively) one-dimensional cases.
A semiclassical treatment of Klein tunneling for smooth one-dimensional graphene heterojunctions was given in Refs.~\cite{Reijnders13,Tudorovskiy12}, and this analysis was later extended to cases with a constant mass~\cite{Zalipaev15}.
Unfortunately, it is not possible to extend this analysis to truly two-dimensional potentials and masses.
At the same time, we expect the matrix structure of the Dirac equation and the semiclassical phase to play a much larger role for two-dimensional setups, due to the additional degree of freedom.

Fortunately, one can also realize focusing of charge carriers in graphene using different setups. In particular, electrons in graphene can be focused using two-dimensional potential wells~\cite{Logemann15} or by applying local strain~\cite{Stegmann16}. In this paper, we consider focusing by above-barrier scattering for two-dimensional potentials and masses.
Since there are no classically forbidden regions in this type of focusing, tunneling does not play a role.
We remark that this requires that we neglect the exponentially small reflection induced by the potential~\cite{Reijnders13}, which is typically justifiable.
Since tunneling does not play a role in this setup, we can study it using semiclassical methods, provided that both the mass and the potential are sufficiently smooth.

The primary goal of this paper is to establish how the semiclassical phase influences focusing of charge carriers in graphene by two-dimensional potentials and masses.
In particular, we consider above-barrier scattering for a parallel bundle of incoming electrons. 
However, we believe that the principles we discuss are more widely applicable and can also improve the understanding of focusing in more complicated setups.
For instance, they may help to improve image reconstruction in the Dirac fermion microscope.
We observe that electrons in this system propagate over a large distance from the junction interface, at which they are refracted, to the point at which they are focused. 
During their propagation, these electrons are under the influence of a mass term, since Dirac fermion microscopes will most likely be made of graphene on a substrate~\cite{Boggild17}. 
Hence, a large semiclassical phase may develop along the trajectories, which can subsequently influence the position of the focus.
In order to properly reconstruct the object from the measured intensity, it is important to understand exactly how the focusing is affected by this semiclassical phase.

The second goal of this paper is to provide a detailed introduction to the semiclassical methods that we use to construct the wavefunction.
Many of these methods have mainly been discussed within the mathematical literature, and we hope to make them more accessible to a wider audience.
The most well-known semiclassical approximation is probably the one-dimensional Wentzel-Kramers-Brillouin (WKB) approximation, which is explained in nearly all introductory textbooks on quantum mechanics, see e.g. Ref.~\cite{Griffiths16}.
For scalar Hamiltonians, this method was generalized to higher dimensions by V. P. Maslov~\cite{Maslov65,Maslov72,Maslov73,Maslov81}. 
Later on, this approximation was extended to matrix Hamiltonians, both by Maslov~\cite{Maslov81} and by Bernstein and Friedland~\cite{Bernstein75,Bernstein84}; see also Ref.~\cite{Kaufman87}.
In particular, these authors obtained an expression for the semiclassical phase that emerges for an arbitrary Hamiltonian.
In Ref.~\cite{Carmier08}, their approach was applied to the Dirac Hamiltonian of graphene.

In this paper, we make use of a different approach to construct a semiclassical approximation for the graphene Hamiltonian. 
Since we are not interested in tunneling phenomena, we first construct effective scalar Hamiltonians for both electrons and holes.
We perform this adiabatic reduction using the method formulated in Refs.~\cite{Berlyand87,Belov06} and independently in Refs.~\cite{Littlejohn91,Littlejohn91b}, see also Refs.~\cite{Clemmow54,Friedland87}.
In this approach, which is asymptotic in nature, we first pass from operators to their symbols~\cite{Maslov65,Maslov81,Martinez02,Zworski12}, which are analogous to classical observables on phase space.
After that, we obtain an effective scalar Hamiltonian to first order in the dimensionless semiclassical parameter $h$ by solving algebraic equations.
Integrating Hamilton's equations for this effective scalar Hamiltonian, we obtain the classical trajectories of the system.
Subsequently, we construct the wavefunction for our effective scalar Hamiltonian using the multi-dimensional WKB approximation.
We find this procedure more insightful, since it separates the two steps that are required: we first diagonalize the matrix Hamiltonian and only then we construct the semiclassical approximation.
In particular, this method provides us with a deeper understanding of the origin of the semiclassical phase.

Unfortunately, the multi-dimensional WKB approximation diverges at points at which the density of classical trajectories diverges. 
These points are known as singular points and together they form a so-called caustic.
As the density of trajectories is high near caustics, these are exactly the places at which focusing occurs.
To obtain the wavefunction near points on the caustic, we first lift the problem from the two-dimensional configuration space $(x_1,x_2)$ to the four-dimensional phase space. 
Instead of the Hamilton-Jacobi equation, we therefore solve Hamilton's equations.
Whereas solutions of the former equation become problematic at singular points, solutions of the latter do not.
The solutions of Hamilton's equations form a two-dimensional surface in phase space, which has the structure of a Lagrangian manifold~\cite{Maslov81,Dubrovin84,Arnold89,Abraham08}.
At points on the caustic, the projection of this surface onto the configuration space is not invertible.
Caustics have been extensively studied in the literature and a complete classification of their possible types has been established~\cite{Whitney55,Thom72,Arnold75,Arnold82,Arnold90,Poston78,Berry80}.
This classification shows that one generally expects two types of singular points to occur in two-dimensional Hamiltonian systems: fold points and cusp points.
We expect the strongest foci to lie in the vicinity of cusp points, since the singularity at a cusp point is of higher order than the singularity at a fold point.

We can construct the wavefunction near caustics using the Maslov canonical operator~\cite{Maslov81,Mishchenko90,Guillemin77,Bates97,Dobrokhotov03}.
In this approach, we express the wavefunction in terms of several geometric objects that are defined on the Lagrangian manifold formed by the solutions of Hamilton's equations.
This is made possible by working with the symbols of the differential operators, which are much easier to manipulate than the original operators themselves.
The construction, which is mostly algebraic and geometric in nature, provides a general expression for the wavefunction that can be applied for many different Hamiltonians.
In two-dimensional problems, the Maslov canonical operator conventionally takes the form of an integral over one of the momentum coordinates.
However, the momentum coordinate over which we integrate may differ from singular point to singular point, which makes it non-trivial to implement this expression numerically.

Recently, a new representation of the Maslov canonical operator near singular points was put forward~\cite{Dobrokhotov14,Dobrokhotov14b}. This representation was specifically designed for problems that admit a parametrization in terms of so-called eikonal coordinates~\cite{Dobrokhotov15}. 
This is a special kind of coordinate system, in which one parametrizes the time along the trajectories by the action.
The second coordinate $\phi$ on the Lagrangian manifold is subsequently determined by the initial condition of the trajectories.
We show that our problem admits such a parametrization and that the second coordinate $\phi$ is equal to the coordinate perpendicular to the propagation direction at minus infinity.
Generally, eikonal coordinates naturally arise in two-dimensional scattering problems for which the classical Hamiltonian can be written as a function of $x$ and $|p|$~\cite{Dobrokhotov15}. 
It turns out that eikonal coordinates have very convenient geometrical properties. First of all, the wavefronts are given by lines of equal time. Second, because of the orthogonality of the trajectories and the wavefronts, the Jacobian factorizes in this coordinate system, which essentially simplifies many of the computations.

Using the new representation~\cite{Dobrokhotov14}, we express the wavefunction in the vicinity of a singular point as an integral over the coordinate $\phi$.
The new representation therefore admits a much more intuitive interpretation than the conventional representation, as it is given by an integral over a coordinate that directly labels the trajectories.
Furthermore, it is independent of the singular point in question, unlike the conventional representation.
In both representations, the integrand contains a rapidly oscillating exponent~\cite{Maslov81,Dobrokhotov14}. This makes it difficult to evaluate our expression for the wavefunction numerically, especially in the deep semiclassical limit.

We therefore employ the stationary phase approximation~\cite{Maslov81,Guillemin77,Dobrokhotov14} to obtain an asymptotic expansion of the integral in powers of the dimensionless semiclassical parameter $h$.
If we confine ourselves to the leading-order approximation, see e.g. Refs.~\cite{Dobrokhotov14,Maslov81}, we obtain an expression for the wavefunction in terms of the Pearcey function~\cite{Pearcey46,Connor81a,Connor82}. Unfortunately, this expression does not capture the influence of the semiclassical phase on the focusing, since the intensity $\lVert \Psi \rVert^2$ that it predicts is independent of this phase.
Using the uniform approximation~\cite{Ursell72,Connor81b}, we obtain an expression for the wavefunction that includes higher-order corrections.
Although we can only construct this expression in the region in which interference occurs, this is precisely the region in which the main focus is located.
We can therefore use the uniform approximation to study the influence of the semiclassical phase on the focusing.
In contrast to the intensity predicted by the leading-order approximation, the intensity predicted by the uniform approximation depends on the semiclassical phase.

The final result of our semiclassical analysis is a collection of approximations for the wavefunction.
Each of these approximations is only valid within its own specific domain. 
The size of these domains is given in terms of the dimensionless semiclassical parameter $h$, and may depend slightly on the details of the problem.
We subsequently obtain a global approximation for the wavefunction by combining the various local approximations.
We emphasize that this procedure does not require matching of the different local approximations by adjusting their coefficients. 
Instead, each of the approximations is a local asymptotic solution to the scattering problem, without free parameters. 

To study the influence of the semiclassical phase on the position and the intensity of the focus, we consider various setups of potential and mass.
In particular, we study a situation in which the semiclassical phase is small as well as a situation in which the semiclassical phase is large. 
For both cases, we obtain numerical values for the wavefunction in the vicinity of the focal point using the uniform approximation.
We subsequently compare these semiclassical results with the results of tight-binding calculations for large graphene samples, which are performed using the Kwant code~\cite{Groth14}.
For the case of a large semiclassical phase, we also consider the trajectories that arise when we incorporate the semiclassical phase into the Hamiltonian~\cite{Littlejohn91,Xiao10}.
The latter approach has recently attracted a lot of interest, as it has been able to successfully explain experimental observations in heterostructures of graphene and hexagonal boron nitride~\cite{Gorbachev14}.

Although we only consider graphene in this paper, most of our results concern the Dirac equation. Therefore, they are also applicable to other two-dimensional materials in which the electrons are governed by the Dirac equation, such as the two-dimensional surfaces of three-dimensional topological insulators~\cite{Qi11,Moore10,Hasan10,Hasan11,Bansil16}.

We have tried to structure the paper in such a way that the results for graphene can be largely understood without detailed knowledge of the semiclassical methods that we use. Likewise, our review of the semiclassical methods can be read without considering the specific application to graphene. In section~\ref{sec:assumptions}, we provide some preliminary considerations. In particular, we discuss the scattering setup and the assumptions that we make, as well as several symmetries of the graphene Hamiltonian. Since this Hamiltonian is a two-dimensional matrix, it describes both electrons and holes. Section~\ref{sec:adiabatic} shows how we can obtain an effective scalar Hamiltonian for each of these modes. We subsequently take a first step towards the construction of our semiclassical approximation in section~\ref{sec:semiclassics}. In this section, we also discuss the difference between the Berry phase and the semiclassical phase. In section~\ref{sec:lagman}, we introduce the important concept of a Lagrangian manifold to gain a deeper understanding of the singular points and their classification. Furthermore, we introduce eikonal coordinates. The results from this section are used in section~\ref{sec:maslov} to construct the semiclassical approximation in both regular and singular points. We discuss the Maslov canonical operator and its representation in the vicinity of singular points, paying particular attention to the Maslov index. In section~\ref{sec:causticsSolution}, we discuss how we can simplify the wavefunction in the vicinity of singular points using the leading-order approximation and the uniform approximation. In section~\ref{sec:examples}, we discuss the numerical implementation of our semiclassical approximations. We consider several examples and show how the semiclassical phase affects the position and intensity of the focus. We also compare the semiclassical approximation with the results of tight-binding calculations for graphene. We present our conclusions and ideas for further research in section~\ref{sec:conclusion}. Readers who are mainly interested in the results for graphene are advised to read sections~\ref{sec:assumptions},~\ref{subsec:separate-Dirac} and~\ref{sec:semiclassics}, before having a look at the results in section~\ref{sec:examples}.

Finally, we would like to make a few notational remarks. Throughout this paper, the index $\alpha$ labels the two valleys in graphene. When we discuss graphene, we explicitly include this index in the notation. For instance, we denote the wavefunction as $\Psi_\alpha$. We suppress this index when we discuss general semiclassical methods. Hence, we generally suppress $\alpha$ in sections~\ref{sec:lagman},~\ref{sec:maslov} and~\ref{sec:causticsSolution}. 
Furthermore, the subscripts $\phi$, $\tau$ and $t$ typically indicate partial derivatives with respect to these variables, i.e., $X_\phi = \partial X/\partial \phi$.
By the inner product $\langle a, b\rangle$, we generally mean the conventional inner product of the $n$-dimensional vectors $a$ and $b$ in $\mathbb{R}^n$, i.e. $\langle a, b\rangle = \sum_j a_j b_j$. 
The only exception to this general rule can be found in the beginning of section~\ref{subsec:separation}, where we use the notation $\langle a, b\rangle_{L^2(\mathbb{R}^m)}$ to denote the standard inner product in the Hilbert space $L^2(\mathbb{R}^m)$. The Fourier transform and its inverse are defined in equation~(\ref{eq:fourier-transform}).

\section{Preliminary considerations} \label{sec:assumptions}

The dynamics of low energy charge carriers in graphene are governed by the two-dimensional Dirac equation~\cite{Wallace47,McClure57,Slonczewski58,Semenoff84,Novoselov05,Zhang05,CastroNeto09,Katsnelson13}. 
Although the Hamiltonians for charge carriers in the valleys $K$ and $K'$ differ slightly, we can consider both of them at the same time by studying the Hamiltonian
\begin{equation}  \label{eq:Ham-dim}
  \hat{H}_\alpha = v_F \sigma_x \hat{p}_1 - \alpha v_F \sigma_y \hat{p}_2 + m(x) \sigma_z + U(x) \mathbb{1} ,
\end{equation}
where $\alpha=-1$ for the $K$-valley and $\alpha=+1$ for the $K'$-valley. The quantities $\sigma_i$ are the Pauli matrices. Since our problem is two-dimensional, the position vector equals $x=(x_1,x_2)$ and the momentum operators equal $\hat{p}_j=-i \hbar \partial/\partial x_j$. When one studies graphene using the nearest-neighbor approximation, one finds that the Fermi velocity $v_F$ is determined by $\hbar v_F = \tfrac{3}{2} t a_{CC}$, where $t$ is the hopping parameter and $a_{CC}=0.142$~nm is the carbon-carbon distance in graphene~\cite{Katsnelson13}. In this paper, we use $t=3$~eV, which leads to a Fermi velocity that approximately equals $c/300$, with $c$ the speed of light.

We consider the scattering problem for this Hamiltonian, that is,
\begin{equation} \label{eq:eigenvalue-HPsi}
  \hat{H}_\alpha \Psi_\alpha = E \Psi_\alpha, \qquad  \Psi_\alpha = \begin{pmatrix} \psi_1 \\ \psi_2 \end{pmatrix},
\end{equation}
where $E$ is the energy of the electron. We assume that the potential $U(x)$ and the mass $m(x)$ are of the same order of magnitude as the energy $E$. Furthermore, we assume that there is a typical length scale $l$ that describes changes in both $U(x)$ and $m(x)$. These assumptions allow us to introduce dimensionless variables in the eigenvalue problem~(\ref{eq:eigenvalue-HPsi}). Let us denote the characteristic energy scale of the problem as $E_0$. Throughout this paper we use $E_0=E$, although one could also use alternative quantities such as $\max U(x)$ or $\min |U(x)-E|$. We can then define the dimensionless semiclassical parameter $h=\hbar v_F/(E_0 l)$ and the dimensionless quantities $\tilde{x} = x/l$, $\hat{\tilde{p}}=-i h \partial/\partial \tilde{x}$, $\tilde{E}=E/E_0$, $\tilde{U}(\tilde{x})=U(x)/E_0$ and $\tilde{m}(\tilde{x})=m(x)/E_0$.
From now on, we only consider these dimensionless variables, unless explicitly stated otherwise. We therefore omit the tildes in the notation. The Hamiltonian then reads:
\begin{equation} \label{eq:Dirac}
  \hat{H}_\alpha =
  \begin{pmatrix}
    U(x) + m(x) & \hat{p}_1 + i \alpha \hat{p}_2 \\
    \hat{p}_1 - i \alpha \hat{p}_2 & U(x) - m(x)
 \end{pmatrix}.
\end{equation}

In this paper, we consider scattering of a plane wave that is incident on a potential $U(x)$ and a mass $m(x)$. Without loss of generality, we study an electron with momentum $p^0 = (p_1^0, 0)$ that comes in from the left, i.e., from $x_1=-\infty$. We assume that both $U(x)$ and $m(x)$ are smooth and localized in a finite domain $D$, in the sense that they are constant outside of $D$. 
We limit ourselves to above-barrier scattering, which means that the potential and mass are chosen in such a way that there are no classically forbidden regions. 
In section~\ref{sec:semiclassics}, we show that this assumption requires that 
\begin{equation} \label{eq:above-assump}
  (U(x)-E)^2 - m(x)^2 > 0 
\end{equation}
for all points $x$. In particular, this means that we do not consider (Klein) tunneling, as discussed in the introduction.
Finally, we consider a setup in which all trajectories of the classical Hamiltonian system corresponding to the Hamiltonian~(\ref{eq:Dirac}) run away to infinity~\cite{Kucherenko69,Vainberg89}. More precisely, every trajectory leaves any closed and bounded set in a finite time. This means that there are no trapped trajectories, which is very important for the construction of the asymptotic solution later on.

Far outside of the domain $D$, the solution $\Psi_\alpha(x)$ can be written as an incoming plane wave plus a scattered wave:
\begin{equation}  \label{eq:form-inc-solution}
  \Psi_\alpha(x) = A^0 e^{i \langle p^0, x \rangle/h} + \Psi_{\text{scat},\alpha}(x) .
\end{equation}
where $A^0$ is the amplitude of the incoming wave. In order to properly define the scattering problem, we require that $\Psi_{\text{scat},\alpha}$ satisfies the Sommerfeld radiation conditions at infinity~\cite{Sommerfeld49,Schot92,Vainberg89}:
\begin{equation} \label{eq:Sommerfeld}
  \lim_{|x|\rightarrow \infty} |x|^{1/2} \left( - i h \frac{\partial}{\partial |x|} - |p^0| \right)\Psi_{\text{scat},\alpha} = 0 .
\end{equation}
In words, this condition states that $\Psi_{\text{scat},\alpha}$ only consists of outgoing waves. Hence, the only incoming wave in our problem is the wave that comes in from the left and there are no waves that come in from other sides.
Although we formally impose this condition, we do not use it in the rest of the paper, since the constructions in the following sections automatically ensure that it is fulfilled.

Now that we have stated the scattering problem, let us discuss its symmetries. Our electron comes in from the left, with an amplitude $A^0$ that was defined in equation~(\ref{eq:form-inc-solution}) and which can in principle depend on $x_2$.
Let us assume that this amplitude is symmetric in $x_2$ and that the potential $U(x)$ and the mass $m(x)$ are symmetric in $x_2$ as well, i.e. $U(x_1,x_2)=U(x_1,-x_2)$.
Subsequently, consider the eigenvalue equation~(\ref{eq:eigenvalue-HPsi}) for the Hamiltonian~(\ref{eq:Dirac}): $\hat{H}_\alpha \Psi_\alpha(x_1,x_2) = E \Psi_\alpha(x_1,x_2)$. When we replace $x_2$ by $-x_2$ and use the symmetries that we just imposed, we see that we arrive at the equation $\hat{H}_{-\alpha} \Psi_\alpha(x_1,-x_2) = E \Psi_\alpha(x_1,-x_2)$. Since all boundary conditions are symmetric in $x_2$, this means that $\Psi_\alpha(x_1,-x_2)$ is an eigenfunction of $\hat{H}_{-\alpha}$ with energy $E$. Since the solution is unique, this in turn means that
\begin{equation} \label{eq:symm-between-valleys}
  \Psi_{\alpha}(x_1,x_2) = \Psi_{-\alpha}(x_1,-x_2) .
\end{equation}
This first symmetry thus connects the solutions for electrons in the two valleys.

Under the same assumptions, i.e. that the potential, mass and initial amplitude are symmetric in $x_2$, we can derive a second symmetry. 
Replacing $x_2$ by $-x_2$ in the eigenvalue equation and multiplying by $\sigma_x$, we arrive at 
\begin{equation}
  (\sigma_x \hat{H}_{-\alpha} \sigma_x) [\sigma_x \Psi_\alpha(x_1,-x_2)] = E [ \sigma_x \Psi_\alpha(x_1,-x_2)].
\end{equation}
Subsequently, we note that the expression $\sigma_x \hat{H}_{-\alpha} \sigma_x$ equals the Hamiltonian $\hat{H}_\alpha$ when we replace $m(x)$ by $-m(x)$. Therefore, reversing the sign of the mass results in a reflection of the wavefunction in the $x_1$-axis, i.e., 
\begin{equation} \label{eq:symm-reverse-mass}
  \Psi_{\alpha,m}(x_1,x_2) = \sigma_x \Psi_{\alpha,-m}(x_1,-x_2) ,
\end{equation}
where we have included the mass $m$ in the notation. This equality is especially important when the mass $m(x)$ is identically zero, in which case it reads
\begin{equation} \label{eq:symm-zero-mass}
  \Psi_{\alpha}(x_1,x_2) = \sigma_x \Psi_{\alpha}(x_1,-x_2) .
\end{equation}
When we define the norm of the wavefunction by $\lVert \Psi_\alpha \rVert = \sqrt{\Psi_\alpha^\dagger \Psi_\alpha}$, we observe that $\lVert \Psi_{\alpha}(x_1,x_2) \rVert = \lVert \Psi_{\alpha}(x_1,-x_2) \rVert$. The second symmetry thus states that when the mass vanishes, the intensity is symmetric in $x_2$. Note that this only holds when both the potential and the initial amplitude are symmetric in $x_2$.

We can combine the symmetries~(\ref{eq:symm-between-valleys}) and~(\ref{eq:symm-zero-mass}) to obtain a third symmetry. When the mass vanishes and when the potential and the initial amplitude are symmetric in $x_2$, we obtain
\begin{equation} \label{eq:symm-between-valleys-zero-mass}
  \Psi_{\alpha}(x_1,x_2) = \Psi_{-\alpha}(x_1,-x_2) = \sigma_x \Psi_{-\alpha}(x_1,x_2) .
\end{equation}
Therefore, $\lVert \Psi_{K}(x_1,x_2) \rVert = \lVert \Psi_{K'}(x_1,x_2) \rVert$, which means that the norm of the wavefunction is equal for both valleys. Hence, there is no symmetry breaking between the two valleys in the absence of a mass $m(x)$.

\section{Adiabatic reduction to scalar equations} \label{sec:adiabatic}

The Hamiltonian~(\ref{eq:Dirac}) simultaneously describes both electron and hole states. However, since we look at above-barrier scattering, we do not need to consider transitions from electron to hole states. Therefore, our first step towards the construction of an asymptotic solution to the eigenvalue equation~(\ref{eq:eigenvalue-HPsi}) consists of reducing the matrix Hamiltonian~(\ref{eq:Dirac}) to two separate scalar Hamiltonians, one for electrons and one for holes. In section~\ref{subsec:separation}, we review how this reduction can be performed order by order in the dimensionless semiclassical parameter $h$, based on~\cite{Berlyand87,Belov06,Littlejohn91}. In this reduction, we mainly make use of the symbols of the quantum operators, which are much easier to manipulate. In section~\ref{subsec:pd-op}, we therefore briefly review the relation between pseudodifferential operators and their symbols. Our exposition is mainly based on Ref.~\cite{Martinez02}, but also draws inspiration from Refs.~\cite{Zworski12,Maslov81}. For a complete account of pseudodifferential operators, we refer the interested reader to the books by H\"ormander~\cite{HormanderBooks} and Ivrii~\cite{Ivrii98}, noting that the former also includes many historical remarks.
Using the theory developed in the first two subsections, we perform the reduction for the Dirac Hamiltonian in section~\ref{subsec:separate-Dirac}.

\subsection{Pseudodifferential operators and symbols} \label{subsec:pd-op}

The goal of this subsection is to review the correspondence between operators $\hat{f}$ and functions $f(x,p)$ of the variables $x$ and $p$, representing position and momentum, respectively. 
The function $f(x,p)$ may be thought of as a classical observable on phase space and is called a symbol. 
Given such a symbol $f(x,p)$, we define an operator $\text{Op}_t(f)$ by specifying how it acts on a function $u(x)$. Specifically, we define the $t$-quantization of $f(x,p)$ as~\cite{Martinez02}
\begin{equation} \label{eq:quantization-t-symbol}
  \text{Op}_t(f) u(x) = \frac{1}{(2\pi h)^n} \int e^{i \langle p , x-y \rangle/h} f\big( (1-t)x + t y, p\big) u(y) \text{d}y \text{d}p ,
\end{equation}
where $\langle a , b \rangle = \sum_i a_i b_i$ denotes the standard inner product, $n$ is the dimensionality of space and $h$ is the dimensionless semiclassical parameter.
The operator $\text{Op}_t(f)$ is also called the semiclassical pseudodifferential operator with symbol $f$ and depends on $t$. For example, a straightforward calculation shows that for $f(x,p) = \langle x, p\rangle$, one has $\text{Op}_0(\langle x, p\rangle) u(x) = -i h \langle x , \partial u(x)/\partial x \rangle$, meaning that $\text{Op}_0(\langle x, p\rangle) = \langle x, \hat{p} \rangle$. On the other hand, one has $\text{Op}_{1/2}(\langle x, p\rangle) = \tfrac{1}{2} (\langle x, \hat{p}\rangle + \langle \hat{p}, x \rangle)$, which is symmetric.

As a more general example, one can consider $f(x,p)= \sum_\beta f_\beta(x) p^\beta$, where $\beta=(\beta_1,\ldots,\beta_n)$ is a multi-index, and $p^\beta = \prod_i p_i^{\beta_i}$.
By a straightforward calculation, one sees that its zero-quantization equals the (semiclassical) differential operator $\text{Op}_{0}(f) = \sum_\beta f_\beta(x) \hat{p}^\beta$. 
However, the application of equation~(\ref{eq:quantization-t-symbol}) is not limited to symbols $f$ that are polynomials in $p$. 
Symbols can have much more complicated functional forms and may also explicitly depend on $h$.
In general, the quantization of such symbols will not give rise to usual (semiclassical) differential operators. Instead, their action on a function $u(x)$ is more complicated, hence the name semiclassical pseudodifferential operators.

In order for equation~(\ref{eq:quantization-t-symbol}) to make sense, one should impose certain constraints on the symbol~$f(x,p)$. Different authors impose slightly different constraints, leading to different classes of pseudodifferential operators. The difference between these classes is not that important for the purpose of this paper, but plays a role when one needs to make precise estimates.
Martinez~\cite{Martinez02} defines a class $S_{2n}((1 + |p|^{2})^{m/2})$, where $2n$ is the dimensionality of phase space and $m$ is called the degree of the associated pseudodifferential operators. A symbol $f(x,p)$ is in this class when it depends smoothly on $x$ and $p$ and $\partial^\beta/\partial x^\beta \, \partial^\gamma/\partial p^\gamma f(x,p) = \mathcal{O}((1 + |p|^{2})^{m/2})$ for any multi-indices $\beta$ and $\gamma$, uniformly in $x$, $p$ and $h$ for $h$ sufficiently small. The latter condition can also be stated as
\begin{equation} \label{eq:symbol-condition-Martinez}
  \left| \frac{\partial^\beta}{\partial x^\beta} \frac{\partial^\gamma}{\partial p^\gamma} f(x,p) \right| \leq \mathcal{C}_{\beta\gamma} (1 + |p|^{2})^{m/2} 
\end{equation}
for a certain constant $\mathcal{C}_{\beta\gamma}$ that is independent of $h$ and the specific point $(x,p)$ that is considered. In words, condition~(\ref{eq:symbol-condition-Martinez}) means that the symbols should not diverge and that their growth at infinity should be bounded by a polynomial in $p$. Furthermore, their growth rate should not be increased when one takes an arbitrary amount of derivatives with respect to either $p$ or $x$. It can be shown~\cite{Martinez02} that for symbols in class~$S_{2n}((1 + |p|^{2})^{m/2})$ there is a unique way to extend the operator $\text{Op}_t(f)$ to a linear continuous operator on Schwartz space.

Instead, Maslov~\cite{Maslov81} only considers $t=0$ and $t=1$ and defines a class $T^m$ in which the variables $x$ and $p$ are treated on equal footing. A symbol $f(x,p)$ that does not depend on $h$ belongs to this class when it is continuous in $x$ and $p$ and when 
\begin{equation} \label{eq:symbol-condition-Maslov}
  \left| \frac{\partial^\beta}{\partial x^\beta} \frac{\partial^\gamma}{\partial p^\gamma} f(x,p) \right| \leq \mathcal{C}_{\beta\gamma} (1 + |x|)^{m} (1 + |p|)^{m} 
\end{equation}
for any multi-indices $\beta$ and $\gamma$. Subsequently, Maslov~\cite{Maslov81} defines a class $T_+^m$ for symbols that depend on $h$. A symbol belongs to this class when its dependence on $x$, $p$ and $h$ is smooth; the symbol can be expanded in a power series in $h$; each of the expansion coefficients is in $T^m$ and an additional constraint on the remainder is satisfied. Other classes of symbols are considered by Zworski~\cite{Zworski12} and H\"ormander~\cite{HormanderBooks}.

Looking at the two-dimensional Dirac Hamiltonian~(\ref{eq:Dirac}), we observe that it is linear in momentum. It turns out that all symbols that we use do not grow faster than $|p|$ at infinity, together with all their derivatives. Furthermore, the potentials $U(x)$ and masses $m(x)$ that we consider are bounded, as are all their derivatives. Therefore, the class $S_{4}((1 + |p|^{2})^{1/2})$ is sufficient for this paper and we do not need to consider wider classes. However, we should keep in mind that the Hamiltonian~(\ref{eq:Dirac}) is a matrix. Matrix valued symbols are explicitly considered by Maslov~\cite{Maslov81}, who defines a class $T_+^m$ for matrix symbols. In his definition, a matrix valued symbol belongs to class $T_+^m$ if all its elements belong to $T_+^m$. Although a similar extension for the class $S_{2n}((1 + |p|^{2})^{m/2})$ is not explicitly discussed by Martinez~\cite{Martinez02}, we believe that this does not pose any fundamental problems. Alternatively, one could think about replacing the absolute value in equation~(\ref{eq:symbol-condition-Martinez}) by an appropriate matrix norm.

Instead of viewing equation~(\ref{eq:quantization-t-symbol}) as a quantization procedure, we can also look at it the other way around: given an operator $\hat{a}$, equation~(\ref{eq:quantization-t-symbol}) defines a unique symbol of index $t$~\cite{Martinez02}. We denote this symbol by $a^{(t)}=\sigma_t(\hat{a})$ and we naturally have $\hat{a} = \text{Op}_t(a^{(t)})$.
It is this point of view that we predominantly take in this paper, since we start with a quantum Hamiltonian $\hat{H}$ and we want to construct its symbol.
We can obtain the zero-symbol $a^{(0)}$ of an arbitrary operator $\hat{a}$ by computing~\cite{Martinez02}
\begin{equation} \label{eq:symbol-zero}
  a^{(0)}(x, p, h) = \sigma_0(\hat{a}) = e^{-i \langle p, x \rangle/h} ( \hat{a} e^{i \langle p, x \rangle/h} ) .
\end{equation}
Subsequently, we can find the $t'$-symbol from the $t$-symbol using the formula
\begin{equation} \label{eq:symbol-change-quant}
  a^{(t')}(x, p, h) = \exp\left(i h (t'-t) \left\langle \frac{\partial}{\partial x} , \frac{\partial}{\partial p} \right\rangle \right) a^{(t)}(x, p, h) .
\end{equation}
For example, for $\hat{a} = \tfrac{1}{2} (\langle x, \hat{p}\rangle + \langle \hat{p}, x \rangle)$ one has $a^{(0)} = \langle x, p \rangle - i n h/2$ and $a^{(1/2)} = \langle x, p \rangle$. The latter result is of course in agreement with the example given at the beginning of this subsection. The symbol $a^{(t)}(x, p, h)$ that one obtains from an operator $\hat{a}$ usually depends on $h$, as illustrated by the first example. In this paper, we only consider classical symbols, which are symbols that are equivalent to a formal power series in $h$ as $h\to 0$~\cite{Martinez02}.
With a slight abuse of notation, we denote this correspondence by an equality sign, i.e. we write 
\begin{equation} \label{eq:symbol-expansion}
  a^{(t)}(x, p, h)=\sum_j a^{(t)}_j(x, p) h^j .
\end{equation}
The zeroth-order term $a_0^{(t)}$ of this expansion is known as the principal symbol~\cite{Martinez02, Maslov81} and is independent of $t$, which can for instance be seen from equation~(\ref{eq:symbol-change-quant}). Since the principal symbol is independent of the quantization, one can really think of $a_0^{(t)}$ as a classical observable on phase space.

In this paper, we consider two specfic values of $t$. When $t=0$, we are dealing with the so-called standard quantization. For this case, we denote the symbol as $\sigma_0(\hat{a}) = a(x,p,h)$ and the operator as $\hat{a}=a(x, \hat{p}, h)$.
In many calculations standard quantization is extremely convenient, since the relation between the symbol and the operator can be expressed using Fourier transforms~\cite{Martinez02,Maslov81,Maslov73,Zworski12}.
Specifically, one can write
\begin{equation} \label{eq:standard-quantization}
  \hat{a} \, u(x) = \text{Op}_0(a) u(x) = a(x, \hat{p}, h) u(x)  
          = \mathcal{F}^{-1}_{p \to x} a(x,p,h) \mathcal{F}_{y \to p} u(y) ,
\end{equation}
where the $n$-dimensional Fourier transform and its inverse are defined by
\begin{equation}
\begin{aligned} \label{eq:fourier-transform}
  \overline{u}(p) &= \mathcal{F}_{x \to p} u(x) = \frac{e^{-i n \pi/4}}{(2 \pi h)^{n/2}} \int e^{- i \langle p, x \rangle/h} u(x) \, \text{d}x , \\
  u(x) &= \mathcal{F}^{-1}_{p \to x} \overline{u}(p) = \frac{e^{i n \pi/4}}{(2 \pi h)^{n/2}} \int e^{i \langle p, x \rangle/h} \overline{u}(p) \, \text{d}p .
\end{aligned}
\end{equation}
At the beginning of this section, we already saw that $\text{Op}_0(\langle x, p\rangle) = \langle x, \hat{p} \rangle$. This example points to an important feature of the standard quantization: it is the quantization that results when one lets the momentum operator $\hat{p}$ act first and the position (multiplication) operator $x$ act second~\cite{Maslov81,Maslov73}. Because of this property, the notation $a(\stackrel{2}{x}, \stackrel{1}{\hat{p}}, h )$ is also used, in which the order of the operators is shown explicitly. Standard quantization is sometimes called left quantization~\cite{Martinez02} and the resulting operator ordering is sometimes called the Feynman-Maslov ordering~\cite{Maslov81,Maslov73}. The operator calculus that results from this ordering is sometimes called the Kohn-Nirenberg calculus~\cite{Kohn65}.
Although we do not consider $t=1$ in this paper, we remark that in this quantization the order of the operators is reversed: the position (multiplication) operator $x$ acts first and the momentum operator $\hat{p}$ acts second~\cite{Maslov81,Maslov73}.

When $t=\frac{1}{2}$, the quantization procedure is called Weyl quantization~\cite{Weyl27}. In this case, we denote the symbol as $\sigma_{1/2}(\hat{a}) = a^W(x, p, h)$ and the operator as $\hat{a}=a^W(x, \hat{p}, h)$. Therefore, one has
\begin{equation} \label{eq:weyl-quantization}
\begin{aligned} 
  \hat{a} \, u(x) &= \text{Op}_{1/2}(a) u(x) = a^W(x, \hat{p}, h) u(x) \\
                  &= \frac{1}{(2\pi h)^2} \int e^{i \langle p, x-y \rangle/h} a\left( \frac{x+y}{2}, p, h\right) u(y) \text{d}y \text{d}p .
\end{aligned}
\end{equation}
At the beginning of this section, we already saw that Weyl-quantizing the symbol $\langle x, p\rangle$ leads to the symmetric operator $\text{Op}_{1/2}(\langle x, p\rangle) = \tfrac{1}{2} (\langle x, \hat{p}\rangle + \langle \hat{p}, x \rangle)$. This points to an important feature of Weyl quantization: when an operator $\hat{a}$ whose symbol is a scalar function is self-adjoint, then its Weyl symbol $a^W = \sigma_{1/2}(\hat{a})$ is real~\cite{Martinez02,Zworski12}. Although Martinez~\cite{Martinez02} does not explicitly consider operators with matrix valued symbols, the results can be easily generalized to accomodate them. We start from the identity $\sigma_{1-t}(\hat{a}^\dagger) = [ \sigma_t(\hat{a}) ]^\dagger$, the scalar version of which can be found in Ref.~\cite{Martinez02}. In this equality, $\hat{a}^\dagger$ denotes the adjoint of $\hat{a}$ and the dagger on the right-hand side denotes complex conjugation and transposition. For a self-adjoint operator $\hat{a}$, we then have $\sigma_{1/2}(\hat{a})=\sigma_{1/2}(\hat{a}^\dagger) = [ \sigma_{1/2}(\hat{a}) ]^\dagger$. Thus, the Weyl symbol of a self-adjoint operator is a Hermitian matrix. In particular, the Weyl symbol $\sigma_{1/2}(\hat{a})$ is real when the symbol of $\hat{a}$ is a scalar function.

When we consider standard quantization, the relation between the symbol of an operator and the symbol of its adjoint is somewhat more complicated. From the aforementioned relation, we obtain $\sigma_{0}(\hat{a}^\dagger) = [ \sigma_1(\hat{a}) ]^\dagger$. Subsequently, we can express $\sigma_1(\hat{a})$ in terms of $\sigma_0(\hat{a})$ using equation~(\ref{eq:symbol-change-quant}), leading to~\cite{Martinez02}
\begin{equation} \label{eq:adjoint-symbol-relation}
\begin{aligned}
  \sigma_0(\hat{a}^\dagger)(x,p,h) &= \exp \left( -i h \left\langle \frac{\partial}{\partial x} , \frac{\partial}{\partial p} \right\rangle \right) [\sigma_0(\hat{a})(x,p,h)]^\dagger \\
    &= \sum_\beta \frac{h^{|\beta|}}{i^{|\beta|} \beta!} \left( \frac{\partial^\beta}{\partial x^\beta} \frac{\partial^\beta}{\partial p^\beta} [\sigma_0(\hat{a})(x,p,h)]^\dagger \right) ,
\end{aligned}
\end{equation}
where $\beta$ is a multi-index, $\beta!=\prod_i \beta_i!$ and $|\beta|=\sum_i \beta_i$. Let us now consider a self-adjoint operator $\hat{a}$ that has a classical symbol, i.e. a symbol that can be expanded in a power series as in equation~(\ref{eq:symbol-expansion}). Entering the series expansion into the right-hand side of equation~(\ref{eq:adjoint-symbol-relation}) and demanding that it equals the original power series, we obtain conditions on the expansion coefficients $a_j$. Collecting terms of order $h^0$, we find that $a_0(x,p)$ is real. Since the principal symbol of an operator does not depend on the quantization, as we discussed before, $a_0(x,p)$ is in fact equal to the principal Weyl symbol $a_0^W(x,p)$.
When we collect the terms of order $h^1$ in equation~(\ref{eq:adjoint-symbol-relation}), we obtain a condition on the so-called subprincipal symbol $a_1(x,p)$ of the self-adjoint operator $\hat{a}$, namely
\begin{equation} \label{eq:self-adjoint-relation-a1}
  \frac{1}{2 i} \big( a_1(x,p) - a_1^\dagger(x,p) \big) = -\frac{1}{2} \left\langle \frac{\partial}{\partial x} , \frac{\partial}{\partial p} \right\rangle a_0(x,p) .
\end{equation}
Note that when $a_1(x,p)$ is a scalar function, the left-hand side equals its imaginary part. Using this equality, we can obtain an expression for the subprincipal Weyl symbol $a_1^W$ in terms of the subprincipal symbol $a_1$. When we construct the asymptotic expansion of equation~(\ref{eq:symbol-change-quant}) for $t'=\tfrac{1}{2}$ and $t=0$, and gather the terms of order $h^1$, we arrive at
\begin{equation} \label{eq:self-adjoint-Weyl-a1}
  a_1^W(x,p) = a_1(x,p) + \frac{i}{2} \left\langle \frac{\partial}{\partial x} , \frac{\partial}{\partial p} \right\rangle a_0(x,p) 
    = \frac{1}{2} \big( a_1(x,p) + a_1^\dagger(x,p) \big) ,
\end{equation}
where the last equality follows from equation~(\ref{eq:self-adjoint-relation-a1}). This relation shows explicitly that the subprincipal Weyl symbol $a_1^W$ is Hermitian, as we already proved in general. In particular, when the subprincipal symbols are scalar functions, $a_1^W$ is simply the real part of $a_1$. 
We can obtain similar conditions on the higher-order expansion coefficients $a_n$ by collecting terms of higher orders in $h$ in equation~(\ref{eq:adjoint-symbol-relation}).

One can show that the product of two pseudodifferential operators is again a pseudodifferential operator~\cite{Martinez02}. In particular, one can express the $t$-symbol of $\hat{a}\hat{b}$ in terms of the $t$-symbols $a^{(t)}$ and $b^{(t)}$ of the operators $\hat{a}$ and $\hat{b}$, respectively. 
When we consider standard quantization, i.e. $t=0$, the symbol of $\hat{a}\hat{b}$ is given by~\cite{Martinez02}
\begin{equation}
\begin{aligned} \label{eq:standard-symbol-product}
  \sigma_{0}( \hat{a} \hat{b} ) 
  &= \left. \exp\left( - i h \left\langle \frac{\partial}{\partial q} , \frac{\partial}{\partial y} \right\rangle \right)  a(x, q, h) b(y, p, h) \right|_{\substack{y=x \\ q=p}} \\
  &= \sum_{\beta} \frac{h^{|\beta|}}{i^{|\beta|}\beta!} 
       \left( \frac{\partial^\beta}{\partial p^\beta} a(x, p, h) \right) \left( \frac{\partial^\beta}{\partial x^\beta}  b(x, p, h) \right) ,
\end{aligned}
\end{equation}
where $\beta$ is again a multi-index. We make extensive use of this formula in the next subsections.
When we consider Weyl quantization, we can express the symbol of the product as~\cite{Martinez02}
\begin{equation}
\begin{aligned} \label{eq:weyl-symbol-product}
  \sigma_{1/2}( \hat{a} \hat{b} ) 
  &= \left. \exp\left( - \frac{i h}{2} \left[ \left\langle \frac{\partial}{\partial q} , \frac{\partial}{\partial x} \right\rangle - \left\langle \frac{\partial}{\partial y} , \frac{\partial}{\partial p} \right\rangle \right] \right) 
       a^W(y, q, h) b^W(x, p, h) \right|_{\substack{y=x \\ q=p}} \\
  &= \sum_{\beta, \gamma} \frac{h^{|\beta+\gamma|}(-1)^{|\beta|}}{(2i)^{|\beta+\gamma|}\beta!\gamma!} 
       \left( \frac{\partial^\beta}{\partial x^\beta}  \frac{\partial^\gamma}{\partial p^\gamma} a^W(x, p, h) \right) 
       \left( \frac{\partial^\beta}{\partial p^\beta}  \frac{\partial^\gamma}{\partial x^\gamma} b^W(x, p, h) \right) ,
\end{aligned}
\end{equation}
where $a^W(x,p,h)$ and $b^W(x,p,h)$ are the Weyl symbols of the operators $\hat{a}$ and $\hat{b}$, respectively, and $\beta$ and $\gamma$ are multi-indices. One may think of the previous two formulas as ways to define a product on the space of symbols. For Weyl symbols, this product, first discovered by Groenewold~\cite{Groenewold46}, is known as the Moyal product~\cite{Moyal49}. Generally, such products are known as star products, and denoted with a star, e.g. $\sigma_{1/2}( \hat{a} \hat{b} ) \equiv a^W \star \, b^W$.

Finally, we briefly discuss the relation between the commutator and the Poisson bracket. Denoting the commutator of $\hat{a}$ and $\hat{b}$ by $[ \hat{a}, \hat{b} ] = \hat{a}\hat{b} - \hat{b}\hat{a}$, one can show that for all $t$~\cite{Martinez02}
\begin{equation} \label{eq:commutator-Poisson-bracket}
  \sigma_t([ \hat{a}, \hat{b} ]) = i h \{ a^{(t)}, b^{(t)}\} + \mathcal{O}(h^2) ,
\end{equation}
where the Poisson bracket $\{ a, b \}$ is defined by
\begin{equation}  \label{eq:def-Poisson-bracket}
  \{ a, b \} = \left\langle \frac{\partial a}{\partial x} , \frac{\partial b}{\partial p} \right\rangle - \left\langle \frac{\partial a}{\partial p} , \frac{\partial b}{\partial x} \right\rangle.
\end{equation}
When we consider Weyl symbols, the terms of order $h^2$ in equation~(\ref{eq:commutator-Poisson-bracket}) cancel~\cite{Groenewold46,Zworski12} and one has $\sigma_{1/2}([ \hat{a}, \hat{b} ]) = i h \{ a^W, b^W\} + \mathcal{O}(h^3)$.
These equalities show the intimate relation between the quantum commutator and the classical Poisson bracket.

\subsection{Operator separation of variables}  \label{subsec:separation}

Now that we have reviewed the basic properties of pseudodifferential operators, we discuss how we can decouple the different modes that are comprised within a matrix Hamiltonian. This mode decoupling is possible when we do not consider transitions between the different modes. 
In our present context, this condition means that we do not consider transitions between electrons and holes, in agreement with the assumptions that we made in section~\ref{sec:assumptions}. We use the scheme devised in Refs.~\cite{Berlyand87,Belov06}, see also Ref.~\cite{Karasev90}, which finds its origin in the ideas of the Foldy-Wouthuysen transformation~\cite{Foldy50,Bjorken64,Bruening11}. For simplicity, we confine ourselves to the case where all the modes are scalar, although this is not a fundamental limitation of the method. The same method was formulated, independently, in Refs.~\cite{Littlejohn91,Littlejohn91b}. Within this scheme, the mode separation can be performed to any order in $h \ll 1$. 

We consider the eigenvalue problem $\hat{H} \Psi = E \Psi$, where $\hat{H}$ is an $n \times n$ matrix and $\Psi$ is an $n$-dimensional vector.
The first step of the mode decoupling is to look for a solution of this equation of the form 
\begin{equation} \label{eq:Psi-decomp}
  \Psi(x) = \hat{\chi} \, \psi(x) ,
\end{equation}
where $\psi$ is an effective scalar wavefunction that corresponds to a single mode. The operator $\hat{\chi}$ reconstructs the $n$-component wavefunction $\Psi$ of the full eigenvalue problem from the scalar wavefunction $\psi$. Denoting the number of components of the vector $x$ by $m$, we can say that $\hat{\chi}$
maps an element of the Hilbert space $L^2(\mathbb{R}^m)$ to an element of the Hilbert space $\bigoplus_n L^2(\mathbb{R}^m)$. We require that this operator is norm-preserving, i.e.
\begin{equation} \label{eq:chi-norm-preserving}
  \langle \Psi , \Psi \rangle_{[\bigoplus_n L^2(\mathbb{R}^m)]} = \langle \hat{\chi} \psi , \hat{\chi} \psi \rangle_{[\bigoplus_n L^2(\mathbb{R}^m)]}
    = \langle \psi , \psi \rangle_{L^2(\mathbb{R}^m)} ,
\end{equation}
where $\langle f , g \rangle_{L^2(\mathbb{R}^m)}$ denotes the standard inner product in $L^2(\mathbb{R}^m)$. Using the definition of the adjoint operator, we see that equation~(\ref{eq:chi-norm-preserving}) is equivalent to the condition $\hat{\chi}^\dagger \hat{\chi} = 1$.

We subsequently demand that the scalar wavefunction $\psi(x)$ satisfies the effective scalar eigenvalue equation
\begin{equation} \label{eq:L-eff}
  \hat{L} \psi = E \psi ,
\end{equation}
where $\hat{L}$ plays the role of the scalar Hamiltonian.
Combining equations~(\ref{eq:Psi-decomp}) and~(\ref{eq:L-eff}) with the original equation $\hat{H}\Psi=E\Psi$, we obtain $( \hat{H}\hat{\chi} - \hat{\chi}\hat{L} ) \psi = 0$. This equation is certainly satisfied when the operator equality
\begin{equation}  \label{eq:defLChiOps}
  \hat{H}\hat{\chi} - \hat{\chi}\hat{L}=0
\end{equation}
is satisfied. Before we construct a solution to this equation, let us take a closer look at the operator $\hat{L}$.
Since it is our effective Hamiltonian, we would like it to be self-adjoint. Using equation~(\ref{eq:defLChiOps}) and the fact that $\hat{H}$ is self-adjoint, we can show that $\hat{L}$ is symmetric:
\begin{multline}
  \!\!\!
  \langle \phi , \hat{L} \psi \rangle_{L^2(\mathbb{R}^m)}
    = \langle \phi , \hat{\chi}^\dagger \hat{\chi} \hat{L} \psi \rangle_{L^2(\mathbb{R}^m)}
    = \langle \phi , \hat{\chi}^\dagger \hat{H} \hat{\chi} \psi \rangle_{L^2(\mathbb{R}^m)}
    = \langle \phi , \hat{\chi}^\dagger \hat{H}^\dagger \hat{\chi} \psi \rangle_{L^2(\mathbb{R}^m)} \\
    = \langle \phi , \hat{L}^\dagger \hat{\chi}^\dagger \hat{\chi} \psi \rangle_{L^2(\mathbb{R}^m)}
    = \langle \phi , \hat{L}^\dagger \psi \rangle_{L^2(\mathbb{R}^m)}
    = \langle \hat{L} \phi , \psi \rangle_{L^2(\mathbb{R}^m)} 
\end{multline}
for two elements $\phi$, $\psi$ of $L^2(\mathbb{R}^m)$. It is then possible to show that $\hat{L}$ is self-adjoint~\cite{Hall13}.

We construct an asymptotic solution to equation~(\ref{eq:defLChiOps}) by passing to symbols. We assume that all of the operators have classical symbols, i.e. symbols that can be expanded in a power series in $h$. Let us first consider standard quantization. We can then obtain asymptotic expansions for the symbols of the products $\hat{H}\hat{\chi}$ and $\hat{\chi}\hat{L}$ using equation~(\ref{eq:standard-symbol-product}). Subsequently, we expand the (classical) symbols in powers of $h$, as in equation~(\ref{eq:symbol-expansion}). By gathering all terms of a given order in $h$ and demanding that their sum vanishes, we can then construct an asymptotic solution to equation~(\ref{eq:defLChiOps}).
Collecting all terms of order $h^0$, we have
\begin{equation} \label{eq:L0}
  H_0(x,p) \chi_0(x,p) = L_0(x,p) \chi_0(x,p) ,
\end{equation}
which means that the principal symbols $L_0$ and $\chi_0$ are the eigenvalues and eigenvectors, respectively, of the principal symbol of the matrix Hamiltonian $\hat{H}$.
Note that $H_0$ is an $n\times n$ matrix and $\chi_0$ is an $n$-dimensional vector.
The fact that the symbol $\chi_0$ depends on $p$ makes this scheme different from other adiabatic schemes that are generally employed. For an extensive discussion of this point, with many examples, we refer to Ref.~\cite{Belov06}. Furthermore, we remark that the principal symbols $L_0(x,p)$ and $\chi_0(x,p)$ will generally not be polynomials in $p$, even when the principal symbol $H_0(x,p)$ of the Hamiltonian is. Therefore, the operators $\hat{L}$ and $\hat{\chi}$ are actual pseudodifferential operators. 

Using equations~(\ref{eq:standard-symbol-product}) and~(\ref{eq:adjoint-symbol-relation}), we can also pass to symbols in the condition $\hat{\chi}^\dagger \hat{\chi} = 1$. Collecting the terms of order $h^0$, we obtain $\chi_0^\dagger(x,p) \chi_0(x,p) = 1$, where the dagger denotes transposition and complex conjugation of the $n$-dimensional vector $\chi_0(x,p,h)$.
Hence, the norm preserving condition dictates that the eigenvectors $\chi_0(x,p)$ are normalized.

When we collect the terms of order $h^1$ after passing to symbols in equation~(\ref{eq:defLChiOps}), we obtain
\begin{equation} \label{eq:LChiOrderOne}
  L_1 \chi_0  = (H_0 - L_0) \chi_1 + H_1 \chi_0
    - i \left\langle \frac{\partial H_0}{\partial p} , \frac{\partial \chi_0}{\partial x} \right\rangle 
    + i \left\langle \frac{\partial L_0}{\partial x} , \frac{\partial \chi_0}{\partial p} \right\rangle ,
\end{equation}
where $\langle a , b \rangle = \sum_i a_i b_i$ once again denotes the standard inner product on Euclidean space.
Multiplying this equation by $\chi_0^\dagger(x,p)$ from the left, we see that the first term on the right-hand side vanishes, and we obtain an equation for the subprincipal symbol $L_1$, namely
\begin{equation} \label{eq:L1}
  L_1 = \chi_0^\dagger H_1 \chi_0 
    - i \chi_0^\dagger \left\langle \frac{\partial H_0}{\partial p} , \frac{\partial \chi_0}{\partial x} \right\rangle 
    + i \chi_0^\dagger \left\langle \frac{\partial L_0}{\partial x} , \frac{\partial \chi_0}{\partial p} \right\rangle .
\end{equation}
The (scalar) subprincipal symbol $L_1$ that we obtain from this expression is generally complex. 
However, since the operator $\hat{L}$ is self-adjoint, its imaginary part satisfies equation~(\ref{eq:self-adjoint-relation-a1}) and we have
\begin{equation} \label{eq:ImL1}
  \text{Im} \, L_1(x,p) = -\frac{1}{2} \left\langle \frac{\partial}{\partial x} , \frac{\partial}{\partial p} \right\rangle L_0(x,p)
\end{equation}
Thus, the fact that $L_1$ is complex does not have any physical significance, but is purely an artefact of the standard quantization. 
We note that equality~(\ref{eq:ImL1}) can also be derived explicitly using equation~(\ref{eq:L0}) and the fact that $\hat{H}$ satisfies equation~(\ref{eq:self-adjoint-relation-a1}) since it is self-adjoint. Since this derivation provides a nice illustration of how we can manipulate symbols, we present it in appendix~\ref{app:subprincipal-symbol}.

Let us also consider Weyl quantization. Since principal symbols are independent of the specific quantization, see section~\ref{subsec:pd-op}, we once again obtain equation~(\ref{eq:L0}) when we pass to symbols in equation~(\ref{eq:defLChiOps}). Hence, we have $L_0^W = L_0$ and $\chi_0^W = \chi_0$ and we do not use the superscript $W$ for these quantities.
However, the subprincipal Weyl symbol $L_1^W$ is different from the subprincipal symbol $L_1$. When we pass to symbols in equation~(\ref{eq:defLChiOps}) using the product formula~(\ref{eq:weyl-symbol-product}), collect the terms of order $h$ and subsequently multiply by $\chi_0^\dagger$, we arrive at
\begin{equation} \label{eq:L1-Weyl}
  L_1^W = \chi_0^\dagger H_1^W \chi_0 + \frac{i}{2} \chi_0^\dagger \{ H_0, \chi_0 \} - \frac{i}{2} \chi_0^\dagger \{ \chi_0 , L_0 \} .
\end{equation}
In the previous subsection, we showed that when a self-adjoint operator has a scalar symbol, its Weyl symbol is purely real. Therefore, also the subprincipal Weyl symbol $L_1^W$ is real. Using equation~(\ref{eq:self-adjoint-Weyl-a1}), we see that
\begin{equation} \label{eq:L1-Weyl-relation-L1}
  L_1^W(x,p) = \text{Re} \, L_1(x,p) .
\end{equation}
In appendix~\ref{app:subprincipal-symbol}, we show this relation explicitly, using equation~(\ref{eq:L0}) and the fact that $\hat{H}$ satisfies equation~(\ref{eq:self-adjoint-relation-a1}).

Following Ref.~\cite{Littlejohn91}, we split the Weyl symbol~(\ref{eq:L1-Weyl}) into two parts and write $L_1^W = L_{1B}^W + L_{1A}^W$. The term $L_{1B}^W$ is called the Berry part and is given by
\begin{equation} \label{eq:L1WB}
  L_{1B}^W = -i \chi_0^\dagger \{ \chi_0 , L_0 \} .
\end{equation}
In section~\ref{sec:semiclassics}, we show how this part gives rise to the Berry phase of the wavefunction. The second term, $L_{1A}^W$, does not have a specific name. It can be written as
\begin{align} \label{eq:L1WA}
  L_{1A}^W &= \chi_0^\dagger H_1^W \chi_0 + \frac{i}{2} \chi_0^\dagger \{ H_0, \chi_0 \} + \frac{i}{2} \chi_0^\dagger \{ \chi_0 , L_0 \} , \\ 
           &= \chi_0^\dagger H_1^W \chi_0 - \frac{i}{2} \sum_{j,k} (H_{jk} - L_0 \delta_{jk}) \{ \chi_{0,j}^*, \chi_{0,k} \} ,
\end{align}
where the subscripts $j$ and $k$ denote vector components. The second form can be found in Ref.~\cite{Littlejohn91}, and can be obtained with the help of equation~(\ref{eq:L0}). Note that both $L_{1B}^W$ and $L_{1A}^W$ are purely real. For the former this is easy to show, as
\begin{equation} \label{eq:L1WB-Im-zero}
  2i \text{Im}\, L_{1B}^W = -i \big( \chi_0^\dagger \{ \chi_0 , L_0 \} + \{ \chi_0^\dagger , L_0 \} \chi_0 \big) 
                          = -i \{ \chi_0^\dagger \chi_0 , L_0 \} = 0 ,
\end{equation}
where the second equality follows from the properties of the Poisson bracket, and the third equality follows from $\chi_0^\dagger \chi_0 = 1$. In a similar way, one can show that $L_{1A}^W$ is real. Alternatively, it follows from the fact that both $L_1^W$ and $L_{1B}^W$ are real.

Although we have been consistently calling the operator $\hat{L}$ the effective scalar Hamiltonian, this term is not entirely adequate. As noted in Ref.~\cite{Littlejohn91}, there is a certain gauge freedom in the choice of $\chi_0$, which affects the subprincipal symbol $L_1^W$. To clarify what this means, suppose that $\chi_0$ is a normalized eigenvector of $H_0$, which satisfies equation~(\ref{eq:L0}). Then the vector
\begin{equation}  \label{eq:chi0-gauge-transformation}
  \widetilde{\chi}_0(x,p) = e^{i g(x,p)} \chi_0(x,p) ,
\end{equation}
where $g(x,p)$ is a smooth scalar function, is also a normalized eigenvector of $H_0$, for the same eigenvalue $L_0$. Hence the principal symbol $L_0$ is not affected by the gauge freedom. However, let us compute the influence of this transformation on two terms that make up $L_1^W$. Inserting the new eigenvector~(\ref{eq:chi0-gauge-transformation}) into equation~(\ref{eq:L1WB}), we find that
\begin{equation} \label{eq:L1-gauge-transformation-effect}
  \widetilde{L}_{1B}^W
    = -i \widetilde{\chi}_0^\dagger \{ \widetilde{\chi}_0 , L_0 \}
    = -i \chi_0^\dagger \{ \chi_0 , L_0 \} + \{ g, L_0 \} 
    = L_{1B}^W + \{ g, L_0 \} .
\end{equation}
Therefore, $L_{1B}^W$ is not gauge invariant. On the contrary, the term $L_{1A}^W$ is gauge invariant, which can be shown with a somewhat more elaborate computation. Hence, the subprincipal Weyl symbol $L_1^W$ is not gauge invariant and depends on the choice of the eigenvectors $\chi_0$. We return to this point in section~\ref{subsec:maslovintro}, where we show that this gauge invariance does not affect the final result for the wavefunction. For an elaborate discussion on the significance of this gauge freedom, we refer to Ref.~\cite{Littlejohn91}.

Collecting terms of order $h^n$ after passing to standard symbols in equation~(\ref{eq:defLChiOps}), we obtain relations involving $L_n$ and $\chi_n$, similar to equations~(\ref{eq:L0}) and~(\ref{eq:LChiOrderOne}). When we supplement these with the relations obtained after passing to symbols in the equality $\hat{\chi}^\dagger \hat{\chi} = 1$, we can in principle obtain all higher-order corrections $L_n$ and $\chi_n$. However, for the asymptotic solution to equation~(\ref{eq:eigenvalue-HPsi}) that we construct in this paper the coefficients $L_0$, $L_1$ and $\chi_0$ suffice. Therefore, we do not construct the higher-order terms of the expansions, but refer to, for instance, Ref.~\cite{Belov06}.

\subsection{Application to the Dirac Hamiltonian} \label{subsec:separate-Dirac}

Now that we have reviewed the scheme to separate the different modes of a matrix Hamiltonian, let us apply it to the Hamiltonian~(\ref{eq:Dirac}).
We compute the classical symbol $H_\alpha(x,p,h)$ of $\hat{H_\alpha}$ using equation~(\ref{eq:symbol-zero}), which gives
\begin{equation} \label{eq:Dirac-symbol}
  H_{\alpha}(x,p,h) = H_{\alpha}(x,p) =
  \begin{pmatrix}
    U(x) + m(x) & p_1 + i \alpha p_2 \\
    p_1 - i \alpha p_2 & U(x) - m(x)
 \end{pmatrix} .
\end{equation}
Thus, the classical symbol $H_{\alpha}$ is equal to the principal symbol $H_{0,\alpha}$ and all higher-order expansion coefficients $H_{n\geq1,\alpha}$ in the symbol expansion are zero.
The Hamiltonian $\hat{H}$ has two scalar eigenmodes, corresponding to electron states, denoted with a plus sign, and hole states, denoted with a minus sign. 
The principal symbols $L_0^\pm$ of the effective scalar Hamiltonians for these modes are given by the eigenvalues of $H_{0,\alpha}$, as indicated by equation~(\ref{eq:L0}). 
Computing these eigenvalues, we immediately see that they are independent of the specific valley, i.e. $L_0^\pm$ does not depend on $\alpha$.
For both valleys, we obtain
\begin{equation} \label{eq:L0Dirac}
  L_0^{\pm}(x,p) = U(x) \pm \sqrt{p^2+m^2(x)} .
\end{equation}
We remark that when the mass $m(x)$ vanishes, the derivative of $L_0^\pm$ with respect to $p$ diverges at $p=0$.
Looking back at the symbol classes that we considered in subsection~\ref{subsec:pd-op}, we see that we therefore have to exclude a small area around this point from the space on which $L^\pm$ is defined. Otherwise, the symbol $L^\pm$ will not be an element of $S_{4}((1 + |p|^{2})^{1/2})$ and hence the pseudodifferential operator $\hat{L}$ will not be well-defined. From a physical point of view, this restriction is very natural, since the electron and hole bands touch at the Dirac point at $p=0$. As we want to separate the different modes of the matrix Hamiltonian, we should stay away from the point where they intersect. 
We remark that, when we come close to the Dirac point, the energy of the electrons also becomes lower, whence the dimensionless semiclassical parameter $h$ becomes larger. Therefore, the asymptotic expansion~(\ref{eq:symbol-expansion}) also becomes less sensible. This provides another reason why we cannot come too close to the Dirac point.

According to equation~(\ref{eq:L0}), the principal symbols $\chi_{0,\alpha}^\pm$ are given by the eigenvectors of the symbol $H_{\alpha}(x,p)$. Therefore,
\begin{equation} \label{eq:chi0Dirac}
  \chi_{0,\alpha}^\pm(x,p) = \frac{1}{\big(2 \sqrt{p^2+m^2}(\sqrt{p^2+m^2} \mp m) \big)^{1/2}} \begin{pmatrix} p_1 + i \alpha p_2 \\ \pm \sqrt{p^2+m^2} - m \end{pmatrix} .
\end{equation}
In contrast to the effective Hamiltonian~$L_0^\pm$, the symbol $\chi_{0,\alpha}^\pm$ is dependent on the valley index $\alpha$.

Within standard quantization, the subprincipal symbol $L_{1,\alpha}^\pm$ is given by equation~(\ref{eq:L1}). However, as we have shown from general considerations in the previous subsections and explicitly in appendix~\ref{app:subprincipal-symbol}, the imaginary part of $L_{1,\alpha}^\pm$ satisfies equation~(\ref{eq:ImL1}). In the next section, we show that this imaginary part $\text{Im} \, L_{1,\alpha}^\pm$ does not have any physical significance, as it only ensures conservation of probability and does not affect the wavefunction. Instead, only the real part $\text{Re} \, L_{1,\alpha}^\pm$ appears in the wavefunction. This real part equals the subprincipal Weyl symbol $L_{1,\alpha}^{W\pm}$, as we have seen in equation~(\ref{eq:L1-Weyl-relation-L1}) in the previous subsection.
We therefore compute $L_{1,\alpha}^{W\pm}$, starting with the two terms $L_{1B,\alpha}^{W\pm}$ and $L_{1A,\alpha}^{W\pm}$ that make up this subprincipal Weyl symbol.

With the help of equations~(\ref{eq:L0Dirac}) and~(\ref{eq:chi0Dirac}), we find that the Berry part~(\ref{eq:L1WB}) equals
\begin{multline}
  L_{1B,\alpha}^{W\pm} = \frac{\alpha}{2\sqrt{p^2+m^2}(\sqrt{p^2+m^2}\mp m)} \left(p_2 \frac{\partial U}{\partial x_1} - p_1 \frac{\partial U}{\partial x_2} \right) \\
    \pm \frac{\alpha m}{2 (p^2+m^2)(\sqrt{p^2+m^2}\mp m)} \left(p_2 \frac{\partial m}{\partial x_1} - p_1 \frac{\partial m}{\partial x_2} \right) .
\end{multline}
Using the definition~(\ref{eq:L1WA}) of $L_{1A,\alpha}^{W\pm}$, we obtain, after an elaborate calculation,
\begin{equation} 
  L_{1A,\alpha}^{W\pm} = \frac{\alpha}{2(p^2+m^2)} \left(p_2 \frac{\partial m}{\partial x_1} - p_1 \frac{\partial m}{\partial x_2} \right) .
\end{equation}
We remark that $L_{1A,\alpha}^{W\pm}$ vanishes when the mass $m(x)$ is constant, whereas $L_{1B,\alpha}^{W\pm}$ does not.
Adding these two contributions, we arrive at an expression for the subprincipal Weyl symbol $L_{1,\alpha}^{W\pm}$, namely
\begin{equation} \label{eq:L1Dirac}
  L_{1,\alpha}^{W\pm} = \frac{\alpha}{2\sqrt{p^2+m^2}(\sqrt{p^2+m^2}\mp m)} \left(p_2 \frac{\partial(U+m)}{\partial x_1} - p_1 \frac{\partial(U+m)}{\partial x_2} \right) .
\end{equation}
Hence, the reduction of the initial matrix equation to an effective scalar equation comes at a price: the effective scalar Hamiltonian has a nonzero subprincipal symbol, i.e. a correction term that is proportional to $h$. 

In this paper, we only consider above-barrier scattering of electrons explicitly. From here on, we therefore only consider the relevant quantities for electrons and omit the superscript ``$+$''. The derivations for holes can be done analogously.

\section{Semiclassical Ansatz} \label{sec:semiclassics}

In this section, we take a first step towards the construction of an asymptotic solution of equation~(\ref{eq:L-eff}). Based on the theory explained in Refs.~\cite{Maslov73,Maslov72,Maslov81,Kucherenko69,Vainberg89,Dobrokhotov03}, we review how the standard semiclassical Ansatz leads to the Hamilton-Jacobi equation and to the transport equation, and solve the latter to find the semiclassical phase. The main goal of this section is to introduce these basic concepts, which we further explore in the following sections.

We would like to construct an asymptotic solution $\psi(x)$, which solves equation~(\ref{eq:L-eff}) to a given order in $h$. To this end, we look for a solution in the form of the standard semiclassical Ansatz~\cite{Maslov81}
\begin{equation} \label{eq:semicl-Ansatz}
  \psi(x) = \varphi(x) e^{i S(x)/h} ,
\end{equation}
where $S(x)$ is known as the action, and the amplitude $\varphi(x)$ is expressed as an asymptotic series in the semiclassical parameter $h$, i.e.
\begin{equation}
  \varphi(x) = \sum_n h^n \varphi_n(x) .
\end{equation}
The action of the pseudodifferential operator $\hat{L}$ with standard symbol $L(x,p,h)$ on the function $\psi(x)$ is given by equation~(\ref{eq:standard-quantization}). However, we do not know the exact form of the symbol $L(x,p,h)$, but only its asymptotic expansion. In particular, we constructed the principal symbol $L_0$ and the subprincipal symbol $L_1$ in the previous section.
Hence, we require an asymptotic expansion (in powers of $h$) for the action of $\hat{L}$ on the Ansatz~(\ref{eq:semicl-Ansatz}).
For an arbitrary pseudodifferential operator $\hat{Q}=Q(x,\hat{p},h)$, one can show that, see e.g. Refs.~\cite{Maslov73,Maslov72,Maslov81},
\begin{equation} \label{eq:commutation-relation-x-gen}
  Q(x,\hat{p},h) \varphi(x) e^{i S(x)/h} = e^{i S(x)/h} \left( Q_0\left(x,\frac{\partial S}{\partial x} \right) \varphi(x) + \mathcal{O}(h) \right) ,
\end{equation}
where $Q_0$ is the principal symbol of $\hat{Q}$. Intuitively, one can justify this relation by realizing that terms of order one can only arise when a momentum operator $\hat{p}$ is applied to the exponential factor. In particular, expression~(\ref{eq:commutation-relation-x-gen}) holds for the pseudodifferential operator $\hat{L}$ and its principal symbol $L_0$, which is defined by equation~(\ref{eq:L0}). We therefore have
\begin{equation} \label{eq:commutation-relation-x-L0}
  L(x,\hat{p},h) \varphi(x) e^{i S(x)/h} = e^{i S(x)/h} \left( L_0\left(x,\frac{\partial S}{\partial x} \right) \varphi(x) + \mathcal{O}(h) \right) .
\end{equation}
Inserting this expression into equation~(\ref{eq:L-eff}), multiplying both sides by the factor $\exp(-i S(x)/h)$ and collecting the terms of order $h^0$, we find 
\begin{equation} \label{eq:preHJ}
  L_0\left(x, \frac{\partial S(x)}{\partial x}\right) \varphi_0 = E \varphi_0 .
\end{equation}
Since we want the Ansatz~(\ref{eq:semicl-Ansatz}) to be an asymptotic solution of equation~(\ref{eq:L-eff}), we require the action $S(x)$ to satisfy the Hamilton-Jacobi equation
\begin{equation} \label{eq:HJ}
  L_0\left(x, \frac{\partial S(x)}{\partial x}\right) = E .
\end{equation}
From classical mechanics, see e.g. Ref.~\cite{Arnold89}, it is well known that this equation is equivalent to the system of Hamilton equations:
\begin{equation} \label{eq:Hamilton}
  \frac{\text{d} x}{\text{d} t} = \frac{\partial L_0}{\partial p} , \qquad \frac{\text{d} p}{\text{d} t} = - \frac{\partial L_0}{\partial x} .
\end{equation}
As we discussed in section~\ref{sec:assumptions}, we would like to solve the scattering problem for a bundle of incoming electrons.
Without loss of generality, we consider electrons incoming along the $x_1$-axis. In the language of classical mechanics, this means that we consider a family of Cauchy problems for the system~(\ref{eq:Hamilton}). The initial conditions for this system are parametrized by the variable $\phi$ and constitute the line
\begin{equation} \label{eq:Lambda1}
  \Lambda^1 = \{(x,p), \quad p_1=p_1^0(\phi),\; p_2=0,\; x_1=x_1^0, \; x_2=\phi, \; \phi \in \mathbb{R}\}
\end{equation}
in four-dimensional phase space. Formally, the electrons in our scattering problem come in from minus infinity. However, in practice, one uses a finite starting point $x_1^0$, independent of $\phi$, for the integration of Hamilton's equations~(\ref{eq:Hamilton}). Since we assumed that both the potential $U(x)$ and mass $m(x)$ are constant outside of the domain $D$, the point $x_1^0$ should ideally be chosen sufficiently far outside of this domain. When one can choose $x_1^0$ in this way, the function $p_1^0(\phi)$ is constant. In fact, it is given by the energy $E$ when both $U(x)$ and $m(x)$ are zero outside of $D$. When one cannot choose $x_1^0$ outside of $D$, the function $p_1^0(\phi)$ should be constructed in such a way that $L_0(x,p)$ has the same value for all points on $\Lambda^1$, since all incoming electrons have the same energy. In other words, one should make sure that $\Lambda^1$ is contained in a level set of $L_0(x,p)$. 

For a given value of $\phi$, we denote the solutions to the Hamiltonian system~(\ref{eq:Hamilton}) with the initial condition~(\ref{eq:Lambda1}) by $\big(X(t,\phi),P(t,\phi)\big)$. For the purpose of the discussion in this section, let us assume that the equation $x=X(t,\phi)$ is invertible, and that we can determine the inverse functions $t(x)$ and $\phi(x)$. In the next section, we come back to this important point and consider the set of solutions $\big(X(t,\phi),P(t,\phi)\big)$ and its geometry in detail. 
Given a solution to the Hamiltonian system with the initial condition $\Lambda^1$, the action $S(x)$ is determined by, see e.g. Ref.~\cite{Arnold89},
\begin{equation} \label{eq:action}
  S(x) = \int^x_{x_0} \langle P, \text{d}X \rangle ,
\end{equation}
where we integrate from an initial point $x_0$ on $\Lambda^1$ to the point $x$. We discuss this integration in greater detail in section~\ref{subsec:eikonal}.

When we consider the Dirac Hamiltonian~(\ref{eq:Dirac}), the principal symbol $L_0(x,p)$ of the effective Hamiltonian~$\hat{L}$ is given by equation~(\ref{eq:L0Dirac}). Rewriting the expression $L_0(x,p)=E$, see equation~(\ref{eq:HJ}), we arrive at
\begin{equation} \label{eq:momenta-U-E-sol}
  p^2 = (U(x) - E)^2 - m^2(x) .
\end{equation}
Since we consider above-barrier scattering, we require that there are no classically forbidden regions. These are characterized by imaginary momenta, i.e. by $p^2<0$. Thus, the right-hand side of equation~(\ref{eq:momenta-U-E-sol}) should always be positive. This leads to condition~(\ref{eq:above-assump}), which we discussed in section~\ref{sec:assumptions}. Note in particular that this condition is independent of the valley index $\alpha$ and is therefore the same for both valleys.
For the principal symbol~(\ref{eq:L0Dirac}), the Hamiltonian system becomes
\begin{equation} \label{eq:Ham-Dirac-1}
  \frac{\text{d} x}{\text{d} t} = \frac{p}{\sqrt{p^2+m^2(x)}} , \qquad \frac{\text{d} p}{\text{d} t} = - \frac{\partial U}{\partial x} - \frac{2 m}{\sqrt{p^2+m^2(x)}} \frac{\partial m}{\partial x} .
\end{equation}
Since $L_0$ does not depend on the valley index $\alpha$, these equations of motion are also independent of $\alpha$.
Hence, the classical trajectories are independent of whether we are in the $K$-valley or in the $K'$-valley. For a given potential $U(x)$ and mass $m(x)$, one typically cannot solve the system~(\ref{eq:Ham-Dirac-1}) of differential equations analytically. Therefore, one has to use numerical integration to obtain a solution.

We now turn back to equation~(\ref{eq:L-eff}) and the Ansatz~(\ref{eq:semicl-Ansatz}). When we collect the terms of order $h^1$ in the asymptotic expansions on both sides, we find that~\cite{Maslov81}
\begin{equation} \label{eq:transport1}
  L_1 \varphi_0 - i \left\langle \frac{\partial L_0}{\partial p}, \frac{\partial \varphi_0}{\partial x} \right\rangle -
  \frac{i}{2} \sum_{i,j} \frac{\partial^2 L_0}{\partial p_i \partial p_j} \frac{\partial^2 S}{\partial x_i \partial x_j} \varphi_0 = (E-L_0)\varphi_1 = 0,
\end{equation}
where all symbols are to be evaluated at the point $(x, \partial S/\partial x)$, and the last equality holds by virtue of equation~(\ref{eq:HJ}). This equation is known as the transport equation~\cite{Maslov73,Maslov81,Guillemin77,Bates97}. It can be essentially simplified along the trajectories of the Hamiltonian system~(\ref{eq:Hamilton}), as detailed in e.g. Ref.~\cite{Maslov81}. To this end, we introduce the Jacobian
\begin{equation} \label{eq:JacobianX}
  J = \det \frac{\partial (X_1, X_2)}{\partial (t, \phi)} ,
\end{equation}
where $(X_1, X_2)$ denotes a vector containing the solutions of the Hamiltonian system. The time evolution of the Jacobian can be computed using the Liouville formula, see e.g. Refs.~\cite{Maslov81,Arnold89}. Using that $\text{d} X/\text{d} t= \partial L_0(x, \partial S/\partial x)/\partial x$, one obtains
\begin{equation} 
  \frac{\text{d}}{\text{d} t} \log J = \sum_j \frac{\partial}{\partial x_j} \left( \frac{\partial L_0(x, \partial S/\partial x)}{\partial p_j}  \right) = 
      \sum_{i,j} \frac{\partial^2 L_0}{\partial p_i \partial p_j} \frac{\partial^2 S}{\partial x_i \partial x_j} + \sum_j \frac{\partial^2 L_0}{\partial x_j \partial p_j} .
\end{equation}
Furthermore, along the trajectories of the Hamiltonian system, one has
\begin{equation} \label{eq:time-deriv-along-traj-x}
  \left\langle \frac{\partial L_0}{\partial p}, \frac{\partial \varphi_0}{\partial x} \right\rangle = 
  \left\langle \frac{\text{d} x}{\text{d} t}, \frac{\partial \varphi_0}{\partial x} \right\rangle = 
  \frac{\text{d} \varphi_0}{\text{d} t} .
\end{equation}
When we subsequently introduce $A_0$ by $A_0 = \varphi_0 \sqrt{J}$, we therefore find that equation~(\ref{eq:transport1}) becomes
\begin{equation} \label{eq:transport2}
  \frac{\text{d} A_0}{\text{d} t} + \bigg(i L_1 - \frac{1}{2} \sum_j \frac{\partial^2 L_0}{\partial x_j \partial p_j} \bigg) A_0  = 0 .
\end{equation}
Looking back at our derivation, we see that we have established a second commutation formula~\cite{Maslov73,Maslov72,Maslov81}. Unlike the first commutation formula~(\ref{eq:commutation-relation-x-gen}), this one does not hold for any pseudodifferential operator, but specifically for the effective Hamiltonian $\hat{L}$. Provided that $S(x)$ is a solution of the Hamilton-Jacobi equation~(\ref{eq:HJ}), we have established that~\cite{Maslov81}
\begin{equation} \label{eq:commutation-relation-x-L1}
  \big( L(x,\hat{p},h) - E \big) \frac{A_0}{\sqrt{J}} e^{\frac{i}{h} S} 
    = -i h \frac{e^{\frac{i}{h} S}}{\sqrt{J}} \bigg( \frac{\text{d} A_0}{\text{d} t} + i L_1 A_0 - \frac{1}{2} \sum_j \frac{\partial^2 L_0}{\partial x_j \partial p_j}A_0 + \mathcal{O}(h) \bigg) ,
\end{equation}
where the time derivative is taken along the solutions of the Hamiltonian system, see equation~(\ref{eq:time-deriv-along-traj-x}).
This implies that when $A_0$ solves equation~(\ref{eq:transport2}), the function
\begin{equation} \label{eq:psi-sol-x-jac}
  \psi(x) = \frac{A_0(x)}{\sqrt{J(x)}} e^{i S(x)/h}
\end{equation}
is an asymptotic solution of equation~(\ref{eq:L-eff}). The remaining terms on the right-hand side of equation~(\ref{eq:commutation-relation-x-L1}) are of order $h^2$, which means that the corrections to the asymptotic solution~(\ref{eq:psi-sol-x-jac}) are of order $h$.
Note that in the above equation both $A_0$ and $J$ can be viewed as functions of the point $x$, since we have assumed that the inverse functions $t(x)$ and $\phi(x)$ exist.
The asymptotic solution~(\ref{eq:psi-sol-x-jac}) is a multidimensional generalization of the Wentzel-Kramers-Brillouin (WKB) approximation that is often used in theoretical physics~\cite{Griffiths16,Hall13}.

One easily sees that the solution to equation~(\ref{eq:transport2}) is a complex exponential. As we discussed in the previous section, the imaginary part of $L_1$ satisfies equation~(\ref{eq:ImL1}), since the effective Hamiltonian~$\hat{L}$ is self-adjoint. Hence, the second derivative of $L_0$ cancels the imaginary part of $L_1$ in equation~(\ref{eq:transport2}). This leaves us with the real part of $L_1$, which equals the subprincipal Weyl symbol $L_1^W$ by virtue of equation~(\ref{eq:L1-Weyl-relation-L1}). Since this subprincipal Weyl symbol is purely real, the exponential factor is a pure phase and we have conservation of probability. Thus, as we already anticipated in section~\ref{subsec:separate-Dirac}, the imaginary part of $L_1$ has no physical significance. Instead, it satisfies relation~(\ref{eq:ImL1}), which ensures conservation of probability.
When we consider a point $x$ that is reached at time $t$ by a trajectory with initial position $\phi$ on $\Lambda^1$, we therefore obtain~\cite{Maslov81}
\begin{equation} \label{eq:Amp-sol}
  A_0 = A_0^0(\phi) \exp \left(i \Phi_{sc}(t,\phi) \right) , \qquad \Phi_{sc}(t, \phi) = -\int_0^t L_1^W(X,P) \, \text{d} t' .
\end{equation}
In this equation, the variables $X$ and $P$ represent a solution $\big(X(t,\phi),P(t,\phi)\big)$ of the Hamiltonian system~(\ref{eq:Hamilton}). Both $t$ and $\phi$ are functions of the point $x$, since we have assumed that the inverse functions exist. The quantity $A_0^0(\phi)$ is the initial amplitude, i.e. the amplitude on $\Lambda^1$. One can for instance consider $A_0^0(\phi) = 1$, or a smooth cutoff function localized in a certain interval.

We call the quantity $\Phi_{sc}$ the semiclassical phase~\cite{Maslov81,Littlejohn91,Bernstein75,Bernstein84,Kaufman87}. In the previous section, we decomposed $L_1^W(X,P)$ into two parts, each of which was purely real. Hence, we can also decompose the semiclassical phase into two parts. We write $\Phi_{sc} = \Phi_B + \Phi_A$, where
\begin{equation} \label{eq:phases-B-A}
  \Phi_B = -\int_0^t L_{1B}^W(X,P) \, \text{d} t' , \qquad \Phi_A = -\int_0^t L_{1A}^W(X,P) \, \text{d} t' .
\end{equation}
The phase $\Phi_B$ is known as the Berry phase~\cite{Berry84,Shapere89,Littlejohn91}. We can show that it equals Berry's original expression by using the definition~(\ref{eq:L1WB}) and the equations of motion~(\ref{eq:Hamilton}). We have
\begin{equation} \label{eq:PhiB-simplified}
  \Phi_B = i \int_0^t \chi_0^\dagger \{ \chi_0 , L_0 \} \, \text{d} t' = i \int_0^t \chi_0^\dagger \left\langle \frac{\partial \chi_0}{\partial x} , \frac{\text{d} x}{\text{d} t'} \right\rangle + 
    \chi_0^\dagger \left\langle \frac{\partial \chi_0}{\partial p} , \frac{\text{d} p}{\text{d} t'} \right\rangle  \, \text{d} t' .
\end{equation}
We subsequently obtain Berry's original expression by combining the phase space coordinates $x$ and $p$ into a single vector. Thus, the Berry phase is obtained from an integral along a path in phase space. For a more extensive discussion about the differences between the Berry phase and the semiclassical phase, we refer to Ref.~\cite{Littlejohn91}.

For the Dirac Hamiltonian, we can now easily compute the semiclassical phase~(\ref{eq:Amp-sol}) using equation~(\ref{eq:L1Dirac}). However, let us rewrite it in a somewhat different form. Using the Hamiltonian system~(\ref{eq:Ham-Dirac-1}), we find that along the solutions $(X,P)$ of the Hamiltonian system:
\begin{equation} \label{eq:BerryPhaseDiracPreCoordfree}
  L_{1,\alpha}^W = \frac{\alpha}{2 (\sqrt{p^2+m^2}-m)} \left( \frac{\partial(U+m)}{\partial x_1}\frac{\text{d} x_2}{\text{d} t} - \frac{\partial(U+m)}{\partial x_2}\frac{\text{d} x_1}{\text{d} t} \right) .
\end{equation}
Finally, using that $E= L_0 = U + \sqrt{p^2+m^2}$, we find that
\begin{equation} \label{eq:BerryPhaseDiracCoordfree} 
  \Phi_{sc,\alpha}(t,\phi) = -\int \frac{\alpha}{2 (E-U(X)-m(X))} \left( \frac{\partial(U+m)}{\partial X_1}\text{d} X_2 - \frac{\partial(U+m)}{\partial X_2}\text{d} X_1 \right) ,
\end{equation}
where the integration is to be performed along the trajectories $X(t,\phi)$. This expression has the advantage that it only depends on the trajectories themselves, and not on their parametrization. We come back to this point in section~\ref{sec:maslov}. 

In section~\ref{subsec:separate-Dirac}, we showed that when the mass is constant, and hence in particular when it vanishes, $L_{1A,\alpha}^W$ vanishes. Therefore, the semiclassical phase equals the Berry phase in this case. 
Furthermore, one directly sees from equation~(\ref{eq:chi0Dirac}) that $\chi_{0,\alpha}$ is independent of $x$ when the mass is constant. Hence, the first term in equation~(\ref{eq:PhiB-simplified}) vanishes in this case. When the mass is identically zero, one can show by a direct calculation~\cite{Carmier08,Tudorovskiy12} that
\begin{equation}
  \Phi_{B,\alpha} 
    = i \int_0^t \chi_0^\dagger \left\langle \frac{\partial \chi_0}{\partial p} , \frac{\text{d} p}{\text{d} t'} \right\rangle  \, \text{d} t' 
    = -\frac{\alpha}{2} \int_0^t \left\langle \frac{\partial \phi_p}{\partial p} , \frac{\text{d} p}{\text{d} t'} \right\rangle  \, \text{d} t'
    = -\frac{\alpha}{2}\Delta \phi_p ,
\end{equation}
where $\alpha$ is the valley degree of freedom and $\phi_p=\arctan(p_2/p_1)$ is the angle in momentum space. 
Thus, when the mass vanishes, the semiclassical phase $\Phi_{sc}$ of an electron in graphene equals the difference between its final and its initial angle in momentum space. 
This example was already considered by Berry in his original paper~\cite{Berry84}, in which he showed that for the massless Dirac equation the Berry phase of a closed trajectory equals half of the solid angle that such a trajectory spans in momentum space.
A more elaborate discussion of the difference between the Berry phase and the semiclassical phase in the context of graphene was presented in Ref.~\cite{Carmier08}.

\section{Classical trajectories and the Lagrangian manifold} \label{sec:lagman}

In the previous section, we showed that the semiclassical approximation gives rise to the Hamilton-Jacobi equation~(\ref{eq:HJ}). We also discussed how this equation can be solved by solving the associated Hamiltonian system~(\ref{eq:Hamilton}) with the initial condition~(\ref{eq:Lambda1}) and the equation~(\ref{eq:action}) for the action. In this section, we study the solutions of these equations in detail. In section~\ref{subsec:caustics}, we introduce the concept of a caustic, and explore its consequences. Subsequently, we introduce eikonal coordinates in section~\ref{subsec:eikonal}, and discuss the properties that they give rise to. Section~\ref{subsec:lagman} introduces the concept of a Lagrangian manifold. Finally, we discuss the classification of the singular points of this Lagrangian manifold in section~\ref{subsec:classification}.

\subsection{Caustics} \label{subsec:caustics}
\begin{figure}[htb]
  \begin{center}
    \hfill
    \includegraphics[width=0.45\textwidth]{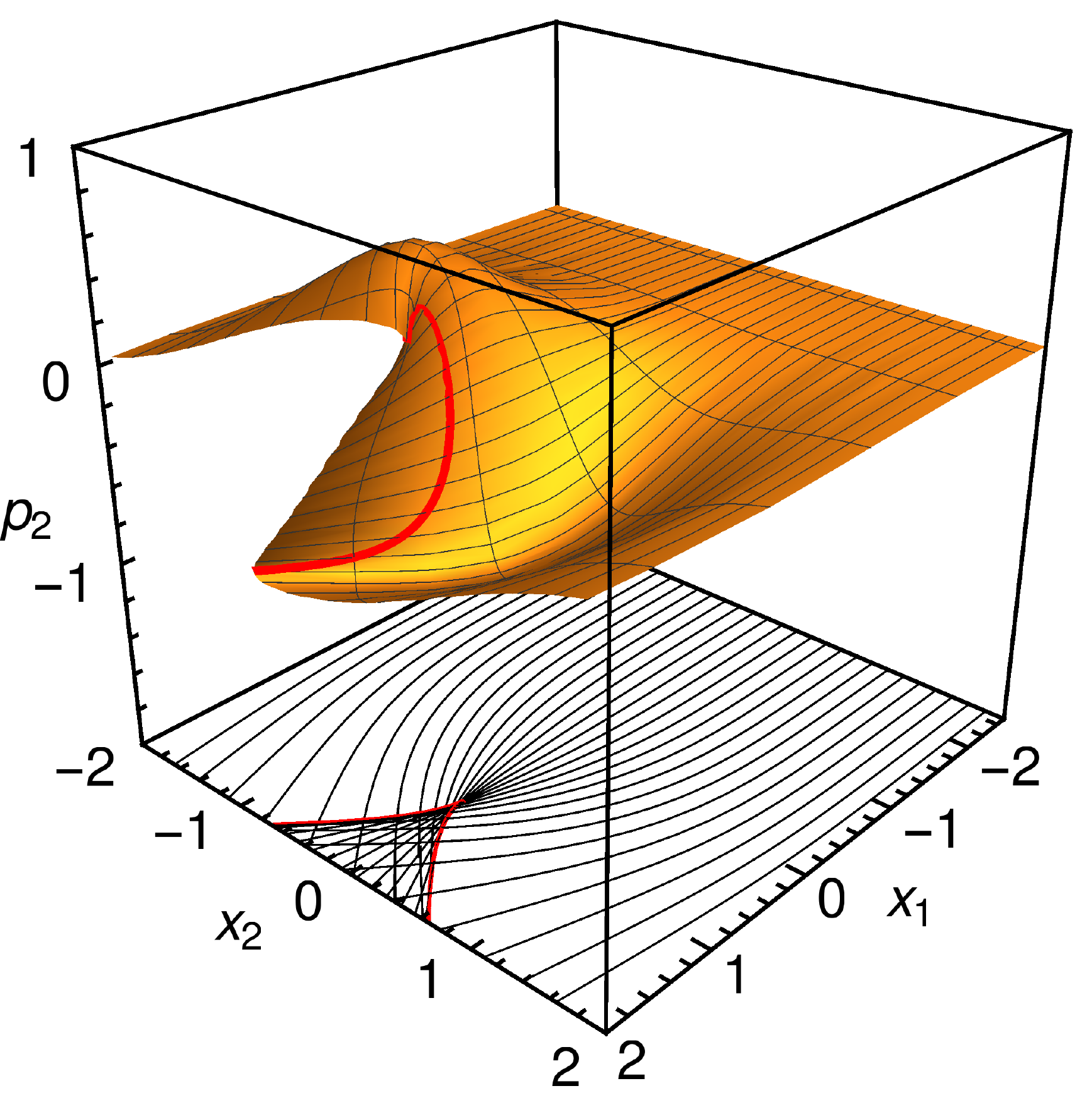}
    \hfill\hfill\hfill
    \includegraphics[width=0.45\textwidth]{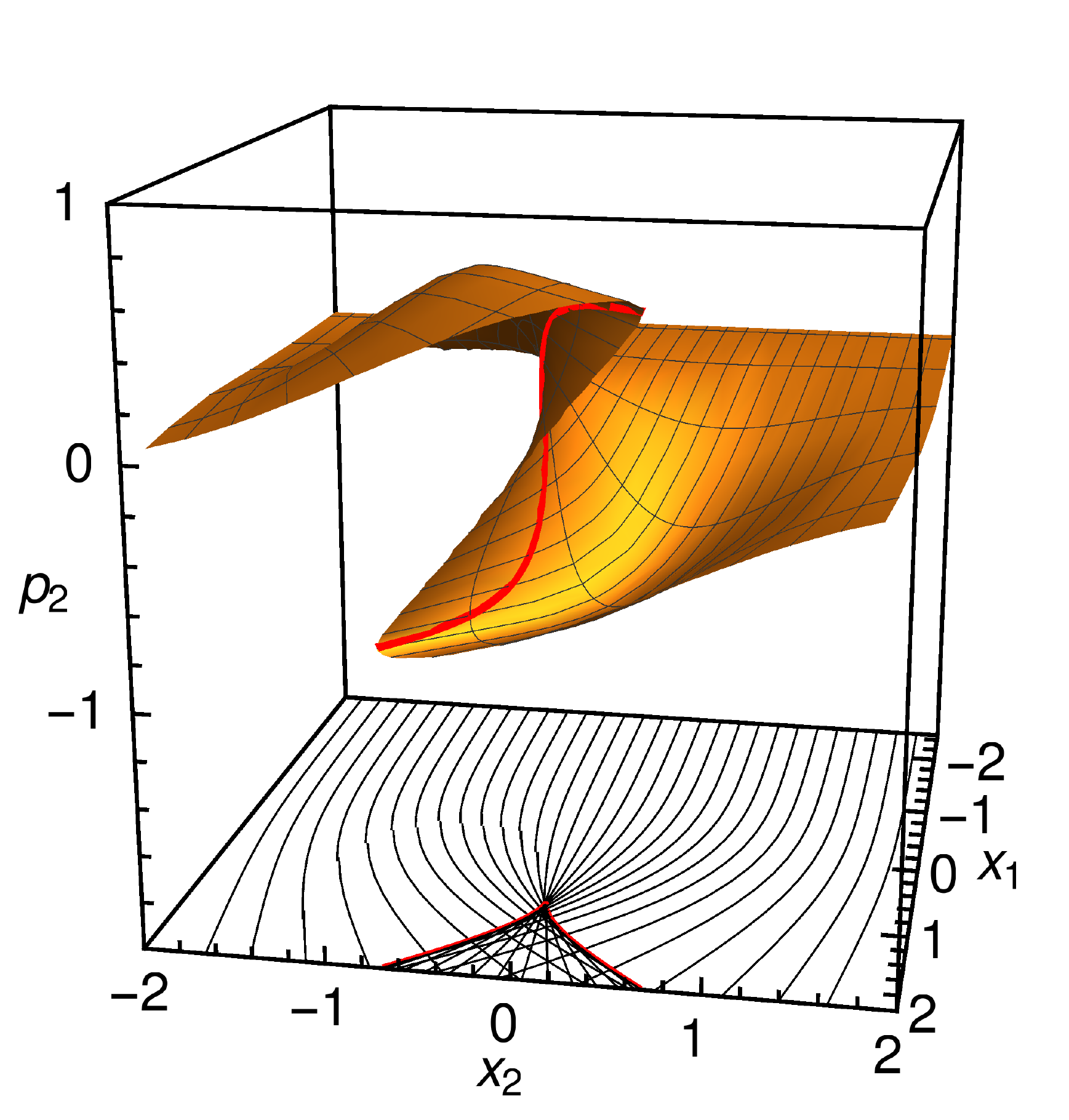}
    \hfill
  \end{center}
  \caption{Projection (two different views) of the solution of the Hamiltonian system~(\ref{eq:Hamilton}) with the initial condition~(\ref{eq:Lambda1}) onto the coordinates $(x_1,x_2,p_2)$, for a Gaussian potential well and vanishing mass. The bottom of the graph shows the trajectories, i.e. the projection of the solution onto the coordinate plane $(x_1,x_2)$. The red line shows the caustic, which consists of the points where $J$ vanishes.}
  \label{fig:LagMan}
\end{figure}

The solution to Hamilton's equations~(\ref{eq:Hamilton}) with the initial condition~(\ref{eq:Lambda1}) consists of the set of curves $\big\{ X(t,\phi),P(t,\phi)\big\}$ in phase space $\mathbb{R}^4_{x,p}$, parametrized by the variables $t$ and $\phi$. The collection of these curves forms a smooth two-dimensional surface $\Lambda^2$ in four-dimensional phase space, in a way that will be made precise in the next subsection. Before we take a closer look at the geometry of this surface, let us first consider a typical example. 
We set the mass to zero and consider a Gaussian potential, an example that will be discussed in greater detail in section~\ref{sec:examples}. We subsequently integrate Hamilton's equations~(\ref{eq:Ham-Dirac-1}) numerically to find the set of solutions $\big\{ X(t,\phi),P(t,\phi)\big\}$. 
In figure~\ref{fig:LagMan}, we plot the projection of this set onto the coordinates $(x_1,x_2,p_2)$. 
In the bottom of the figure, we also plot a number of the trajectories of the system. By the term trajectory we mean the projection of the solution $\big\{ X(t,\phi),P(t,\phi)\big\}$ for a given value of $\phi$ onto the coordinate plane $(x_1,x_2)$, i.e. the projection $x=X(t,\phi)$.

Looking at the surface in figure~\ref{fig:LagMan}, we immediately see that we can distinguish two regions. In the first region, the projection of the surface $\Lambda^2$ onto the $(x_1,x_2)$ plane is a one-to-one map, i.e. the system of equations $x = X(t,\phi)$ has a unique solution $(t(x),\phi(x))$. In the second region, the projection of the surface $\Lambda^2$ onto the $(x_1,x_2)$ plane is ``three-to-one'', i.e. the equation $x = X(t,\phi)$ has three solutions $(t_i(x),\phi_i(x))$. In this case, we say that the manifold $\Lambda^2$ has three leaves. The boundary between these two regions is given by the red line in figure~\ref{fig:LagMan}. One sees that on this line the surface $\Lambda^2$ has folds. In the neighborhood of these folds, the projection of the surface $\Lambda^2$ onto the plane $(x_1,x_2)$ is not invertible, i.e. the system of equations $x = X(t,\phi)$ has no unique solution $(t(x),\phi(x))$. By the implicit function theorem, this means that the Jacobian $J$, given by equation~(\ref{eq:JacobianX}), vanishes on this line. The points on $\Lambda^2$ with coordinates $(t,\phi)$ where the Jacobian equals zero are known as singular points or focal points, as opposed to regular points that have non-zero Jacobian~\cite{Arnold82,Poston78,Whitney55}. The connected components of the set on which the Jacobian vanishes are known as caustics or Lagrangian singularities.

Looking at the set of trajectories, i.e. the projection of $\Lambda^2$ onto the $(x_1,x_2)$ plane, it is clear that the caustic separates the region where each point lies on three trajectories from the region where each point lies on a single trajectory. 
We also observe that the density of trajectories increases as we move towards the caustic. We may think of the inverse of the Jacobian as a measure for the density of trajectories, which vanishes at the caustic.
We therefore expect a larger intensity near the caustic: focusing occurs. This effect should be strongest near the cusp, or, as one may say, the `tip' of the caustic, where the density of trajectories is highest~\cite{Arnold82,Poston78}.

Going back to our asymptotic solution~(\ref{eq:psi-sol-x-jac}), we see that it diverges on the caustic, since the Jacobian $J$ vanishes on the caustic. This indicates that something is wrong with our asymptotic solution and that it is no longer a good approximation to the real solution. The origin of this divergence lies in the fact that the projection of the surface $\Lambda^2$ onto the $(x_1,x_2)$ plane is no longer invertible. However, looking at figure~\ref{fig:LagMan}, we are led to a possible solution, first suggested by Maslov~\cite{Maslov72,Maslov81}: near the caustic we could try to consider the projection onto the $(x_1,p_2)$ plane, since this projection seems to be invertible. We could even try to use different coordinates, as long as the projections are invertible. 

When we made the transition from the Hamilton-Jacobi equation~(\ref{eq:HJ}) to the system of Hamilton equations~(\ref{eq:Hamilton}), we already lifted the problem from the configuration space $(x_1,x_2)$ to the phase space~\cite{Maslov81}. The above reasoning makes it plausible that this is a necessary step, and that we should study the properties of the surface $\Lambda^2$ before we continue with the development of an asymptotic solution to the Dirac equation.

\subsection{Eikonal coordinates and the Jacobi -- Maupertuis principle} \label{subsec:eikonal}

In the previous section, we considered an example of the surface $\Lambda^2$ and discussed some of its properties. We now take a closer look at the solutions $\big\{ X(t,\phi),P(t,\phi)\big\}$ of the Hamiltonian system~(\ref{eq:Ham-Dirac-1}). If we think of a solution for a given value of $\phi$ as a curve in phase space, then we can reparametrize the time $t$ with which we follow theis curve. In this section, we show how such a change in parametrization can be generated by a change in the classical Hamiltonian. It will turn out that this new parametrization leads to very convenient properties. 

Let us consider the Hamiltonian $L_0(x,p)$, given by equation~(\ref{eq:L0Dirac}), for a certain energy $E$. Then we have
\begin{equation} \label{eq:L0Dirac2}
  L_0(x,p) = \sqrt{p^2 + m^2(x)} + U(x) = E .
\end{equation}
This equation can be rewritten as 
\begin{equation} \label{eq:L0Eikonal}
  \mathcal{L}_0(x,p) \equiv C(x) |p| = 1, \qquad \text{where} \;\; C(x) = \frac{1}{\sqrt{(E-U(x))^2-m^2(x)}} ,
\end{equation}
where we have defined the function $\mathcal{L}_0$. 
The correspondence between equations~(\ref{eq:L0Dirac2}) and~(\ref{eq:L0Eikonal}) is one-to-one because the function is $C(x)$ is non-singular. The latter is a consequence of the fact that we consider above-barrier scattering.
We can consider the function $\mathcal{L}_0$ as our new Hamiltonian and write down the corresponding Hamiltonian system:
\begin{equation} \label{eq:Ham-Dirac-2}
  \frac{\text{d} x}{\text{d} \tau} = C(x) \frac{p}{|p|} , \qquad \frac{\text{d} p}{\text{d} \tau} = -\frac{\partial C}{\partial x} |p| .
\end{equation}
We denote the solutions to this system with initial data on the curve $\Lambda^1$ by $\big\{ \mathcal{X}(\tau,\phi),\mathcal{P}(\tau,\phi)\big\}$. In the next paragraph, we show, based on Ref.~\cite{Dobrokhotov15}, that the solutions $\big\{ X(t,\phi),P(t,\phi)\big\}$ of the Hamiltonian system~(\ref{eq:Ham-Dirac-1}) with energy $E$ coincide with the solutions $\big\{ \mathcal{X}(\tau,\phi),\mathcal{P}(\tau,\phi)\big\}$ of the Hamiltonian system~(\ref{eq:Ham-Dirac-2}) with energy 1, up to a reparametrization of time. 
This correspondence can be generalized to a wider class of Hamiltonians, and is known as the Maupertuis-Jacobi principle. A detailed exposition can be found in Refs.~\cite{Dobrokhotov10,Dobrokhotov11,Dobrokhotov15}, see also theorem 3.7.7 in Ref.~\cite{Abraham08}. Furthermore, the Hamiltonian $\mathcal{L}_0$ is related to the so-called Finsler metric~\cite{Katok73}.

Let us consider a solution $\big\{ X(t,\phi),P(t,\phi)\big\}$ which satisfies the Hamiltonian system~(\ref{eq:Ham-Dirac-1}) for a given energy $E$. Using equations~(\ref{eq:L0Dirac2}) and~(\ref{eq:L0Eikonal}), we can rewrite this Hamiltonian system as
\begin{equation} \label{eq:Ham-Dirac-2-equiv}
  \begin{aligned}
  \frac{\text{d} X}{\text{d} t} &=
    \frac{P}{\sqrt{P^2+m^2(X)}} = R(X) C(X) \frac{P}{|P|} \\
  \frac{\text{d} P}{\text{d} t} &=
    -\frac{m(X)}{\sqrt{P^2+m^2(X)}} \frac{\partial m(X)}{\partial X} - \frac{\partial U(X)}{\partial X} = -R(X) \frac{\partial C(X)}{\partial X} |P| ,
  \end{aligned}
\end{equation}
where
\begin{equation} \label{eq:Rtransition}
  R(X) = \frac{1/C^2(X)}{E-U(X)} = \frac{(E-U(X))^2-m^2(X)}{E-U(X)} .
\end{equation}
Subsequently, we can change the time variable from $t$ to $\tau(t,\phi)$, where $\tau(t,\phi)$ satisfies
\begin{equation} \label{eq:timereparametrization}
  \frac{\text{d} \tau}{\text{d} t} = R(X(t,\phi)) , \qquad \tau|_{t=0} = 0 ,
\end{equation}
When we perform this change of variables, the system~(\ref{eq:Ham-Dirac-2-equiv}) becomes the Hamiltonian system~(\ref{eq:Ham-Dirac-2}). We therefore conclude that $(\mathcal{X}(\tau(t,\phi),\phi),\mathcal{P}(\tau(t,\phi),\phi))$ and $(X(t,\phi),P(t,\phi))$ satisfy the same system of ordinary differential equations. By the uniqueness of the solution, this means that~\cite{Dobrokhotov15}
\begin{equation} \label{eq:trajectoriesequal}
  \big( X(t,\phi),P(t,\phi)\big) = \big( \mathcal{X}(\tau,\phi),\mathcal{P}(\tau,\phi)\big)|_{\tau=\tau(t,\phi)} .
\end{equation}
Hence the solutions of the Hamiltonian system~(\ref{eq:Ham-Dirac-2}) coincide with those of the Hamiltonian system~(\ref{eq:Ham-Dirac-1}), up to a reparametrization of time. Therefore, we conclude that both Hamiltonians define the same smooth surface $\Lambda^2$, and that the only difference is the coordinate system that is used. We can thus perform all classical computations with the Hamiltonian~(\ref{eq:L0Eikonal}), as well as with the classical Hamiltonian~(\ref{eq:L0Dirac}). In the remainder of this section, we discuss the properties of $\Lambda^2$ with the eikonal coordinate system.

Let us start by considering the action on the surface $\Lambda^2$. It is defined with respect to the so-called central point $t=\tau=\phi=0$ on the surface $\Lambda^2$ and is given by
\begin{align}
  S(\tau,\phi) &= \int_{(0,0)}^{(t,\phi)} \langle P(t,\phi) , \text{d} X(t,\phi) \rangle 
                = \int_{(0,0)}^{(\tau,\phi)} \langle \mathcal{P}(\tau,\phi) , \text{d} \mathcal{X}(\tau,\phi) \rangle \nonumber \\
  &= \int_{(0,0)}^{(0,\phi)} \langle \mathcal{P}(\tau,\phi) , \text{d} \mathcal{X}(\tau,\phi) \rangle + 
                  \int_{(0,\phi)}^{(\tau,\phi)} \langle \mathcal{P}(\tau,\phi) , \text{d} \mathcal{X}(\tau,\phi) \rangle , \nonumber
\end{align}
where the first integral is performed along the line $\Lambda^1$ and the second part along the trajectory with initial condition parametrized by $\phi$. Now we note that $\langle \mathcal{P} , \text{d} \mathcal{X} \rangle$ vanishes on $\Lambda^1$, since $\langle \mathcal{P} , \mathcal{X}_\phi \rangle$ vanishes. Furthermore, using the Hamiltonian system~(\ref{eq:Ham-Dirac-2}), we find that~\cite{Dobrokhotov15}
\begin{equation} \label{eq:actionTau}
  S(\tau,\phi) = \int_{(0,\phi)}^{(\tau,\phi)} \left\langle \mathcal{P}(\tau,\phi) , \frac{\text{d} \mathcal{X}}{\text{d} \tau} \right\rangle  \text{d} \tau = \int_{(0,\phi)}^{(\tau,\phi)} C(\mathcal{X}(\tau,\phi)) |\mathcal{P}(\tau,\phi)| \, \text{d} \tau = \tau ,
\end{equation}
where we have used that the solutions lie on the level set $\mathcal{L}_0(x,p)=1$. This means that in our new coordinates the action has a particularly simple form.

For the second important property, we consider the variational system that corresponds to the Hamiltonian system~(\ref{eq:Ham-Dirac-2}). It is given by
\begin{equation} \label{eq:varsys}
\begin{aligned} 
  \frac{\text{d} V_i}{\text{d} \tau} &= \sum_j \frac{\partial^2 \mathcal{L}_0}{\partial p_i \partial x_j} V_j + \sum_j \frac{\partial^2 \mathcal{L}_0}{\partial p_i \partial p_j} W_j 
  = \sum_j \frac{\partial C}{\partial x_j} \frac{p_i}{|p|} V_j + \sum_j C \left( \frac{\delta_{i j}}{|p|} - \frac{p_i p_j}{|p|^3} \right) W_j, \\
  \frac{\text{d} W_i}{\text{d} \tau} &= -\sum_j \frac{\partial^2 \mathcal{L}_0}{\partial x_i \partial x_j} V_j - \sum_j \frac{\partial^2 \mathcal{L}_0}{\partial x_i \partial p_j} W_j 
  = -\sum_j \frac{\partial^2 C}{\partial x_i \partial x_j} |p| V_j - \sum_j \frac{\partial C}{\partial x_i} \frac{p_j}{|p|} W_j .
\end{aligned}
\end{equation}
This system arises by considering the derivatives of $\mathcal{X}_\phi$ and $\mathcal{P}_\phi$ with respect to~$\tau$. Therefore, one easily sees that $(V,W) = (\mathcal{X}_\phi,\mathcal{P}_\phi)$ and $(V,W) = (\mathcal{X}_\tau,\mathcal{P}_\tau)$ are solutions of this system. However, for the Hamiltonian $\mathcal{L}_0$, it also has the important solution $(V,W)=(0,\mathcal{P})$. This can be verified by direct insertion into the above equations, upon which the second equation becomes Hamilton's equation for the derivative of $\mathcal{P}$ with respect to $\tau$ and the first equation becomes trivial. In fact, this is a consequence of the fact that $\mathcal{L}_0$ is first-order homogeneous in $|p|$ and can also be derived by using Euler's equality for homogeneous functions.
Subsequently, we use the fact that the skew-scalar product of two solutions $(V^{(1)},W^{(1)})$ and $(V^{(2)},W^{(2)})$ of the variational system is conserved along the trajectories, i.e.
\begin{equation} \label{eq:conservedvarsol}
  \frac{\text{d}}{\text{d} \tau}\Big(\langle V^{(1)} , W^{(2)} \rangle - \langle V^{(2)},W^{(1)} \rangle \Big) = 0 ,
\end{equation}
which can again be verified by direct computation. By applying this to the two solutions $(\mathcal{X}_\phi,\mathcal{P}_\phi)$ and $(0,\mathcal{P})$, we find that $\langle \mathcal{P}, \mathcal{X}_\phi \rangle$ is conserved along the trajectories. Taking into account that it is zero on $\Lambda^1$, we conclude that $\langle \mathcal{P}, \mathcal{X}_\phi \rangle = 0$ on the surface $\Lambda^2$.

We have thus established the following two properties on $\Lambda^2$ with coordinate system $(\tau,\phi)$:
\begin{equation} \label{eq:eikonalCoordinates}
  \langle \mathcal{P}, \mathcal{X}_\tau \rangle = 1, \qquad \langle \mathcal{P}, \mathcal{X}_\phi \rangle = 0 .
\end{equation}
Such a coordinate system has recently been denoted by the term eikonal coordinate system in Ref.~\cite{Dobrokhotov14}, and henceforth we call the coordinates $(\tau,\phi)$ eikonal coordinates. When we take the derivative of the first equality with respect to $\phi$ and of the second equality with respect to $\tau$ and subsequently subtract the results, we obtain a third important property of $\Lambda^2$, that is,
\begin{equation} \label{eq:eikonalLagrangeBrackets}
  \langle \mathcal{P}_\phi, \mathcal{X}_\tau \rangle = \langle \mathcal{P}_\tau, \mathcal{X}_\phi \rangle  .
\end{equation}
In the next subsection, we will see that this property implies that the surface $\Lambda^2$ is a so-called Lagrangian manifold.

We finish this section by having another look at the projection of the surface $\Lambda^2$ onto the plane $(x_1,x_2)$. In the previous section, we established that the focal points, i.e. the singular points of the projection, are given by the points where the Jacobian vanishes. It turns out that the Jacobian in eikonal coordinates has a particular simple form. In the remainder of this section, we establish that it is given by~\cite{Dobrokhotov15,Dobrokhotov14}
\begin{equation} \label{eq:JacobianXEikonal}
  \mathcal{J} = \det \frac{\partial (\mathcal{X}_1,\mathcal{X}_2)}{\partial(\tau,\phi)} = \pm C(\mathcal{X}) |\mathcal{X}_\phi| .
\end{equation}
To this end, we first look at the inner product $\langle \mathcal{X}_\tau, \mathcal{X}_\phi \rangle$. Since $\mathcal{X}_\tau$ is proportional to the momentum, see equation~(\ref{eq:Ham-Dirac-2}), the second equality in~(\ref{eq:eikonalCoordinates}) gives $\langle \mathcal{X}_\tau, \mathcal{X}_\phi \rangle = 0$. This simplifies the calculation of the Jacobian considerably, since it implies that
\begin{equation} \label{eq:JacobianXEikonalComputed}
  \mathcal{J}^2 = | \mathcal{X}_\tau |^2 |\mathcal{X}_\phi|^2 , \qquad \text{or} \;\; \mathcal{J} = \pm C(\mathcal{X}) |\mathcal{X}_\phi| ,
\end{equation}
where we have once again used the Hamiltonian system~(\ref{eq:Ham-Dirac-2}). Since we consider above-barrier scattering, see section~\ref{sec:assumptions}, $C(\mathcal{X})$ does not vanish. We therefore conclude that the focal points correspond to the points where $\mathcal{X}_\phi$ vanishes. Note that on $\Lambda^1$ we have $|\mathcal{X}_\phi| = 1$, and hence all its points are regular.

We remark that the equality $\langle \mathcal{X}_\tau, \mathcal{X}_\phi \rangle = 0$ also has a geometrical meaning. Let us consider the smooth curve on $\Lambda^2$ formed by the points that correspond to a given value $\tau_0$ of the action $S(\tau,\phi)=\tau$. Its projection onto the plane $(x_1,x_2)$ is known as a wavefront, and is not necessarily smooth. Since it consists of the points $\mathcal{X}(\phi,\tau_0)$, with a fixed value of $\tau_0$, the vector $\mathcal{X}_\phi$ is tangent to the wavefront. Because $\mathcal{X}_\tau$ is tangent to the trajectories, the equality $\langle \mathcal{X}_\tau, \mathcal{X}_\phi \rangle = 0$ implies that the trajectories and the wavefronts are orthogonal.

Finally, we note that the Jacobian~(\ref{eq:JacobianXEikonal}) in eikonal coordinates is related to the Jacobian defined in equation~(\ref{eq:JacobianX}) by~\cite{Dobrokhotov15,Dobrokhotov14}
\begin{equation} \label{eq:JacobianXrelationEikonal}
  J = \det \frac{\partial (X_1,X_2)}{\partial(t,\phi)} = \det \frac{\partial (\mathcal{X}_1,\mathcal{X}_2)}{\partial(\tau,\phi)} \det \frac{\partial (\tau,\phi)}{\partial(t,\phi)} = R(X(t,\phi)) \mathcal{J} ,
\end{equation}
where $R(X(t,\phi))$ was defined in equation~(\ref{eq:Rtransition}).

\subsection{Lagrangian manifolds} \label{subsec:lagman}

In the previous sections, we considered an example of the surface $\Lambda^2$ and took a closer look at its structure. We introduced eikonal coordinates on it, and found particularly simple expressions for the action, the Jacobian and the focal points in these coordinates. In this section, we introduce the concept of a Lagrangian manifold and show that $\Lambda^2$ has this structure. We do not present the full derivation of all properties that we present here. Instead, we refer the interested reader to the textbooks~\cite{Dubrovin84,Arnold89,Maslov81,Abraham08}, on which our exposition is based.

We start by defining the Lagrange bracket of $\sigma_1$ and $\sigma_2$ as~\cite{Abraham08}
\begin{equation} \label{eq:Lagbrackets}
\left[\sigma_1,\sigma_2\right]_{X,P} \equiv 
\left\langle \frac{\partial X}{\partial \sigma_1} , \frac{\partial P}{\partial \sigma_2} \right\rangle -
\left\langle \frac{\partial P}{\partial \sigma_1} , \frac{\partial X}{\partial \sigma_2} \right\rangle ,
\end{equation}
where we consider $P$ and $X$ as functions of $\sigma_1$ and $\sigma_2$. Let us now consider a manifold $M$ of dimension $m \leq n$ embedded in $2n$-dimensional phase space. We call $M$ an isotropic manifold when the Lagrange brackets of its local coordinates are identically zero~\cite{Abraham08}. We call $M$ a Lagrangian manifold when it is an isotropic manifold and when its dimension equals $n$.
Using somewhat more abstract terminology, the vanishing of the Lagrange brackets is equivalent to the fact that the restriction of the symplectic form $\text{d} x \wedge \text{d} p$ to an isotropic manifold yields zero~\cite{Abraham08}.

As an example~\cite{Maslov81}, we note that any one-dimensional surface in phase space is an isotropic manifold: it has only one coordinate $\sigma$, and the Lagrange bracket $[\sigma,\sigma]_{X,P}$ vanishes by antisymmetry. In particular, the surface $\Lambda^1$, given in equation~(\ref{eq:Lambda1}), is an isotropic manifold. A straightforward example of a Lagrangian manifold embedded in four-dimensional phase space is given by the coordinate Lagrangian plane $(x_1,x_2)$, with $p_1=p_2=0$. More generally, given a partition of the set ${1, \ldots, N}$ into two disjoint subsets $\{\alpha\}$ and $\{\beta\}$, we define a coordinate Lagrangian plane as the plane $(p_{\{\alpha\}},x_{\{\beta\}})$, with $p_{\{\beta\}}=0$ and $x_{\{\alpha\}}=0$. All of these coordinate Lagrangian planes are Lagrangian manifolds~\cite{Maslov81}. In four-dimensional phase space, there are four coordinate Lagrangian planes, namely $(x_1,x_2)$, $(x_1,p_2)$, $(p_1,x_2)$ and $(p_1,p_2)$. On the other hand, the plane $(x_1,p_1)$, which contains a conjugate coordinate and momentum pair, is not a Lagrangian manifold. 

Now let us consider the surface $\Lambda^2$, that we discussed in the previous subsections. By equation~(\ref{eq:eikonalLagrangeBrackets}), the Lagrange bracket $[\tau,\phi]_{\mathcal{X},\mathcal{P}}$ vanishes. Furthermore, the Lagrange brackets $[\tau,\tau]_{\mathcal{X},\mathcal{P}}$ and $[\phi,\phi]_{\mathcal{X},\mathcal{P}}$ vanish by antisymmetry. Therefore, we conclude that the surface $\Lambda^2$ is a Lagrangian manifold, as we already anticipated in the previous subsection.

Alternatively, we can look at an isotropic manifold in terms of the action. Suppose that an $m$-dimensional surface $M$ in $2n$-dimensional phase space is (locally) given in the form $p=f(x)$. Then $M$ is an isotropic manifold if and only if there exists an action function $S(x)$ such that $p=\partial S/\partial x$. Since the proof illustrates some important properties of isotropic manifolds, we give it explicitly, based on the exposition in Refs.~\cite{Dubrovin84,Maslov81}. First, suppose that there is a function $S(x)$ such that $p=\partial S/\partial x$. Taking the $x$-coordinates as local coordinates on $M$, we obtain
\begin{equation}
  [ x_j, x_k ]_{X,P} = \sum_i \left( \frac{\partial x_i}{\partial x_j} \frac{\partial p_i}{\partial x_k} - \frac{\partial p_i}{\partial x_j} \frac{\partial x_i}{\partial x_k} \right)
                     = \frac{\partial p_j}{\partial x_k} -  \frac{\partial p_k}{\partial x_j}
                     = \frac{\partial^2 S}{\partial x_k \partial x_j} -  \frac{\partial S}{\partial x_j \partial x_k}
                     = 0 ,
\end{equation}
where the last equality is implied by the equality of mixed partials. Since the Lagrange brackets of the local coordinates vanish, we conclude that $M$ is isotropic.
Second, suppose that $M$ is isotropic. Then we define an action function $S(x)$ as
\begin{equation} \label{eq:actionLagman}
  S = \int_{\sigma_0}^\sigma \langle p(x), \text{d} x \rangle .
\end{equation}
Since $M$ is isotropic, we have $- \text{d} \langle p , \text{d} x \rangle = \text{d} x \wedge \text{d} p = 0$. By the generalized Stokes theorem, this means that the integral over any sufficiently small closed path is zero. Therefore, the integral~(\ref{eq:actionLagman}) is locally path independent, i.e. it only depends on the endpoint $\sigma$ when it is sufficiently close to the fixed initial point $\sigma_0$. Hence, we have $p=\partial S/\partial x$, which proves the theorem. In section~\ref{sec:semiclassics}, we already saw that this action function $S(x)$ plays a crucial role in the construction of the asymptotic solution through the Hamilton-Jacobi equation.

In the beginning of this section, we saw from a direct computation that the surface $\Lambda^2$ is a Lagrangian manifold. This can not only be verified explicitly, but also follows from a more general theorem~\cite{Maslov81}. To this end, we look at the Hamiltonian $L_0$ as the generator of the time evolution $g^t_{L_0}$ of the points on $\Lambda^1$.
This time-evolution preserves the symplectic form, and therefore also the Lagrange brackets. Hence, the time-evolution of an isotropic manifold $M$ generates new isotropic manifolds. Furthermore, when $L_0(x,p)$ is constant on the isotropic manifold $M$, then the union
\begin{equation}
  \tilde M = \bigcup_{t}\,  g^t_{L_0} M
\end{equation}
is again an isotropic manifold. For the proof of this statement we refer to Ref.~\cite{Maslov81}. 
Since our surface $\Lambda^1$ is one-dimensional, it automatically satisfies the requirements of a Lagrangian manifold. Furthermore, since $L_0(x,p)$ equals the constant $E$ on $\Lambda^1$, we conclude from the theorem above that $\Lambda^2$ is an isotropic manifold. Since $\Lambda^2$ is two-dimensional, we subsequently conclude that it is a Lagrangian manifold. Alternatively, we can view $\Lambda^2$ as the union of the one-dimensional isotropic manifolds that arise from the time-evolution generated by $\mathcal{L}_0(x,p)$:
\begin{equation}
  \Lambda^2 = \bigcup_{t}\,  g^t_{L_0} \Lambda^1 = \bigcup_{\tau}\,  g^\tau_{\mathcal{L}_0} \Lambda^1
\end{equation}
We remark that the manifold $\Lambda^2$ constructed in this way is invariant with respect to the time-evolution, i.e. $g^t_{L_0} \Lambda^2 = g^\tau_{\mathcal{L}_0} \Lambda^2 = \Lambda^2$.

In section~\ref{subsec:caustics}, we looked at a typical example of the surface $\Lambda^2$, shown in figure~\ref{fig:LagMan}. We suggested that in the neighborhood of the folds it might be possible to construct an asymptotic solution using the projection onto the coordinate Lagrangian plane $(x_1,p_2)$, since this projection is a one-to-one map.
It turns out that the fact that our surface is a Lagrangian manifold is crucial for such a one-to-one map to exist. 
In fact, it can be shown that, for any point $(x,p)$ on a Lagrangian manifold, it is always possible to find a coordinate Lagrangian plane onto which a neighborhood can be projected with a one-to-one map. For the proof of this theorem we refer e.g. to Refs.~\cite{Maslov81,Arnold89}.

Because of this theorem, we can introduce a special kind of atlas on the Lagrangian manifold, which consists of so-called regular and singular charts~\cite{Maslov81}. An atlas $\Omega = \{\Omega_n , n=1 \ldots N\}$ is a set of $N$ charts, which together cover the entire Lagrangian manifold.
In regular charts $\Omega_i$, we require that the system of equations $x = X(t,\phi)$ has a unique solution $(t(x),\phi(x))$, which means that we can use the coordinates $x$ as local coordinates on such charts. In particular, the Jacobian~(\ref{eq:JacobianX}) does not vanish in a regular chart, so it consists of regular points.
For the example presented in figure~\ref{fig:LagMan}, this means that we need at least three regular charts. 
In section~\ref{sec:assumptions}, we assumed that the potential $U(x)$ and the mass $m(x)$ are constant outside a certain domain. Furthermore, we stated that all trajectories of the Hamiltonian system run away to infinity. In the present context, this means that the number of leaves of our Lagrangian manifold is finite~\cite{Kucherenko69,Vainberg89}, and hence that we need a finite number of charts.

In the singular charts $\Omega_i^s$, which we mark with the upper index $s$, there are focal points at which the Jacobian~(\ref{eq:JacobianX}) vanishes. Hence the system of equations $x = X(t,\phi)$ does not have a unique solution. However, by the theorem above, we can find a different Lagrangian plane onto which such a chart can be projected in a one-to-one way. This can for instance be the plane $(x_1,p_2)$, in which case we require that the Jacobian $\partial (x_1, p_2)/\partial (\tau,\phi)$ does not vanish. We can then use the coordinates $(x_1,p_2)$ as local coordinates on these charts.
In the next subsection, we investigate the focal points in more detail.

\subsection{The classification of singular points} \label{subsec:classification}

In section~\ref{subsec:caustics}, we defined a caustic as the set of singular points of the projection of the Lagrangian manifold $\Lambda^2$ onto the plane $(x_1,x_2)$. In this section, we look deeper into the nature of these focal points. We are mainly interested in what happens to the set of singular points when the surface $\Lambda^2$ changes slightly. Such changes can be caused by small changes in the potential $U(x)$ or the mass $m(x)$. This question was first considered by Whitney~\cite{Whitney55}, who established the properties of smooth maps from two-dimensional manifolds to two-dimensional manifolds. He found that the shape of the caustic shown in figure~\ref{fig:LagMan} is generic, in a sense that we will specify further on. His analysis was the starting point for the study of singularities of differentiable maps, discussed in detail in the textbook~\cite{Arnold82}. This subject is closely related to the field of catastrophe theory, developed in Refs.~\cite{Thom72,Arnold75}. For a broader introduction into these subjects, we refer to the textbook~\cite{Poston78} and to Ref.~\cite{Berry80}.

In section~\ref{subsec:eikonal}, we saw that $\langle \mathcal{X}_\tau , \mathcal{X}_\phi \rangle=0$. Consequently, the Jacobian factorizes, i.e. $|\mathcal{J}| = |\mathcal{X}_\tau|  |\mathcal{X}_\phi|$. Since $|\mathcal{X}_\tau| = C^2(\mathcal{X})$ by the Hamiltonian system~(\ref{eq:Ham-Dirac-2}), the velocity vector is always nonzero. Hence, the singular points, at which $\mathcal{J}$ vanishes, correspond to the points with $\mathcal{X}_\phi=0$. The rank of the matrix $(\mathcal{X}_\tau , \mathcal{X}_\phi)$ therefore equals two for regular points, and one for singular points. With these observations, one can show that the derivative $\mathcal{J}_\tau$ does not vanish at the focal points. Since the proof is rather elaborate, we postpone it to the very end of this section and instead first look at its consequences. Our discussion follows the general line of Ref.~\cite{Whitney55}, making use of the properties of eikonal coordinates, see also Ref.~\cite{Dobrokhotov14}.

Let us consider the set of singular points in the space of coordinates $(\tau, \phi)$. Since $\mathcal{J}_\tau \neq 0$, the implicit function theorem tells us that these points form a smooth curve in this space. Let us consider a smooth parametrization $g(s)$ of this set in the space of coordinates $(\tau, \phi)$. Following Ref.~\cite{Whitney55}, we call a singular point a fold point if
\begin{equation} \label{eq:foldDefgeneral}
  \frac{\text{d} \mathcal{X}(g(s))}{\text{d} s} \neq 0 
\end{equation}
at that point, and we call a singular point a cusp point if at that point
\begin{equation} \label{eq:cuspDefgeneral}
  \frac{\text{d} \mathcal{X}(g(s))}{\text{d} s} = 0 , \qquad \frac{\text{d}^2 \mathcal{X}(g(s))}{\text{d} s^2} \neq 0 .
\end{equation}
These definitions are independent of the specific parametrization, and we exploit this fact to considerably simplify these conditions in our present context. In what follows, we denote quantities that are evaluated at the focal point with coordinates $(\tau^*,\phi^*)$ with a star, e.g. $\mathcal{X}^*=\mathcal{X}(\tau^*,\phi^*)$.

First, we note that the vector $V(\tau,\phi) = (-\mathcal{J}_\phi,\mathcal{J}_\tau)$ does not vanish anywhere. Second, it is tangent to the level sets of $\mathcal{J}$, since the directional derivative
\begin{equation} \label{eq:tangentVecJ}
  \nabla_V \mathcal{J} = -\mathcal{J}_\phi \frac{\partial \mathcal{J}}{\partial \tau} + \mathcal{J}_\tau \frac{\partial \mathcal{J}}{\partial \phi} = 0 .
\end{equation}
Therefore, it is in particular tangent to the set of singular points in the space of coordinates $(\tau, \phi)$. Hence, we can choose a smooth parametrization $\tilde g(s)$ such that 
\begin{equation} \label{eq:gtildeparam}
  \frac{\text{d} \tilde g(s)}{\text{d} s} = V(\tilde g(s)).
\end{equation}
With this parametrization, condition~(\ref{eq:foldDefgeneral}) becomes
\begin{equation}
  0 \neq \frac{\text{d} \mathcal{X}(\tilde g(s))}{\text{d} s} = -\mathcal{J}_\phi^* \mathcal{X}_\tau^* + \mathcal{J}_\tau^* \mathcal{X}_\phi^* = -\mathcal{J}_\phi^* \mathcal{X}_\tau^* .
\end{equation}
Therefore, we conclude that at a fold both the Jacobian $\mathcal{J}^*$ and $\mathcal{X}_\phi^*$ vanish, but that $\mathcal{J}_\phi^*$ does not vanish. This condition can be further simplified by taking into account that 
\begin{equation}
  \mathcal{J}_\phi^* = \det(\mathcal{X}_{\tau\phi}^*,\mathcal{X}_\phi^*) + \det(\mathcal{X}_{\tau}^*,\mathcal{X}_{\phi\phi}^*) = \det(\mathcal{X}_{\tau}^*,\mathcal{X}_{\phi\phi}^*)    
\end{equation}
Now we take the derivative of the equality $\langle \mathcal{X}_{\tau} , \mathcal{X}_{\phi} \rangle = 0$ with respect to $\phi$. After confining our attention to the focal points, we obtain $\langle \mathcal{X}_{\tau}^* , \mathcal{X}_{\phi\phi}^* \rangle=0$, which means that the Jacobian factorizes:
\begin{equation}
  |\mathcal{J}_\phi^*|^2 = | \mathcal{X}_\tau^* |^2 |\mathcal{X}_{\phi\phi}^*|^2 , \qquad \text{or} \;\; \mathcal{J_\phi^*} = \pm C(\mathcal{X^*}) |\mathcal{X}_{\phi\phi}^*| .
\end{equation}
We therefore conclude that the condition that $\mathcal{J}_\phi^*$ does not vanish at a fold point is equivalent to the condition that $\mathcal{X}_{\phi\phi}^*$ does not vanish there.

Let us now consider a cusp point. From definition~(\ref{eq:cuspDefgeneral}) and our previous considerations, we immediately see that both $\mathcal{J}_\phi^*$ and $\mathcal{X}_{\phi\phi}^*$ vanish at a cusp point. With these equalities, the second condition in equation~(\ref{eq:cuspDefgeneral}) becomes
\begin{equation}
  0 \neq \frac{\text{d}^2 \mathcal{X}}{\text{d} s^2} = -\mathcal{J}_\phi^* (-\mathcal{J}_\phi^* \mathcal{X}_\tau^* + \mathcal{J}_\tau^* \mathcal{X}_\phi^*)_\tau +
   \mathcal{J}_\tau^* (-\mathcal{J}_\phi^* \mathcal{X}_\tau^* + \mathcal{J}_\tau^* \mathcal{X}_\phi^*)_\phi
   = - \mathcal{J}_\tau^* \mathcal{J}_{\phi\phi}^* \mathcal{X}_\tau^* 
\end{equation}
Therefore, we conclude that at a cusp point the derivative $\mathcal{J}_{\phi\phi}^*$ does not vanish. Proceeding in a similar fashion as in the case of a fold point, we find that at a cusp point $\mathcal{J}_{\phi\phi}^* = \det(\mathcal{X}_{\tau}^*,\mathcal{X}_{\phi\phi\phi}^*)$. Taking the second derivative of the equality $\langle \mathcal{X}_{\tau} , \mathcal{X}_{\phi} \rangle = 0$ with respect to $\phi$, and confining our attention to the cusp points with $\mathcal{X}_{\phi}^*= \mathcal{X}_{\phi\phi}^* = 0$, we obtain that $\langle \mathcal{X}_{\tau}^* , \mathcal{X}_{\phi\phi\phi}^* \rangle = 0$ at these points. Therefore, we find that the Jacobian factorizes, i.e.
\begin{equation}
  |\mathcal{J}_{\phi\phi}^*|^2 = | \mathcal{X}_\tau^* |^2 |\mathcal{X}_{\phi\phi\phi}^*|^2 , \qquad \text{or} \;\; \mathcal{J_{\phi\phi}^*} = \pm C(\mathcal{X^*}) |\mathcal{X}_{\phi\phi\phi}^*| .
\end{equation}
Hence the condition that $\mathcal{J}_{\phi\phi}^*$ does not vanish at a cusp point is equivalent to the condition that $\mathcal{X}_{\phi\phi\phi}^*$ does not vanish there.

Now let us consider the behavior of $\mathcal{X(\tau,\phi)}$ in the vicinity of a cusp point. From definition~(\ref{eq:cuspDefgeneral}), we see that cusp points are isolated points on the curve of singular points. For a cusp point at $s=s_0$, we have
\begin{equation}
  \left.\frac{\text{d} \mathcal{X}(g(s))}{\text{d} s}\right|_{s=s_0} = 0 , \qquad W = \left.\frac{\text{d}^2 \mathcal{X}(g(s))}{\text{d} s^2}\right|_{s=s_0} \neq 0, \qquad
  \frac{\text{d} \mathcal{X}(g(s))}{\text{d} s} \approx W (s-s_0),
\end{equation}
where the last equality holds for $s$ near $s_0$. Hence, the change of $\mathcal{X}$ along the set of singular points is in the direction $-W$ for $s<s_0$, and in the opposite direction $W$ for $s>s_0$. We may therefore say that the set of singular points ``makes a 180-degree turn'' at the cusp point. Looking back at figure~\ref{fig:LagMan}, we immediately see that the caustic on the plane $(x_1,x_2)$ indeed shows this behavior. This suggests this caustic consists of a single cusp point, and that the other singular points are fold points.

It was shown by Whitney~\cite{Whitney55} that folds and cusps are the only stable singularities that can occur in a mapping from a two-dimensional manifold to a two-dimensional manifold, in the sense that arbitrarily close to any mapping there is a map for which there are only folds and cusps. We therefore conclude that the shape of the caustic shown in figure~\ref{fig:LagMan} is generic, i.e. for an arbitrary potential $U(x)$ and mass $m(x)$ we expect only folds and cusps. An example of an unstable singular point is a sharp focus, where all trajectories come together in a single point. By an arbitrarily small perturbation, such a sharp focus splits up into folds and cusps. 
If the only singularities that occur in the system are folds and cusps, we say that the problem is in general position.

Now we return to the derivative $\mathcal{J}_\tau$ of the Jacobian, and prove, following Refs.~\cite{Dobrokhotov14,Dobrokhotov14b}, that it does not vanish at focal points. This means that all the zeroes of the Jacobian $\mathcal{J}$ on the trajectories are simple. In this proof, we need the determinant 
\begin{equation} \label{eq:JtildeDef}
  \widetilde{\mathcal{J}}(\tau,\phi) = \det(\mathcal{P},\mathcal{P}_\phi) ,
\end{equation}
i.e. the determinant of the matrix composed of the vectors $\mathcal{P}$ and $\mathcal{P}_\phi$. We now show that this determinant, which will play a crucial role in the constructions in section~\ref{sec:maslov}, does not vanish at a singular point.

Let us first show that both vectors $\mathcal{P}$ and $\mathcal{P}_\phi$ are nonvanishing at a focal point, which means that the determinant can only vanish if the two vectors are parallel. First, the Hamiltonian system~(\ref{eq:Ham-Dirac-2}) shows that the vector $\mathcal{P}$ is parallel to $\mathcal{X}_\tau$ and does not vanish anywhere on the Lagrangian manifold $\Lambda^2$. To see that the vector $\mathcal{P}_\phi^*$ does not vanish either, we use the fact that the dimension of the tangent space to $\Lambda^2$ equals the dimension of $\Lambda^2$, which is two. Therefore, the tangent vector $(\mathcal{X}_\phi^T,\mathcal{P}_\phi^T)^T$ cannot vanish. Since $\mathcal{X}_\phi^*=0$, $\mathcal{P}_\phi^*$ has to be nonzero to fulfill this condition. Thus, both $\mathcal{P}$ and $\mathcal{P}_\phi$ are nonvanishing at a singular point.

There are multiple ways to show that the vectors $\mathcal{P}$ and $\mathcal{P}_\phi$ are not parallel, which then implies that the determinant~(\ref{eq:JtildeDef}) is nonzero. We discuss two of them.
We first give a proof by contradiction. To this end, let us assume that $\mathcal{P}^* = \gamma \mathcal{P}_\phi^*$, where $\gamma$ is the constant of proportionality. Using equations~(\ref{eq:eikonalCoordinates}) and~(\ref{eq:eikonalLagrangeBrackets}), we obtain
\begin{equation}
  1 = \langle \mathcal{P}^*, \mathcal{X}_\tau^* \rangle = \gamma \langle \mathcal{P}_\phi^*, \mathcal{X}_\tau^* \rangle = \gamma \langle \mathcal{P}_\tau^*, \mathcal{X}_\phi^* \rangle .
\end{equation}
However, this implies that $\mathcal{X}_\phi^*$ does not vanish, which means that we are not at a singular point. Therefore, the vectors $\mathcal{P}$ and $\mathcal{P}_\phi$ are not parallel.

The second proof shows that the vectors $\mathcal{P}$ and $\mathcal{P}_\phi$ are perpendicular at a singular point. By taking the derivative of the relation $\mathcal{P}^2 = 1/C^2(\mathcal{X})$ with respect to $\phi$, we obtain
\begin{equation}
  2 \langle \mathcal{P}, \mathcal{P}_\phi \rangle = -\frac{2}{C^3(\mathcal{X})} \left\langle \frac{\partial C}{\partial \mathcal{X}} , \mathcal{X}_\phi \right\rangle .
\end{equation}
Since $\mathcal{X}_\phi^*=0$, this means that $\langle \mathcal{P}^*, \mathcal{P}_\phi^* \rangle=0$. Hence, the determinant at the focal point factorizes, i.e. $|\widetilde{\mathcal{J}}^*| = |\mathcal{P}^*| |\mathcal{P}_\phi^*|$. In particular, it does not vanish at a singular point.

Now we return to the derivative $\mathcal{J}_\tau$. We use the fact that $\widetilde{\mathcal{J}}^*$ does not vanish to show that $\mathcal{J}_\tau$ does not vanish at a singular point. First, we note that $\mathcal{J}_\tau^* = \det(\mathcal{X}_{\tau}^*,\mathcal{X}_{\phi\tau}^*)$, since $\mathcal{X}_\phi^*$ vanishes. Using the Hamiltonian system~(\ref{eq:Ham-Dirac-2}) and the fact that $(V,W) = (\mathcal{X}_\phi,\mathcal{P}_\phi)$ is a solution to the variational system~(\ref{eq:varsys}), we find that
\begin{align}
  \mathcal{J}_\tau^* &= \det(\mathcal{X}_{\tau}^*,\mathcal{X}_{\phi\tau}^*) 
    = C^2 \det \left( \mathcal{P}^* , \left\langle \frac{\partial C}{\partial x}, \mathcal{X}_\phi^* \right\rangle \frac{\mathcal{P}^*}{|\mathcal{P}^*|} + \frac{C}{|\mathcal{P}^*|} \mathcal{P}_\phi^* - \frac{\langle \mathcal{P}^*,\mathcal{P}_\phi^* \rangle}{|\mathcal{P}^*|^3} \mathcal{P}^* \right) \nonumber \\
  &= C^4 (\mathcal{X}^*) \widetilde{\mathcal{J}}^*
  \label{eq:JtauCaustic}
\end{align}
Since $\widetilde{\mathcal{J}}^*$ does not vanish, we conclude that $\mathcal{J}_\tau^*$ does not vanish. 
With this proof we complete our geometrical considerations.

\section{Asymptotic solution via the canonical operator} \label{sec:maslov}

In the previous section, we studied the properties of the Lagrangian manifold $\Lambda^2$. In particular, we identified regular and singular points. In section~\ref{subsec:maslovregular}, we use this knowledge to construct a local asymptotic solution to equation~(\ref{eq:L-eff}) for regular points, building on the results of section~\ref{sec:semiclassics}. Subsequently, we discuss various approaches to construct an asymptotic solution for singular points in section~\ref{subsec:maslovapproaches}. In section~\ref{subsec:maslovsingular}, we discuss one of these constructions in detail. We consider a recently proposed representation~\cite{Dobrokhotov14,Dobrokhotov14b} of the asymptotic solution for singular points in which one integrates over the coordinate $\phi$. Section~\ref{subsec:maslovindex} covers the Maslov index, which connects the various local asymptotic solutions. In section~\ref{subsec:maslovintro}, we discuss how we can combine the various local asymptotic solutions to obtain a global asymptotic solution to equation~(\ref{eq:L-eff}). To this end, we introduce the canconical operator, originally proposed by Maslov~\cite{Maslov72,Maslov73,Maslov81}. Finally, we use the asymptotic solution to equation~(\ref{eq:L-eff}) to obtain an asymptotic solution to equation~(\ref{eq:eigenvalue-HPsi}). This section is mainly based on Refs.~\cite{Maslov81,Dobrokhotov14,Dobrokhotov14b,Dobrokhotov15}.

\subsection{Asymptotic solution for regular points} \label{subsec:maslovregular}

In section~\ref{sec:semiclassics}, we made a first attempt to construct an asymptotic solution to equation~(\ref{eq:L-eff}). We found that the asymptotic solution~(\ref{eq:psi-sol-x-jac}), with the amplitude given by equation~(\ref{eq:Amp-sol}), satisfies the commutation relations~(\ref{eq:commutation-relation-x-L0}), (\ref{eq:commutation-relation-x-gen}) and~(\ref{eq:commutation-relation-x-L1}).
However, as we saw in section~\ref{sec:lagman}, the asymptotic solution~(\ref{eq:psi-sol-x-jac}) is not valid near caustics, since the Jacobian vanishes at focal points. This means that we have not yet found a global asymptotic solution, which is valid on the entire configuration space $(x_1,x_2)$. However, the asymptotic solution that we found can still be used locally. In this section, we make this notion more precise. Furthermore, we simplify our asymptotic solution, using the eikonal coordinates that were introduced in section~\ref{subsec:eikonal}.

Let us consider a regular chart $\Omega_i$ on the Lagrangian manifold $\Lambda^2$. This manifold is formed by the solutions $\big(X(t,\phi), P(t,\phi)\big)$ of the Hamiltonian system~(\ref{eq:Ham-Dirac-1}) with inital condition $\Lambda^1$, where the effective Hamiltonian $L_0(x,p)$ equals $E$ on all points of $\Lambda^1$. Since $\Omega_i$ is regular, the Jacobian $J(t,\phi)$ does not vanish anywhere on this chart. Hence, the projection of the chart $\Omega_i$ onto the plane $(x_1,x_2)$ is one-to-one. In other words, the equation $x=X(t,\phi)$ has a unique solution $(t_i(x),\phi_i(x))$ in this chart. We can therefore define~\cite{Maslov81}
\begin{equation} \label{eq:precanonicalX}
  (K_{\Lambda^2(t,\phi)}^{\Omega_i} A_0)(x) = \left. \frac{A_0(t, \phi)}{\sqrt{|J(t, \phi)|}} \exp\left(-\frac{i \pi}{2} \mu_{\Omega_i} \right)  \exp\left(\frac{i}{h} S(t, \phi) \right) \right|_{\begin{subarray}{l}t=t_i(x) \\ \phi=\phi_i(x)\end{subarray}} ,
\end{equation}
cf. equation~(\ref{eq:psi-sol-x-jac}). The operator $K_{\Lambda^2(t,\phi)}^{\Omega_i}$ is called a precanonical operator. The function $S(t, \phi)$ is the action on the Lagrangian manifold $\Lambda^2$, given by equation~(\ref{eq:action}), and solves the Hamilton-Jacobi equation~(\ref{eq:HJ}).
From the computation in section~\ref{sec:semiclassics}, we see that the precanonical operator~(\ref{eq:precanonicalX}) satisfies the commutation relations~(\ref{eq:commutation-relation-x-L0}) and~(\ref{eq:commutation-relation-x-L1}), see also Ref.~\cite{Maslov81}. 
The precanonical operator $K_{\Lambda^2(t,\phi)}^{\Omega_i}$ is therefore an asymptotic solution corresponding to the chart $\Omega_i$ when $A_0(t,\phi)$ is given by expression~(\ref{eq:Amp-sol}), i.e. when $A_0$ is a solution of equation~(\ref{eq:transport2}). As such, it constitutes a ``local asymptotic solution''. The corrections to the leading-order asymptotic solution~(\ref{eq:precanonicalX}) are one order in $h$ higher, meaning that they are $\mathcal{O}(h)$.

Compared to the asymptotic solution presented in equation~(\ref{eq:psi-sol-x-jac}), the precanonical operator~(\ref{eq:precanonicalX}) contains an additional phase factor $\exp(-i \pi \mu_{\Omega_i}/2)$. This factor is necessary because the Jacobian $J$ vanishes at singular points. More precisely, it changes sign when we pass through a singular point along a trajectory, 
since $\mathcal{J}_\tau$ does not vanish. In order to be able to combine our local asymptotic solutions into a global asymptotic solution later on, we have to choose the arguments of $J$ and $\sqrt{J}$ in a consistent way in the different charts. To ensure such a consistent choice, we have to include the phase factor $\exp(-i \pi\mu_{\Omega_i}/2)$. In this paper, we call $\mu_{\Omega_i}$ the Maslov index of the regular chart $\Omega_i$. Strictly speaking, this terminology is somewhat misleading, since the Maslov index is only properly defined for a chain of charts~\cite{Maslov81}. This reflects the fact that we have to make a consistent choice for the argument of the Jacobian across all charts. In practice, we always set the Maslov index to zero for the points on $\Lambda^1$. This automatically fixes the Maslov index for all other charts, which justifies our terminology.
It turns out that the Maslov index is a topological characteristic of the Lagrangian manifold itself~\cite{Arnold67,Maslov72,Maslov81,Mishchenko90,Guillemin77} and that it can be defined without reference to the Hamiltonian system.
We come back to the Maslov index and its computation in section~\ref{subsec:maslovindex}.

We can simplify the precanonical operator~(\ref{eq:precanonicalX}) by performing a coordinate change from $(t,\phi)$ to the eikonal coordinates $(\tau,\phi)$. First, we saw in equation~(\ref{eq:JacobianXrelationEikonal}) that the Jacobians $J$ and $\mathcal{J}$ are related by $R(X)$, the determinant that connects the coordinate systems $(t,\phi)$ and $(\tau,\phi)$. This factor does not vanish, which means that when the inverse functions $(t_i(x),\phi_i(x))$ exist on a chart $\Omega_i$, the inverse functions $(\tau_i(x),\phi_i(x))$ also exist. Second, by equation~(\ref{eq:trajectoriesequal}) the solutions $\big( X(t,\phi),P(t,\phi)\big)$ of the Hamiltonian system~(\ref{eq:Ham-Dirac-1}) equal the solutions $\big( \mathcal{X}(\tau,\phi),\mathcal{P}(\tau,\phi)\big)$ of the Hamiltonian system~(\ref{eq:Ham-Dirac-2}) at the time $\tau=\tau(t,\phi)$. Furthermore, the action~(\ref{eq:actionLagman}) only depends on the point of the Lagrangian manifold. Therefore, equation~(\ref{eq:actionTau}) tells us that once we have made the transformation to eikonal coordinates, the action with respect to the central point $(0,0)$ equals $\tau$. Finally, we can find the amplitude in eikonal coordinates by changing the integration over $t$ in equation~(\ref{eq:Amp-sol}) to an integration over $\tau$, at the expense of a Jacobian factor. We obtain
\begin{equation} \label{eq:BerryPhaseEikonal}
  \begin{aligned}
    \Phi_{sc}(t, \phi) &= -\int_0^t L_1^W(X(t',\phi),P(t',\phi)) \, \text{d} t'  \\
    &= -\int_0^\tau \frac{1}{R(\mathcal{X}(\tau',\phi))} L_1^W(\mathcal{X}(\tau',\phi),\mathcal{P}(\tau',\phi)) \, \text{d} \tau' \equiv \Phi_{sc}(\tau,\phi) ,
  \end{aligned}
\end{equation}
where we have made use of equation~(\ref{eq:timereparametrization}). Of course, the result~(\ref{eq:BerryPhaseEikonal}) should equal our previous expression~(\ref{eq:BerryPhaseDiracCoordfree}) for the semiclassical phase, which only depends on the trajectories themselves and not on their parametrization. Using manipulations similar to those used at the end of section~\ref{sec:semiclassics}, it is easy to show that this is indeed the case.

Taking all of the above simplifications into account, the expression for the precanonical operator corresponding to the regular chart $\Omega_i$ becomes~\cite{Dobrokhotov15,Dobrokhotov14}
\begin{equation} \label{eq:precanonicalXsimplified}
  (K_{\Lambda^2(t,\phi)}^{\Omega_i} A_0)(x) = \left. \frac{A_0^0(\phi_i)}{\sqrt{R(\mathcal{X}) C(\mathcal{X}) |\mathcal{X}_\phi|}} \exp\left(i \Phi_{sc}(\tau_i,\phi_i) -\frac{i \pi}{2} \mu_{\Omega_i} + \frac{i}{h} \tau_i \right) \right|_{\begin{subarray}{l}\tau_i=\tau_i(x) \\ \phi_i=\phi_i(x)\end{subarray}} ,
\end{equation}
where we have also used equation~(\ref{eq:JacobianXEikonalComputed}). Note that for graphene the semiclassical phase $\Phi_{sc,\alpha}$ depends on $\alpha$ since $L_{1,\alpha}^W$ does, see equation~(\ref{eq:L1Dirac}).

\subsection{Alternative approaches for singular charts} \label{subsec:maslovapproaches}

In the previous subsection, we constructed an asymptotic solution to equation~(\ref{eq:L-eff}) corresponding to a regular chart~$\Omega_i$. In this subsection, we discuss several alternative methods to construct an asymptotic solution corresponding to a singular chart. Although this discussion may seem rather abstract, it will give us some essential tools for the explicit construction of the precanonical operator corresponding to singular charts in the next subsection.

The conventional way~\cite{Maslov72,Maslov81} to construct an asymptotic solution corresponding to a singular chart uses the idea that we set forth in section~\ref{subsec:caustics}: although for a singular chart $\Omega_i^s$ the projection of the Lagrangian manifold $\Lambda^2(t,\phi)$ onto the plane $(x_1,x_2)$ is not one-to-one, the projection onto one of the planes $(p_1,x_2)$ or $(x_1,p_2)$ is one-to-one. We remark that we only need a single momentum coordinate, since the rank of the matrix $(\mathcal{X}_\tau,\mathcal{X}_\phi)$ equals one at singular points, see section~\ref{subsec:classification}. Therefore, we can perform a Fourier transform of the operator $\hat L$ with respect to one of the momentum coordinates. For definiteness, we henceforth assume that we transform with respect to $x_2$. Then, we can use an Ansatz similar to the one used in equation~(\ref{eq:semicl-Ansatz}), namely~\cite{Maslov81}
\begin{equation} \label{eq:Ansatz-mixed-rep}
  \overline{\psi}(x_1,p_2) = \overline{\varphi}(x_1,p_2) e^{i \overline{S}(x_1,p_2)/h} .
\end{equation}
Inserting this Ansatz into the Fourier transform of the effective scalar equation~(\ref{eq:L-eff}), we obtain another Hamilton-Jacobi equation. This equation gives rise to the same system of Hamilton equations and therefore generates the same Lagrangian manifold as we had before. The connection between the Hamilton-Jacobi equation and the system of Hamilton equations is made through the action~\cite{Maslov81} $\overline{S}(x_1,p_2) = S(x_1,X_2(x_1,p_2)) - p_2 X_2(x_1,p_2)$, which is the Legendre transform of the action $S(x_1,x_2)$.

Collecting terms of order $h$, we obtain a new transport equation~\cite{Maslov81}, similar to equation~(\ref{eq:transport1}). However, when we introduce the Jacobian 
\begin{equation} \label{eq:Jacobian-fourier-p1}
  \overline{J}(t,\phi)=\det \frac{\partial(X_1, P_2)}{\partial(t,\phi)}
\end{equation}
and set $A_0(t,\phi) = \overline{\varphi}_0 \sqrt{\overline{J}}$, we once again obtain equation~(\ref{eq:transport2}). Therefore, we can view the transport equation as a geometrical object associated to the Lagrangian manifold. In the mathematical literature, this notion is formalized using the concept of a half-density~\cite{Guillemin77,Bates97}, which is however beyond the scope of this text.

Following the above discussion, we define the precanonical operator corresponding to the singular chart $\Omega_i^s$ as the inverse Fourier transform of equation~(\ref{eq:Ansatz-mixed-rep})~\cite{Maslov81}
\begin{multline} \label{eq:precanonical-singular-conventional}
  (K_{\Lambda^2(t,\phi)}^{\Omega_i^s} A_0)(x) = \mathcal{F}^{-1}_{p_2 \to x_2} \overline{\psi}(x_1,p_2) \\
  = \frac{e^{i\pi/4}}{\sqrt{2\pi h}} \int_{-\infty}^\infty \text{d} p_2
  \left. \frac{A_0(t,\phi)}{| \overline{J}(t,\phi) |^{1/2} } 
  e^{-\frac{i \pi}{2} \mu_{\Omega_i^s} }
  e^{\frac{i}{h} (\overline{S}(x_1,p_2) + p_2 x_2)}
  \right|_{\begin{subarray}{l}t=t_i(x_1,p_2) \\ \phi=\phi_i(x_1,p_2)\end{subarray}} 
  .
\end{multline}
We call $\mu_{\Omega_i^s}$ the Maslov index of a singular chart and discuss it in greater detail later on.
Like the precanonical operator corresponding to regular charts, the precanonical operator~(\ref{eq:precanonical-singular-conventional}) satisfies two commutation formulas. First, for a pseudodifferential operator $\hat{Q}$, one has (cf. equation~(\ref{eq:commutation-relation-x-gen}))
\begin{equation} \label{eq:commutation-relation-precanop-sing-gen}
  Q(x, \hat{p},h) K_{\Lambda^2(t,\phi)}^{\Omega_i^s} A_0 = K_{\Lambda^2(t,\phi)}^{\Omega_i^s} \big( Q_0\left(x,p \right) A_0 + \mathcal{O}(h) \big) ,
\end{equation}
where $Q_0$ is principal symbol of $\hat{Q}$. Intuitively, this formula can be understood by realizing that terms of order $h^0$ only arise when the differential operators $\hat{p}$ act on the exponential term containing the action. Since the derivatives of the action generate the Lagrangian manifold with coordinates $(x,p)$, we obtain equation~(\ref{eq:commutation-relation-precanop-sing-gen}). Note that, by virtue of equation~(\ref{eq:commutation-relation-x-gen}), the precanonical operator~(\ref{eq:precanonicalX}) corresponding to regular charts satisfies the same commutation formula.

The second commutation formula~\cite{Maslov81,Vainberg89b} holds specifically for the effective Hamiltonian $\hat{L}$. Since the action satisfies the Hamilton-Jacobi equation, one has (cf. equation~(\ref{eq:commutation-relation-x-L1}))
\begin{equation} \label{eq:commutation-relation-precanop-sing-L1}
  \big( L(x, \hat{p}, h) - E \big) K_{\Lambda^2(t,\phi)}^{\Omega_i^s} A_0 = -i h K_{\Lambda^2(t,\phi)}^{\Omega_i^s} \bigg( \frac{\text{d} A_0}{\text{d} t} + i L_1 A_0 - \frac{1}{2} \sum_j \frac{\partial^2 L_0}{\partial x_j \partial p_j} A_0 + \mathcal{O}(h) \bigg) .
\end{equation}
Hence, when $A_0(t,\phi)$ is given by expression~(\ref{eq:Amp-sol}), i.e. when it is a solution of equation~(\ref{eq:transport2}), the precanonical operator~(\ref{eq:precanonical-singular-conventional}) is an asymptotic solution of equation~(\ref{eq:L-eff}) corresponding to the singular chart $\Omega_i^s$. The corrections to this asymptotic solution, which form the higher-order terms in the asymptotic expansion, come from the term with $\mathcal{O}(h)$ on the right-hand side of equation~(\ref{eq:commutation-relation-precanop-sing-L1}). When the precanonical operator itself is nonzero, these corrections are of order $h K_{\Lambda^2(t,\phi)}^{\Omega_i^s} A_0$.
As before, the commutation formula~(\ref{eq:commutation-relation-precanop-sing-L1}) is also satisfied by the precanonical operator~(\ref{eq:precanonicalX}) corresponding to regular charts.

The asymptotic solution~(\ref{eq:precanonical-singular-conventional}) that we have constructed is given in the form of an integral representation. Since $h$ is small, this integral contains a rapidly oscillating exponent, which makes it hard to tackle it numerically. Therefore, this integral should be simplified in the vicinity of fold and cusp points. In the previous section, we saw that we could essentially simplify several Jacobians and the defining expressions for caustics by introducing eikonal coordinates on the Lagrangian manifold. In the previous subsection, we saw how these coordinates also lead to simplifications in the asymptotic solution corresponding to regular charts. Therefore, we may also be able to simplify our expression for the precanonical operator corresponding to singular charts by introducing eikonal coordinates.
introducing eikonal coordinates in singular charts may also simplify our expressions for the precanonical operator near folds and cusps. We can introduce such a new parametrization of the Lagrangian manifold in two ways, manipulating either the classical symbol $L_0$ or the quantum operator $\hat{L}$.

In the first method~\cite{Dobrokhotov15}, we introduce the eikonal coordinates on the Lagrangian manifold. As we saw in section~\ref{subsec:eikonal}, this can be done by manipulating the classical Hamiltonian.
Now suppose that we have constructed a precanonical operator $K_{\Lambda^2(\tau,\phi)}^{\Omega_i^s} A_0$ starting from a Lagrangian manifold with eikonal coordinates. Our goal is to find the precanonical operator $K_{\Lambda^2(t,\phi)}^{\Omega_i^s} A_0$, where the Lagrangian manifold is parametrized by the coordinates $(t,\phi)$, since only this object is an asymptotic solution to our original equation. In equation~(\ref{eq:JacobianXrelationEikonal}), we already saw that the Jacobians $J$ and $\mathcal{J}$ are related by a simple Jacobian factor. Equation~(\ref{eq:precanonicalX}) then suggests the following relation, which clearly holds for precanonical operators corresponding to regular charts:
\begin{equation} \label{eq:precanonicalTimeChange}
  (K_{\Lambda^2(t,\phi)}^{\Omega_i} A_0)(x) = \left(K_{\Lambda^2(\tau,\phi)}^{\Omega_i} \left(\det\frac{\partial (\tau,\phi)}{\partial (t,\phi)}\right)^{-1/2} A_0\right)(x) .
\end{equation} 
It can be shown that this relation also holds for precanonical operators corresponding to singular charts~\cite{Dobrokhotov15,Maslov81}, meaning that it holds for all precanonical operators.
Therefore, we can easily transform an asymptotic solution for a Lagrangian manifold with eikonal coordinates into an asymptotic solution for a Lagrangian manifold with our initial coordinates.

In the second method, we consider the following decomposition of the effective Hamiltonian $\hat{L}$:
\begin{equation} \label{eq:operator-relation-Hams-main}
  L_0(x,\hat{p}) + h L_1(x,\hat{p}) - E = 
    \hat{B}^\dagger \big( \mathcal{L}_0(x,\hat{p}) + h \mathcal{L}_1(x,\hat{p}) - 1 \big) \hat{B} + \mathcal{O}(h^2) .
\end{equation}
The operators in this expression are related to their symbols by standard quantization, see equation~(\ref{eq:standard-quantization}). In particular, the symbols $L_0$ and $L_1$ are defined by equations~(\ref{eq:L0}) and~(\ref{eq:L1}), respectively. The symbol $\mathcal{L}_0$ was defined in equation~(\ref{eq:L0Eikonal}). 
Because of the decomposition~(\ref{eq:operator-relation-Hams-main}), we can construct an asymptotic solution for equation~(\ref{eq:L-eff}) by first constructing an asymptotic solution for the new equation
\begin{equation} \label{eq:eigenvalue-eikonal-main}
  \big( \mathcal{L}_0(x,\hat{p}) + h \mathcal{L}_1(x,\hat{p}) \big) \tilde{\psi} = \tilde{\psi} .
\end{equation}
Subsequently, given an asymptotic solution $\tilde{\psi}$ of this equation, an asymptotic solution $\psi$ for equation~(\ref{eq:L-eff}) can be found by computing
\begin{equation}
  \psi = ( \hat{B} )^{-1} \tilde{\psi} .
\end{equation}
It is important to note that one can only perform the decomposition~(\ref{eq:operator-relation-Hams-main}) because the equality $\mathcal{L}_0=1$ can be reached by algebraically manipulating the equality $L_0=E$, as we have seen in section~\ref{subsec:eikonal}.
In that section, we also saw that the classical Hamiltonian $\mathcal{L}_0$ automatically gives rise to eikonal coordinates on the Lagrangian manifold $\Lambda^2$. Hence, by constructing an asymptotic solution for equation~(\ref{eq:eigenvalue-eikonal-main}), we automatically introduce eikonal coordinates on the Lagrangian manifold, without reparametrizing time.
Since introducing eikonal coordinates on the Lagrangian manifold by a reparametrization of time is technically simpler, we do not pursue this operator decomposition in the main text of this paper. However, we compute the symbols $B$ and $\mathcal{L}_1$ explicitly in appendix~\ref{app:operatorReduction} and show that this method leads to the same asymptotic solution for the wavefunction $\psi$.

Now that we have introduced eikonal coordinates on the Lagrangian manifold, we want to use them to simplify the asymptotic solution~(\ref{eq:precanonical-singular-conventional}) near folds and cusps. However, in practice, it is not at all straightforward to perform such simplifications. In particular, we need to carefully consider which singular charts we use. In section~\ref{subsec:caustics}, we considered a singular chart with the coordinates $(x_1,p_2)$, based on the graphical representation of the Lagrangian manifold in figure~\ref{fig:LagMan}. However, for other singular points the Jacobian $\overline{J}$, see equation~(\ref{eq:Jacobian-fourier-p1}), may vanish, in which case we need to construct an asymptotic solution using the momentum coordinate $p_1$ instead. Note that we never need to use the coordinates $(p_1,p_2)$, since the rank of the matrix $(\mathcal{X}_\tau,\mathcal{X}_\phi)$ equals one for singular points. 
Thus, we need to carefully inspect the various Jacobians at all singular points to decide which of them are nonvanishing. We can then choose our singular charts based on the regions in which these Jacobians are nonzero. This has to be done for each problem separately, and may lead to a large number of singular charts.

Instead, we can take full advantage of the eikonal coordinates on the Lagrangian manifold by considering a new representation of the asymptotic solution in the vicinity of singular points.
In section~\ref{subsec:caustics}, we stated that we can also identify points in singular charts using coordinates different from $(x_1, p_2)$. For instance, we can supplement the coordinate $x$ with a third coordinate $\zeta$ and parametrize points with this triple. The simplest choice for $\zeta$ would be $\phi$, the coordinate that labels the trajectories. In the remainder of this subsection, we  explore how, in general, a parametrization $(x,\zeta)$ can give rise to a new representation of the precanonical operator corresponding to singular charts. In the next subsection, we specialize to the case of eikonal coordinates and explain how one can obtain an alternative representation of the asymptotic solution in the vicinity of singular points.

Let us therefore consider the general Fourier integral:
\begin{equation} \label{eq:Fourier-integral-general}
  I(x) = \int \text{d}\zeta \, F(x,\zeta) \exp\left( \frac{i}{h} \Phi(x,\zeta) \right) ,
\end{equation}
where $F(x,\zeta)$ is called the amplitude function and $\Phi(x,\zeta)$ is called the phase function. These integrals are the building blocks of the theory of Fourier integral operators~\cite{Hormander71,Duistermaat96}. There is an intimate relation between Fourier integral operators and the canonical operator, which is discussed in (e.g.) Refs.~\cite{Duistermaat74,Yoshikawa75,Danilov81,Nazaikinskii81,Mishchenko90,Dobrokhotov14b}. Our discussion of the relation between the Fourier integral~(\ref{eq:Fourier-integral-general}) and the canonical operator closely follows the discussion in Ref.~\cite{Dobrokhotov14b}.

We want the Fourier integral~(\ref{eq:Fourier-integral-general}) to be a new representation of the precanonical operator corresponding to singular charts. Therefore, it should coincide with the conventional representation of the precanonical operator~(\ref{eq:precanonical-singular-conventional}) up to higher-order terms.
Since $h$ is small, the integral~(\ref{eq:Fourier-integral-general}) contains a rapidly oscillating exponent and its leading-order term only depends on the stationary points. The stationary points of the phase function comprise the set
\begin{equation}
  Z_\Phi = \{ \, (x,\zeta) \, | \, \Phi_\zeta(x,\zeta)=0 \, \} ,
\end{equation}
where the subscript $\zeta$ denotes the partial derivative, i.e., $\Phi_\zeta = \partial \Phi/\partial \zeta$.
We demand that the phase function is non-degenerate, which means that the matrix $(\Phi_{\zeta x}, \Phi_{\zeta\zeta})$ of second derivatives has maximal rank on $Z_\Phi$. In our simple example, where $\zeta$ is a scalar, this translates to the condition that one of the derivatives is nonzero. In this case, the implicit function theorem guarantees that $Z_\Phi$ is a smooth two-dimensional manifold. 
Let us then consider the mapping
\begin{equation}
  j_\Phi : Z_\Phi \to \mathbb{R}^4_{x,p}: (x,\zeta) \mapsto (x,\Phi_x) ,
\end{equation}
which is an immersion of $Z_\Phi$ into four-dimensional phase space~\cite{Dobrokhotov14b,Duistermaat96}, with the momentum given by $p=\Phi_x$. In fact, one can show that the image of $Z_\Phi$ under this mapping is a two-dimensional Lagrangian manifold~\cite{Dobrokhotov14b,Duistermaat96}. We have therefore seen that a non-degenerate phase function $\Phi(x,\zeta)$ defines a Lagrangian manifold through its set of stationary points.

In order for the Fourier integral~(\ref{eq:Fourier-integral-general}) to represent an asymptotic solution corresponding to the chart $\Omega_i^s$, it is necessary that, on the chart $\Omega_i^s$, the Lagrangian manifold generated by the phase function coincides with the Lagrangian manifold $\Lambda^2$. In fact, we can formulate a more precise statement~\cite{Dobrokhotov14b,Duistermaat96}: there exists an amplitude function $F(x,\zeta)$ such that $I(x)$ equals the precanonical operator corresponding to the chart $\Omega_i^s$ up to higher-order terms, if and only if, on the chart $\Omega_i^s$, the Lagrangian manifold generated by $\Phi(x,\zeta)$ coincides with $\Lambda^2$. When one has made a choice for a phase function $\Phi(x,\zeta)$, one can subsequently compute the corresponding amplitude function $F(x,\zeta)$, as shown in Ref.~\cite{Dobrokhotov14b}.

From the above discussion, it is apparent that there are in principle many equivalent representations of the precanonical operator corresponding to singular charts. However, particular representations may be considerably simpler to construct and to implement numerically. Furthermore, they may be much easier to simplify near fold and cusp caustics. Recently, such a new representation was proposed for problems in which eikonal coordinates can be introduced~\cite{Dobrokhotov14,Dobrokhotov14b}. This new representation uses an integral over the coordinate $\phi$, which has a clear physical interpretation: it labels the trajectories on the Lagrangian manifold. 
Compared to the conventional representation, this means that we no longer have to consider whether we have to choose the coordinates $(p_1, x_2)$ or $(x_1, p_2)$ on the Lagrangian manifold. Instead, we now have a representation that has the same form for all singular points, which considerably simplifies the construction. Furthermore, the new representation makes it easier to simplify the precanonical operator in the vicinity of folds and cusps. As shown in Ref.~\cite{Dobrokhotov14}, this new representation is equal to the conventional representation up to higher-order terms.

We remark that a precursor of the new representation~\cite{Dobrokhotov14,Dobrokhotov14b} can be found in Ref.~\cite{Dobrokhotov08}. Furthermore, in Ref.~\cite{Dobrokhotov17}, this new representation was generalized to Hamiltonians that do not admit a parametrization in terms of eikonal coordinates. 
For completeness, we mention that there are also other ways to express the wavefunction as an integral over the Lagrangian manifold. In some of these cases, the integrand is still a rapidly oscillating function~\cite{Buslaev69,Buslaev69b,Buslaev70}. However, in other cases, one integrates over Gaussian coherent states~\cite{Karasev92,Zheng13}.

In the next section, we show how to construct the new representation~\cite{Dobrokhotov14,Dobrokhotov14b} corresponding to singular charts, discussing both the phase function and the amplitude function in detail. In section~\ref{sec:causticsSolution}, we discuss how to simplify this expression in the vicinity of fold points and cusp points.
In order to provide the reader with a complete picture, some additional details on how one can implement the conventional representation of the canonical operator~(\ref{eq:precanonical-singular-conventional}) corresponding to singular charts are given in appendix~\ref{app:conventionalCanonicalOp}. In particular, we show that the leading-order approximation near singular points coincides with the leading-order approximation that is obtained from the new representation~\cite{Dobrokhotov14}. These computations turn out to be fairly involved and less convenient than the computations for the new representation.

Figure~\ref{fig:comm-diag-can-op} summarizes our discussion on the alternative ways to obtain the precanonical operator $K_{\Lambda^2(t,\phi)}^{\Omega_i^s} A_0$ corresponding to singular charts. It also shows the sections in which the respective steps are discussed in greater detail. 

\begin{figure}[bt]
\begin{displaymath}
\hspace*{-1cm}
\begin{CD}
L(x,\hat{p},h) @>\text{\parbox{1.8cm}{Sections \ref{sec:semiclassics}, \ref{subsec:lagman}}}>> \Lambda^2(t,\phi) @>\phantom{\hspace*{2.75cm}}>> K_{\Lambda^2(t,\phi)}^{\Omega_i^s} A_0 \\
@VV\text{\ref{app:operatorReduction}}V @VV\text{ \parbox{2cm}{Equation~(\ref{eq:timereparametrization});\\Sections \ref{subsec:eikonal}, \ref{subsec:lagman}}}V @AA\text{ \parbox{2.0cm}{Equation~(\ref{eq:precanonicalTimeChange});\\Section \ref{subsec:maslovsingular}}}A \\
\mathcal{L}(x,\hat{p},h) @>\phantom{\hspace*{1.8cm}}>\text{\ref{app:operatorReduction}}> \Lambda^2(\tau,\phi) @>>\text{\parbox{2.75cm}{New: Sections \ref{subsec:maslovsingular}, \ref{sec:causticsSolution};\\Old: \ref{app:conventionalCanonicalOp}}}> K_{\Lambda^2(\tau,\phi)}^{\Omega_i^s} A_0 \\
\end{CD}
\end{displaymath}
\caption{Commutative diagram that shows the alternative ways to obtain the precanonical operator $K_{\Lambda^2(t,\phi)}^{\Omega_i^s} A_0$ corresponding to singular charts. It also shows the section in which the respective step is discussed.}
\label{fig:comm-diag-can-op}
\end{figure}

\subsection{Asymptotic solution near singular points} \label{subsec:maslovsingular}

In this subsection, we introduce the recently proposed new representation~\cite{Dobrokhotov14} for the precanonical operator corresponding to singular charts with eikonal coordinates. Since the problem that we discuss in this paper is two-dimensional, we confine ourselves to this case, noting that an extension to higher dimensions was presented in Ref.~\cite{Dobrokhotov14b}.

Let us therefore consider the Lagrangian manifold $\Lambda^2$ with eikonal coordinates $(\tau,\phi)$. We want to obtain a new representation of the precanonical operator corresponding to the singular chart $\Omega_i^s$, in the form of the Fourier integral~(\ref{eq:Fourier-integral-general}) with $\zeta=\phi$. In the previous subsection, we have seen that this requires that, on the singular chart $\Omega_i^s$, the Lagrangian manifold generated by the phase function $\Phi(x,\phi)$ coincides with $\Lambda^2$.
Therefore, we begin our analysis by choosing an appropriate phase function, following the exposition in Ref.~\cite{Dobrokhotov14}. Note that the requirement stated there, namely that the one-form $\langle P, \text{d} X \rangle$ does not vanish, is automatically satisfied in our case because of equation~(\ref{eq:eikonalCoordinates}).
Let us consider the equation~\cite{Dobrokhotov14}
\begin{equation} \label{eq:defTauSingular}
  \langle \mathcal{P}(\tau,\phi), x - \mathcal{X}(\tau,\phi) \rangle = 0 ,
\end{equation}
where $\mathcal{X}(\tau,\phi)$ and $\mathcal{P}(\tau,\phi)$ are the solutions of the Hamiltonian system~(\ref{eq:Ham-Dirac-2}).
By the implicit function theorem, this equation defines a smooth function $\tau=\tau(x,\phi)$. Our first step is to show that this function generates the Lagrangian manifold $\Lambda^2$. Therefore, we consider the set $Z_\tau$ of stationary points, which consist of points with $\tau_\phi=0$. Computing the partial derivatives, we obtain
\begin{equation} \label{eq:tauStationaryLagMan}
  \tau_\phi = \frac{\langle\mathcal{P}_\phi, x-\mathcal{X}\rangle}{1 - \langle\mathcal{P}_\tau, x-\mathcal{X}\rangle} , \qquad
  \tau_x = \frac{\mathcal{P}}{1 - \langle\mathcal{P}_\tau, x-\mathcal{X}\rangle} .
\end{equation}
In section~\ref{subsec:classification}, we showed that $\widetilde{\mathcal{J}}=\det(\mathcal{P},\mathcal{P}_\phi)$ does not vanish at the singular points. By continuity, there is a certain neighborhood of the singular points in which this also holds. In this neighborhood, the vectors $\mathcal{P}$ and $\mathcal{P}_\phi$ are not parallel. 
Hence, equation~(\ref{eq:defTauSingular}) and the requirement $\tau_\phi=0$ can only be satisfied simultaneously when $x-\mathcal{X}=0$. The set $Z_\tau$ therefore consists of the points $x$ with $x=\mathcal{X}(\tau(x,\phi),\phi)$.
The phase function $\tau(x,\phi)$ is nondegenerate on $Z_\tau$, since we have $\tau_{\phi x}=\mathcal{P}_\phi \neq 0$, where the last inequality follows from the fact that $\det(\mathcal{P},\mathcal{P}_\phi)$ does not vanish on $Z_\tau$.
From the second equality in equation~(\ref{eq:tauStationaryLagMan}), we then immediately see that $\tau_x=\mathcal{P}$. Therefore, the Lagrangian manifold that is defined by the phase function~(\ref{eq:defTauSingular}) coincides with our manifold $\Lambda^2$ in a neighborhood of the singular points in which $\det(\mathcal{P},\mathcal{P}_\phi) \neq 0$. The maximal size of this neighborhood is the maximal size of the singular chart $\Omega_i^s$.

Despite the similarities in notation, we emphasize that the nondegenerate phase function $\tau(x,\phi)$ is not the same as the eikonal coordinate $\tau$ and the previously defined inverse function $\tau(x)$. However, there is an important relationship between these three quantities. The function $\tau(x)$ is defined in nonsingular charts, in which the Jacobian $\mathcal{J}$ does not vanish, and gives the value of the eikonal coordinate $\tau$ for a given nonsingular point $x$. For a singular point, at which $\mathcal{J}$ vanishes, such an inverse function does not exist. However, we can consider the nondegenerate phase function $\tau(x,\phi)$, defined by equation~(\ref{eq:defTauSingular}). This equation admits a clear geometric interpretation, as can be seen by considering a fixed point $x_s$. By equation~(\ref{eq:Ham-Dirac-2}), the vector $\mathcal{P}$ is parallel to $\mathcal{X}_\tau$, which is the vector tangent to the trajectories. For a given value of $\phi$, this tangent vector is perpendicular to $x_s-\mathcal{X}(\tau,\phi)$ when we are at the point $\mathcal{X}(\tau,\phi)$ on the trajectory that is closest to the point $x_s$. Thus, $\tau(x_s,\phi)$ represents the time $\tau$ at which we reach the point closest to $x_s$ on the trajectory $\mathcal{X}(\tau,\phi)$ for a given value of $\phi$.
The function $\tau(x_s,\phi)$ has a stationary point for at least one value $\phi_s$, 
which means that $\tau_{\phi}(x_s,\phi_s)$ vanishes. If we define $\tau_s=\tau(x_s,\phi_s)$, then we have $x_s=\mathcal{X}(\tau_s,\phi_s)$, as we showed above. Therefore, we see that, at this point, the value of the nondegenerate phase function $\tau(x,\phi)$ coincides with the eikonal coordinate $\tau$. This is our main motivation to use the letter $\tau$ for the nondegenerate phase function.

Having verified that the nondegenerate phase function $\tau(x,\phi)$ defines the correct Lagrangian manifold, we define the precanonical operator corresponding to a singular chart by~\cite{Dobrokhotov14}
\begin{multline}
  \label{eq:precanonicalXsingular}
  (K_{\Lambda^2(\tau,\phi)}^{\Omega_i^s} A_0)(x) = \frac{e^{i\pi/4}}{\sqrt{2\pi h}} \exp\left(\frac{-i \pi \mu_{\Omega_i^s}}{2} \right) \\
  \times \int_{-\infty}^\infty \text{d} \phi \sqrt{| \det(\mathcal{P},\mathcal{P}_\phi) |} A_0(\tau(x,\phi),\phi) e^{\frac{i}{h}\tau(x,\phi)} .
\end{multline}
The integral has the general form considered in equation~(\ref{eq:Fourier-integral-general}), with the phase function defined by equation~(\ref{eq:defTauSingular}). The amplitude is given by $A_0 \sqrt{\widetilde{\mathcal{J}}}$. Note that this is different from equation~(\ref{eq:precanonical-singular-conventional}), where we divide by the Jacobian $\overline{J}$ instead. 
As we already discussed, the Jacobian $\widetilde{\mathcal{J}} = \det(\mathcal{P},\mathcal{P}_\phi)$ should be nonvanishing on the singular chart $\Omega_i^s$. The sign of this determinant is absorbed in the phase factor $\exp(-i \pi \mu_{\Omega_i^s}/2 )$. This phase factor should be chosen in such a way that the argument of $\sqrt{\det(\mathcal{P},\mathcal{P}_\phi)}$ is consistent with the choice for the argument of $\sqrt{J}$ that was made in section~\ref{subsec:maslovregular}.
One can prove~\cite{Dobrokhotov14} that the precanonical operator~(\ref{eq:precanonicalXsingular}) coincides with the conventional representation of the precanonical operator corresponding to a singular chart, up to higher-order terms. In particular, the quantity $\mu_{\Omega_i^s}$, defined in equation~(\ref{eq:precanonicalXsingular}), is the Maslov index of a singular chart. Furthermore, in section~\ref{subsec:maslovindex}, we show explicitly that, for regular points, the precanonical operator~(\ref{eq:precanonicalXsingular}) coincides with the precanonical operator corresponding regular charts, up to higher-order terms.

However, we should be very careful when we compare the precanonical operator~(\ref{eq:precanonicalXsingular}) with the previously defined precanonical operator~(\ref{eq:precanonical-singular-conventional}) corresponding to a singular chart, since these two expressions use a different parametrization of the Lagrangian manifold. Whereas the former is specifically constructed for eikonal coordinates, the latter is constructed for the conventional coordinate system.
As discussed in the previous subsection, we can obtain the precanonical operator $ (K_{\Lambda^2(t,\phi)}^{\Omega_i^s} A_0)(x)$ from the precanonical operator~(\ref{eq:precanonicalXsingular}) with the help of a Jacobian factor, see equation~(\ref{eq:precanonicalTimeChange}) and figure~\ref{fig:comm-diag-can-op}. With the help of equation~(\ref{eq:timereparametrization}), which relates eikonal coordinates and conventional coordinates, we find that~\cite{Dobrokhotov15}
\begin{equation} \label{eq:precanonicalXsingularProperTime} 
  (K_{\Lambda^2(t,\phi)}^{\Omega_i^s} A_0)(x) = \frac{e^{i\pi/4}}{\sqrt{2\pi h}} e^{-i \pi \mu_{\Omega_i^s}/2 }
    \int_{-\infty}^\infty \text{d} \phi \sqrt{| \det(\mathcal{P},\mathcal{P}_\phi) |} \frac{A_0(\tau(x,\phi),\phi)}{\sqrt{R(\mathcal{X})}} e^{\frac{i}{h}\tau(x,\phi)} ,
\end{equation}
with $R(\mathcal{X})$ given by equation~(\ref{eq:Rtransition}). By the results of Ref.~\cite{Dobrokhotov14}, see also Ref.~\cite{Dobrokhotov15}, the precanonical operator $K_{\Lambda^2(t,\phi)}^{\Omega_i^s} A_0$ defined in this way coincides with expression~(\ref{eq:precanonical-singular-conventional}). In particular, it satisfies the commutation formulas~(\ref{eq:commutation-relation-precanop-sing-gen}) and~(\ref{eq:commutation-relation-precanop-sing-L1}).
Hence, the precanonical operator~(\ref{eq:precanonicalXsingularProperTime}) constitutes an asymptotic solution corresponding to the singular chart $\Omega_i^s$ when the amplitude function $A_0(\tau(x,\phi),\phi)$ is given by equation~(\ref{eq:Amp-sol}), the solution of equation~(\ref{eq:transport2}). This once again indicates that the transport equation allows an interpretation as a geometrical object associated with the Lagrangian manifold~\cite{Maslov81,Guillemin77,Bates97}.

However, equation~(\ref{eq:precanonicalXsingularProperTime}) is not our final representation for the asymptotic solution corresponding to singular charts. In section~\ref{sec:causticsSolution}, we show how it can be essentially simplified in the vicinity of fold points and cusp points.
Furthermore, we remark that one can just as well use the precanonical operator~(\ref{eq:precanonicalXsingularProperTime}) for regular points, as long as the determinant $\det(\mathcal{P},\mathcal{P}_\phi)$ does not vanish. In this case, one needs to pay careful attention to the value of Maslov index, which we discuss in the next subsection. In section~\ref{subsec:numerics-large-h}, we show an explicit example of the implementation of the precanonical operator~(\ref{eq:precanonicalXsingularProperTime}) for regular points.

\subsection{The Maslov index} \label{subsec:maslovindex}

In the previous subsections, we encountered the Maslov indices $\mu_{\Omega_i}$ and $\mu_{\Omega_j^s}$.
Although we called these objects the Maslov indices of a regular and a singular chart, respectively, we already mentioned that this terminology is somewhat misleading, since the Maslov index is only properly defined for a chain of charts. We nevertheless use these terms to simplify our terminology and justify them later on.

There is an extensive body of literature on the Maslov index, see e.g. Refs.~\cite{Arnold67,Maslov72,Maslov81,Mishchenko90,Guillemin77,Bates97,Dobrokhotov03}, which shows that the Maslov index can be expressed as a topological characteristic of the Lagrangian manifold. In particular, it can be defined without reference to the Hamiltonian system. 
On the other hand, from a more practical point of view~\cite{Maslov72,Dobrokhotov03,Dobrokhotov08,Dobrokhotov14}, the Maslov index ensures a consistent choice of the argument of the Jacobian $J$ and its square root. Equivalently, it defines the analytic continuation of $\sqrt{J}$ in the complex plane and makes sure that we select the correct branch.
In this section, we mainly discuss this more practical point of view. We explicitly compute the Maslov index for our problem by computing the signs of the relevant determinants. This also illustrates the relationship between the precanonical operator~(\ref{eq:precanonicalXsingularProperTime}) corresponding to a singular chart, and the precanonical operator~(\ref{eq:precanonicalX}) corresponding to a regular chart.
However, we emphasize that there are many alternative ways to look at the Maslov index and that it can also be computed without explicitly matching different precanonical operators, using geometrical considerations~\cite{Arnold67,Mishchenko90,Guillemin77,Bates97}.

Let us consider a singular point $(\tau^*,\phi^*)$ on a singular chart $\Omega_k^s$ of the Lagrangian manifold $\Lambda^2$. In the neighborhood of this point, there are points $(\tau^{st},\phi^{st})$ that lie both in the singular chart $\Omega_k^s$ and in a regular chart $\Omega_i$. When we project such a point $(\tau^{st},\phi^{st})$ onto configuration space, we obtain the point $x^{st}=\mathcal{X}(\tau^{st},\phi^{st})$. At the point $x^{st}$, we then have
two local asymptotic solutions, given by the precanonical operators~(\ref{eq:precanonicalX}) and~(\ref{eq:precanonicalXsingularProperTime}). To be able to construct a global asymptotic solution, we require both these representations to be equal up to higher-order terms. We can investigate their relation by evaluating the precanonical operator~(\ref{eq:precanonicalXsingularProperTime}) corresponding to a singular chart using the stationary phase approximation~\cite{Maslov81,Guillemin77,Dobrokhotov14}. Given the point $x^{st}$, the action $\tau(x^{st},\phi)$, given by equation~(\ref{eq:defTauSingular}), has a stationary point at $\phi^{st}$, and we have $\tau^{st}=\tau(x^{st},\phi^{st})$.
We remark that the action~(\ref{eq:defTauSingular}), as a function of $\phi$ for a given point $x^{st}$, may have more than one stationary point. In this case, the stationary phase evaluation will give rise to a sum over the stationary points. However, in the given intersection between $\Omega_k^s$ and $\Omega_i$ there will only be one stationary point, as all stationary points lie on different regular charts. Since our interest lies in this intersection, we discard the other stationary points. This can be formalized using a partition of unity, as discussed in the next section.

We therefore evaluate the precanonical operator~(\ref{eq:precanonicalXsingularProperTime}) at the point $(\tau^{st},\phi^{st})$, which lies in the region where $\Omega_i$ and $\Omega_k^s$ overlap. Using the stationary phase approximation~\cite{Maslov81,Guillemin77,Dobrokhotov14}, we obtain
\begin{multline}  \label{eq:stationaryphase-regpoint}  
  \frac{e^{i\pi/4}}{\sqrt{2\pi h}} e^{-i \pi \mu_{\Omega_k^s}/2 }
    \int_{-\infty}^\infty \text{d} \phi \sqrt{| \det(\mathcal{P},\mathcal{P}_\phi) |} \frac{A_0(\tau(x,\phi),\phi)}{\sqrt{R(\mathcal{X})}} e^{\frac{i}{h}\tau(x,\phi)} 
  = e^{-\frac{i \pi}{2}\mu_{\Omega_k^s}} \\
  \times e^{\frac{i\pi}{4}(1+\text{sign}(\tau_{\phi\phi}^{st}))} \sqrt{\frac{| \det(\mathcal{P}^{st},\mathcal{P}_\phi^{st}) |}{R(\mathcal{X}^{st}) \big| \tau_{\phi\phi}^{st} \big|}}
  A_0(\tau^{st},\phi^{st}) e^{\frac{i}{h}\tau^{st}} + \mathcal{O}(h) ,
\end{multline}
where quantities that are to be evaluated at the stationary point are marked with a superscript \emph{st}. Taking the derivative of the first equality in equation~(\ref{eq:tauStationaryLagMan}) with respect to $\phi$, and specializing to the stationary point, at which $x^{st}=\mathcal{X}(\tau^{st},\phi^{st})$, we find that
\begin{equation} \label{eq:tauphiphi-stationary}
  \tau_{\phi\phi}^{st} = - \langle \mathcal{P}_\phi^{st}, \mathcal{X}_\phi^{st} \rangle .
\end{equation}
Note that when $\mathcal{X}_\phi=0$, i.e. when we are at a singular point, the stationary point is degenerate, that is, $\tau_{\phi\phi}=0$, and we cannot use the stationary phase approximation to evaluate the precanonical operator~(\ref{eq:precanonicalXsingularProperTime}), as one would naturally expect.
When we compute the product of the two relevant determinants, we obtain
\begin{equation}
\begin{aligned} \label{eq:determinantsMultJtildeJ}
  \widetilde{\mathcal{J}} \mathcal{J} &= \det(\mathcal{P}, \mathcal{P}_\phi) \det(\mathcal{X}_\tau, \mathcal{X}_\phi) = \det(\mathcal{P}, \mathcal{P}_\phi)^T \det(\mathcal{X}_\tau, \mathcal{X}_\phi) \\
  &= \det \begin{pmatrix} \langle \mathcal{P}, \mathcal{X}_\tau \rangle & \langle \mathcal{P}, \mathcal{X}_\phi \rangle \\ \langle \mathcal{P}_\phi, \mathcal{X}_\tau \rangle & \langle \mathcal{P}_\phi, \mathcal{X}_\phi \rangle \end{pmatrix} = \langle \mathcal{P}_\phi, \mathcal{X}_\phi \rangle,
\end{aligned}
\end{equation}
where we have used equation~(\ref{eq:eikonalCoordinates}). Therefore, we have $\widetilde{\mathcal{J}}^{st} \mathcal{J}^{st} = -\tau_{\phi\phi}^{st}$, and we find that 
\begin{equation}
\begin{aligned}
  K_{\Lambda^2(t,\phi)}^{\Omega_k^s} A_0
  &= e^{\frac{i\pi}{4}(1+\text{sign}(\tau_{\phi\phi}^{st}))} e^{-\frac{i \pi}{2}\mu_{\Omega_k^s}} \frac{A_0(\tau^{st},\phi^{st})}{\sqrt{R(\mathcal{X}^{st}) | \mathcal{J}^{st} |}} e^{\frac{i}{h}\tau^{st}} + \mathcal{O}(h) \\
  &= e^{\frac{i\pi}{4}(1+\text{sign}(\tau_{\phi\phi}^{st})-2\mu_{\Omega_k^s}+2\mu_{\Omega_i})} K_{\Lambda^2(t,\phi)}^{\Omega_i} A_0 + \mathcal{O}(h) ,
\end{aligned}
\end{equation}
where we have used equations~(\ref{eq:JacobianXrelationEikonal}) and~(\ref{eq:precanonicalX}). Since we require the leading-order terms of the two precanonical operators to be equal, the phase factor should be equal to one. We therefore have the requirement
\begin{equation} \label{eq:MaslovIndexRelationGeneral}
  1+\text{sign}(\tau_{\phi\phi}^{st})-2\mu_{\Omega_k^s}+2\mu_{\Omega_i} = 0 ,
\end{equation}
from which we can obtain a relation between the Maslov indices of regular and singular charts by analyzing the sign of $\tau_{\phi\phi}^{st}$.

From our previous considerations, we know that both $\mathcal{J}$ and $\mathcal{X}_\phi$ vanish as we pass through a singular point on a trajectory. Let us therefore consider a singular point $(\tau^*,\phi^*)$ on the chart $\Omega_k^s$ of the Lagrangian manifold $\Lambda^2$, together with two points $(\tau^\pm,\phi^*)$, with $\tau^- < \tau^* < \tau^+$. We choose these in such a way that $(\tau^-,\phi^*)$ lies both on the singular chart $\Omega_k^s$ and the regular chart $\Omega_j$ and that $(\tau^+,\phi^*)$ lies both on the singular chart $\Omega_k^s$ and the regular chart $\Omega_l$. We can then approximate $\mathcal{J}^\pm \equiv \mathcal{J}(\tau^\pm,\phi^*) \approx \mathcal{J}_\tau^* (\tau^\pm-\tau^*)$. Using equation~(\ref{eq:JtauCaustic}), we subsequently obtain
\begin{equation}
  \widetilde{\mathcal{J}}^\pm \mathcal{J}^\pm \approx \widetilde{\mathcal{J}}^\pm C^4 (\mathcal{X}^*) \widetilde{\mathcal{J}}^* (\tau^\pm-\tau^*) .
\end{equation}
Since $\widetilde{\mathcal{J}} \neq 0$ on the singular chart $\Omega_k^s$, the signs of $\widetilde{\mathcal{J}}^\pm$ and $\widetilde{\mathcal{J}}^*$ coincide. Furthermore, even though the linear approximation for $\mathcal{J}^\pm$ may not be entirely accurate, it does show how the sign of the Jacobian changes at the caustic. Therefore, the sign of $\widetilde{\mathcal{J}} \mathcal{J}$ is negative for $\tau^-<\tau^*$ and positive for $\tau^+>\tau^*$. Hence, we conclude from equation~(\ref{eq:determinantsMultJtildeJ}) that the sign of $\langle \mathcal{P}_\phi, \mathcal{X}_\phi \rangle$ changes from negative to positive as we pass through the caustic along a trajectory.

Let us now return to equation~(\ref{eq:MaslovIndexRelationGeneral}). For the region in which $\Omega_j$ and $\Omega_k^s$ overlap, and in which the point $(\tau^-,\phi^*)$ lies, we have $\langle \mathcal{P}_\phi, \mathcal{X}_\phi \rangle < 0$, whence $\text{sign}(\tau_{\phi\phi}^{st})>0$ by virtue of equation~(\ref{eq:tauphiphi-stationary}). Therefore, $\mu_{\Omega_j} = \mu_{\Omega_k^s} - 1$. For the region in which $\Omega_k^s$ and $\Omega_l$ overlap, and in which the point $(\tau^+,\phi^*)$ lies, we have $\langle \mathcal{P}_\phi, \mathcal{X}_\phi \rangle > 0$. Hence, $\text{sign}(\tau_{\phi\phi}^{st})<0$ and $\mu_{\Omega_l} = \mu_{\Omega_k^s}$. Combining these results, we obtain
\begin{equation}  \label{eq:MaslovIndexRelation}
  \mu_{\Omega_j} = \mu_{\Omega_k^s} - 1 , \quad \mu_{\Omega_l} = \mu_{\Omega_k^s} , \quad \text{and } \mu_{\Omega_l} = \mu_{\Omega_j} + 1 .
\end{equation}
This result clearly shows that the Maslov index is only defined for a chain of charts, since the Maslov index of a chart is defined relative to another chart. 
We observe from equation~(\ref{eq:MaslovIndexRelation}) that, when we pass through a singular point along a trajectory, the Maslov index of the regular chart after the caustic is one larger than the Maslov index of the regular chart before the caustic~\cite{Dobrokhotov14,Maslov81}. The Maslov index of the singular chart is always equal to the Maslov index of the regular chart after the caustic~\cite{Dobrokhotov14}.
These results imply that we can fix the Maslov index of all charts by fixing the Maslov index for one of them, which can indeed be shown to hold~\cite{Maslov81}. 
Since we consider a scattering problem, it appears to be a logical choice to fix the Maslov index for the incoming particles. Because the isotropic manifold $\Lambda^1$ only consists of regular points, it can be covered with a single chart. We set the Maslov index of this chart to zero, thereby fixing the Maslov index of all other charts and thus justifying our terminology.

Based on the result~(\ref{eq:MaslovIndexRelation}), we can subsequently introduce the Maslov index $\mu(\tau,\phi)$ of a regular point~\cite{Maslov81,Maslov73}. For the regular point $(\tau_0,\phi_0)$, which lies on the trajectory with initial value $\phi_0$, its value equals the number of singular points on this trajectory between $\tau=0$ and $\tau=\tau_0$. Since there are no nontrivial cycles on our Lagrangian manifold $\Lambda^2$, i.e. every path can be contracted to a point, we do not have any (Bohr-Sommerfeld) quantization conditions. Therefore, the Maslov index of a regular point is independent~\cite{Maslov81,Arnold67} of the path between this point and the central point on $\Lambda^1$. From this definition, it is clear that the Maslov index is the same for all points on a regular chart and that it equals the previously defined Maslov index of this regular chart.

A concept related to the Maslov index of a regular point is the Morse index of a point on a trajectory, which equals the number of roots of the Jacobian, counted with their multiplicity, between the starting point and the point under consideration. In our problem, all roots of the Jacobian $\mathcal{J}$ on the trajectories are simple, as we showed in section~\ref{subsec:classification}. Hence,  the Maslov index of a regular point equals the Morse index of that point on its trajectory. We do not provide a rigorous proof of this statement here, but refer the interested reader to Refs.~\cite{Maslov81,Dobrokhotov08}. 

As we already stated, we can also look at the Maslov index as a way to ensure a consistent choice of the argument of the Jacobian $\mathcal{J}$ and its square root. 
To make this more precise, let us define the matrix $\mathcal{J}_\varepsilon$ as
\begin{equation}
  \mathcal{J}_\varepsilon = \det \left( \frac{\partial (X_1,X_2)}{\partial(\tau,\phi)} - i \varepsilon \frac{\partial (P_1,P_2)}{\partial(\tau,\phi)} \right) .
\end{equation}
We can then define the Maslov index of a regular point as~\cite{Maslov73,Dobrokhotov03,Dobrokhotov14b}
\begin{equation} \label{eq:maslov-index-regular-point-computer}
  \mu(\tau,\phi) = \frac{1}{\pi} \lim_{\varepsilon \to 0} \text{Arg} \, \big. \mathcal{J}_\varepsilon \big|_{(0,0)}^{(\tau,\phi)}
    = \frac{1}{\pi} \lim_{\varepsilon \to 0} \text{Im} \int_{\gamma(\tau,\phi)} \frac{\text{d} \mathcal{J}_\varepsilon}{\mathcal{J}_\varepsilon} ,
\end{equation}
where $\gamma(\tau,\phi)$ indicates a path from the point $(0,0)$ to the point $(\tau,\phi)$. In the first expression in equation~(\ref{eq:maslov-index-regular-point-computer}), one should consider the difference in argument between the points $(0,0)$ and $(\tau,\phi)$. The second expression is especially convenient for the computation of the Maslov index of a regular point on a computer. To this end, one computes the integral in equation~(\ref{eq:maslov-index-regular-point-computer}) for a small value of $\varepsilon$. The result should subsequently be rounded to the nearest integer, as the Maslov index is always an integer number.
Definition~(\ref{eq:maslov-index-regular-point-computer}) explicitly shows the connection between the Maslov index and the analytic continuation of $\sqrt{\mathcal{J}}$ in the complex plane. It also shows a way to obtain the Maslov index without explicitly matching different precanonical operators. We remark that one can obtain the Maslov index of a singular chart using a similar method, using different determinants, see e.g. Ref.~\cite{Dobrokhotov03}.

\subsection{Maslov's canonical operator}  \label{subsec:maslovintro}

In the previous sections, we constructed precanonical operators in regular and singular charts, which provide local asymptotic solutions to equation~(\ref{eq:L-eff}). The final results, given in equations~(\ref{eq:precanonicalXsimplified}) and~(\ref{eq:precanonicalXsingularProperTime}), satisfy the commutation relations~(\ref{eq:commutation-relation-precanop-sing-gen}) and~(\ref{eq:commutation-relation-precanop-sing-L1}). In this subsection, we show how we can patch these local asymptotic solutions together to obtain a global asymptotic solution.

To this end, we need the atlas $\Omega$ on the Lagrangian manifold that we introduced in section~\ref{sec:lagman}. Furthermore, we need to introduce a partition of unity on the Lagrangian manifold. This is a set of smooth functions $\{e_n, n=1,\dots,N\}$, each of which is supported on a single chart, with the additional property that at each point of the manifold their sum equals one, i.e. $\sum_n e_n = 1$.
With these preliminaries, we can define the canonical operator $K_{\Lambda^2(t,\phi)} A_0$, which was introduced in Refs.~\cite{Maslov72,Maslov73,Maslov81}, see also Refs.~\cite{Mishchenko90,Dobrokhotov03,Guillemin77}, as
\begin{equation} \label{eq:canonicalOperatorDef}
  (K_{\Lambda^2(t,\phi)} A_0)(x) = \sum_{n=1}^N (K_{\Lambda^2(t,\phi)}^{\Omega_n} e_n A_0)(x) ,
\end{equation}
where we sum over all $N$ charts contained in the atlas $\Omega$.

The canonical operator $K_{\Lambda^2(t,\phi)} A_0$ satisfies the first commutation formula~(\ref{eq:commutation-relation-precanop-sing-gen}), because the precanonical operators do. We can show this by writing~\cite{Maslov81}
\begin{equation} \label{eq:commutation-relation-canop-gen}
\begin{aligned}
  Q(x, \hat{p},h) K_{\Lambda^2(t,\phi)} A_0 
    &= \sum_{n=1}^N Q(x, \hat{p}, h) K_{\Lambda^2(t,\phi)}^{\Omega_n} \big( e_n A_0 \big) \\
    &= \sum_{n=1}^N K_{\Lambda^2(t,\phi)}^{\Omega_n} \big( Q_0\left(x,p \right) e_n A_0 + \mathcal{O}(h) \big) \\
    &= K_{\Lambda^2(t,\phi)} \big( Q_0\left(x,p \right) A_0 + \mathcal{O}(h) \big)  .
\end{aligned}
\end{equation}
Since this equality holds for any pseudodifferential operator $\hat{Q}$, it also holds for the pseudodifferential operator $\hat{L}$. In particular, we have $L_0(x,p)=E$ on the Lagrangian manifold $\Lambda^2$.

The canonical operator also satisfies the second commutation formula~(\ref{eq:commutation-relation-precanop-sing-L1}). The first part of the proof is similar to the proof of the first commutation formula. Following Ref.~\cite{Maslov81}, we write
\begin{equation} \label{eq:commutation-relation-canop-L1-prep}
\begin{aligned}
& \big(L(x, \hat{p},h) - E \big) K_{\Lambda^2(t,\phi)} A_0 
    = \sum_{n=1}^N \big(L(x, \hat{p},h) - E \big) K_{\Lambda^2(t,\phi)}^{\Omega_n} \big( e_n A_0 \big) \\
    & \hspace*{1.0cm} = -i h \sum_{n=1}^N K_{\Lambda^2(t,\phi)}^{\Omega_n} \bigg[ \bigg( \frac{\text{d}}{\text{d} t} + i L_1 - \frac{1}{2} \sum_j \frac{\partial^2 L_0}{\partial x_j \partial p_j} \bigg) (e_n A_0) + \mathcal{O}(h) \bigg]\\
    & \hspace*{1.0cm} = -i h K_{\Lambda^2(t,\phi)} \bigg [ \bigg( \frac{\text{d}}{\text{d} t} + i L_1 - \frac{1}{2} \sum_j \frac{\partial^2 L_0}{\partial x_j \partial p_j} \bigg) A_0 + \mathcal{O}(h) \bigg] \\ 
        & \hspace*{3.5cm} - i h \sum_{n=1}^N K_{\Lambda^2(t,\phi)}^{\Omega_n} \left( \frac{\text{d} e_n}{\text{d} t} A_0 \right)  ,
\end{aligned}
\end{equation}
where the last equality follows from the application of the product rule and the definition of the canonical operator~(\ref{eq:canonicalOperatorDef}). The time derivative in this equation is to be taken along the projection of the solution of the Hamiltonian system~\cite{Maslov81}. 
For instance, suppose that we consider a chart $\Omega_i$ that is projected onto the coordinate Lagrangian plane $(p_{\{\alpha\}},x_{\{\beta\}})$, then 
\begin{equation}
  \frac{\text{d}}{\text{d} t} = \sum_{j \in \{\beta\}} \frac{\partial L_0}{\partial p_j} \frac{\partial}{\partial x_j} - \sum_{k \in \{\alpha\}} \frac{\partial L_0}{\partial x_k} \frac{\partial}{\partial p_k} .
\end{equation}
In order to show that the canonical operator satisfies the second commutation formula~(\ref{eq:commutation-relation-precanop-sing-L1}), we need to show that the second term in the last line of equation~(\ref{eq:commutation-relation-canop-L1-prep}) vanishes. To this end, let us consider a point on the Lagrangian manifold $\Lambda^2$. For simplicity, let us first assume that this point and a neighborhood of it lie in a single regular chart $\Omega_i$. Then we have $e_i=1$ and $e_j=0$ for $j\neq i$. Hence, $\text{d}e_j/\text{d}t=0$ for all $j$, and the second term is trivially zero.
Now suppose that this point lies in both a regular chart $\Omega_i$ and a singular chart $\Omega_j^s$. In the previous subsection, we showed that, with a proper definition of the Maslov index, $K_{\Lambda^2(t,\phi)}^{\Omega_i} B = K_{\Lambda^2(t,\phi)}^{\Omega_j^s} B$ up to higher-order terms, for any amplitude function $B$.
Using similar arguments, this equality can be extended to the case where the point lies in a chart $\Omega_i$ as well as in multiple other charts. For such a point, we obtain
\begin{equation}
  \sum_{n=1}^N K_{\Lambda^2(t,\phi)}^{\Omega_n} \left( \frac{\text{d} e_n}{\text{d} t} A_0 \right) = K_{\Lambda^2(t,\phi)}^{\Omega_i} \left( \sum_{n=1}^N \frac{\text{d} e_n}{\text{d} t} A_0 \right) = 0,
\end{equation}
where the last equality holds since $\sum_{n=1}^N e_n=1$ implies $\sum_{n=1}^N \text{d} e_n/\text{d} t=0$. 
Therefore, we obtain the second commutation formula for the canonical operator~\cite{Maslov81,Vainberg89b}
\begin{equation} \label{eq:commutation-relation-canop-L1}
  \big(L(x, \hat{p},h) - E \big) K_{\Lambda^2(t,\phi)} A_0 = -i h K_{\Lambda^2(t,\phi)} \bigg( \frac{\text{d} A_0}{\text{d} t} + i L_1 A_0 - \frac{1}{2} \sum_j \frac{\partial^2 L_0}{\partial x_j \partial p_j} A_0 + \mathcal{O}(h) \bigg) .
\end{equation}
In the derivation of this commutation relation, we have used that $\Lambda^2$ lies in the level set of $L_0(x,p)$ with energy $E$. 
Equation~(\ref{eq:commutation-relation-canop-L1}) shows that when the amplitude function $A_0$ is given by equation~(\ref{eq:Amp-sol}), the canonical operator $K_{\Lambda^2(t,\phi)} A_0$ is an asymptotic solution of equation~(\ref{eq:L-eff}). It is defined on the entire configuration space $(x_1,x_2)$, and is therefore a global asymptotic solution. 
The corrections to this asymptotic solution, which form the higher-order terms in the asymptotic expansion, come from the term with $\mathcal{O}(h)$ on the right-hand side of equation~(\ref{eq:commutation-relation-canop-L1}). When the canonical operator is nonzero, these corrections are of order $h K_{\Lambda^2(t,\phi)} A_0$.
We remark that it can be shown that the canonical operator does not depend on the choice of the atlas, the local coordinates in the charts, and the partition of unity~\cite{Maslov81}.

The canonical operator is a global asymptotic solution of equation~(\ref{eq:L-eff}). However, our original goal was to obtain an asymptotic solution of equation~(\ref{eq:eigenvalue-HPsi}). Such a solution can be obtained using equation~(\ref{eq:Psi-decomp}). Using the first commutation formula~(\ref{eq:commutation-relation-canop-gen}) for the canonical operator, we obtain
\begin{equation} \label{eq:asymptotic-sol-chi-canop}
  \Big( \chi(x, \hat{p},h) K_{\Lambda^2(t,\phi)} A_0 \Big)(x) = \Big( K_{\Lambda^2(t,\phi)} \big( \chi_0\left(x,p \right) A_0 + \mathcal{O}(h) \big) \Big)(x)  .
\end{equation}
Hence, the right-hand side of this equation is an asymptotic solution to equation~(\ref{eq:eigenvalue-HPsi}).

Although the canonical operator is a global asymptotic solution, it is actually more of an algorithm than an actual formula. Given a point~$x$, equation~(\ref{eq:canonicalOperatorDef}) instructs us to consider all charts of the Lagrangian manifold $\Lambda^2$ that are projected onto this point and to add the precanonical operators on these charts. Note that we do not have to match asymptotic solutions in different regions, as all of this has already been taken care of in the construction of the precanonical operators and the Maslov index. Instead, we only have to simplify the canonical operator in different neighborhoods.
As we already discussed in section~\ref{subsec:lagman}, the number of leaves of our Lagrangian manifold is finite, since we have assumed that the potential $U(x)$ and the mass $m(x)$ are constant outside a certain domain. Therefore, we only need to consider a finite number of charts and we have only finitely many terms in the sum~(\ref{eq:canonicalOperatorDef}). From a physical point of view, this sum expresses the well-known physical phenomenon of interference, as we sum over all trajectories that reach the point $x$.

Using equations~(\ref{eq:canonicalOperatorDef}) and~(\ref{eq:asymptotic-sol-chi-canop}), we obtain a representation of the asymptotic solution $\Psi(x)$ in the various domains.
In regular points, we can use our expression~(\ref{eq:precanonicalXsimplified}) for the precanonical operator in regular charts. We thereby obtain the asymptotic solution, which gives, up to $\mathcal{O}(h)$,
\begin{multline}  \label{eq:PsiRegular}
  \Psi(x) = \left. \sum_i \frac{\chi_0(\mathcal{X},\mathcal{P})A_0^0(\phi_i)}{\sqrt{R(\mathcal{X}) C(\mathcal{X}) |\mathcal{X}_\phi|}} \exp\left(\frac{i}{h} \tau_i + i \Phi_{sc}(\tau_i,\phi_i) -\frac{i \pi}{2} \mu(\tau_i,\phi_i) \right) \right|_{\begin{subarray}{l}\tau_i=\tau_i(x) \\ \phi_i=\phi_i(x)\end{subarray}} \\
  \times \big( 1 + \mathcal{O}(h) \big),
\end{multline}
where we have used that $\partial \tau_i/\partial x = \mathcal{P}(\tau_i,\phi_i)$. The semiclassical phase~$\Phi_{sc}$ in this expression is given by equation~(\ref{eq:BerryPhaseEikonal}). For the graphene Hamiltonian~(\ref{eq:Dirac}), we can make further simplifications using equations~(\ref{eq:L0Dirac}), (\ref{eq:chi0Dirac}), (\ref{eq:L0Eikonal}) and~(\ref{eq:Rtransition}). We then arrive at
\begin{multline}  \label{eq:PsiRegular-graphene}
  \Psi_\alpha(x) = \left. \sum_i 
  \frac{A_0^0(\phi_i) \exp(\frac{i}{h} \tau_i + i \Phi_{sc,\alpha} -\frac{i \pi}{2} \mu(\tau_i,\phi_i))}{\sqrt{2(E-U-m)|\mathcal{X}_\phi|} \sqrt[4]{(E-U)^2-m^2}}
  \left( \!\!\!\begin{array}{c} \mathcal{P}_1 + i \alpha\mathcal{P}_2 \\ E-U-m \end{array} \!\!\!\right)
  \right|_{\begin{subarray}{l}\tau_i=\tau_i(x) \\ \phi_i=\phi_i(x)\end{subarray}} \\
  \times \big( 1 + \mathcal{O}(h) \big) .
\end{multline}
We can use this expression for the wavefunction for a very large part of the configuration space $(x_1,x_2)$. We cannot use it in a neighborhood of the caustic, since our expression diverges at these points.

At this point, let us come back to the gauge freedom that we discussed in section~\ref{subsec:separate-Dirac}. When we perform a gauge transformation~(\ref{eq:chi0-gauge-transformation}), the subprincipal Weyl symbol $L_1^W$ changes according to equation~(\ref{eq:L1-gauge-transformation-effect}). However, the asymptotic solution $\Psi(x)$ should not depend on this gauge freedom. To prove that this is indeed the case, let us consider how the semiclassical phase is affected by the gauge transformation. Using equation~(\ref{eq:BerryPhaseEikonal}), we observe that
\begin{equation} \label{eq:Phi-sc-gauge-transformation}
\begin{aligned}
  \widetilde{\Phi}_{sc}
    &= -\int_0^t L_1^W + \{ g, L_0 \} \, \text{d} t'
    = \Phi_{sc} - \int_0^t \left\langle \frac{\partial g}{\partial x} , \frac{\partial L_0}{\partial p} \right\rangle - \left\langle \frac{\partial g}{\partial p} , \frac{\partial L_0}{\partial x} \right\rangle \, \text{d} t' \\
    & = \Phi_{sc} - \int_0^t \frac{\text{d} g}{\text{d} t'} \, \text{d} t' 
    = \Phi_{sc} - g ,
\end{aligned}
\end{equation}
where we have used Hamilton's equations~(\ref{eq:Hamilton}) in the third equality. Looking at equation~(\ref{eq:PsiRegular}), we now observe that the additional phase that arises from the gauge transformation~(\ref{eq:chi0-gauge-transformation}) is exactly cancelled by an opposite phase that arises from the corresponding change~(\ref{eq:Phi-sc-gauge-transformation}) in the semiclassical phase. Thus, the wavefunction $\Psi(x)$ is indeed independent of the choice of $\chi_0$, as one would naturally expect.

In a neighborhood of the caustic, we have to use the precanonical operator~(\ref{eq:precanonicalXsingularProperTime}) corresponding to singular charts to construct the asymptotic solution. 
Let $x_\text{cusp}$ be the point in configuration space corresponding to the cusp of the caustic. To this point corresponds a cusp point on the Lagrangian manifold, see also figure~\ref{fig:LagMan}.
Using equation~(\ref{eq:asymptotic-sol-chi-canop}), we obtain the asymptotic solution for points $x$ in the vicinity of the point $x_\text{cusp}$ as
\begin{multline} \label{eq:PsiSingular}
  \Psi(x) = \frac{e^{i\pi/4}}{\sqrt{2\pi h}} e^{-i \pi \mu_{\Omega_i^s}/2 }
    \int_{-\infty}^\infty \text{d} \phi \sqrt{| \det(\mathcal{P},\mathcal{P}_\phi) |} \frac{A_0^0(\phi)e^{i \Phi_{sc}}}{\sqrt{R(\mathcal{X})}} \chi_0(\mathcal{X},\mathcal{P}) e^{\frac{i}{h}\tau(x,\phi)} \\
    \times \big( 1 + \mathcal{O}(h) \big)
\end{multline}
Like the wavefunction in regular points, this wavefunction is invariant with respect to the gauge transformation. For the graphene Hamiltonian~(\ref{eq:Dirac}), the asymptotic solution~(\ref{eq:PsiSingular}) becomes
\begin{multline} \label{eq:PsiSingular-graphene}
  \Psi_\alpha(x) = \frac{e^{i\pi/4}}{\sqrt{2\pi h}}
    \int_{-\infty}^\infty \text{d} \phi \sqrt{| \det(\mathcal{P},\mathcal{P}_\phi) |}
    \frac{A_0^0(\phi) \exp(\frac{i}{h} \tau(x,\phi) + i \Phi_{sc,\alpha} -\frac{i \pi}{2} \mu_{\Omega_i^s})}{\sqrt{2 (E-U-m)^2 (E-U+m)}} \\
    \times \left( \!\!\!\begin{array}{c} \mathcal{P}_1 + i \alpha\mathcal{P}_2 \\ E-U-m \end{array} \!\!\!\right)
    \big( 1 + \mathcal{O}(h) \big)
\end{multline}
where we have one again used equations~(\ref{eq:L0Dirac}), (\ref{eq:chi0Dirac}) and~(\ref{eq:Rtransition}).
When we consider a point $x_\text{fold}$ on the fold line of the caustic, we need to add two contributions to obtain the full asymptotic solution for the wavefunction.
The first contribution corresponds to the singular chart on the Lagrangian manifold on which the fold point is located.
It is given by equation~(\ref{eq:PsiSingular}). The second contribution, given by expression~(\ref{eq:PsiRegular}), corresponds to a regular point on the third leaf of the Lagrangian manifold, as can be seen in figure~\ref{fig:LagMan}. 
In the next section, we show how we can further simplify the asymptotic solution~(\ref{eq:PsiSingular}) in the vicinity of fold and cusp points.

\section{The wavefunction near caustics} \label{sec:causticsSolution}

In section~\ref{sec:lagman}, we defined a caustic as the set of singular points of the projection of the surface $\Lambda^2$ onto the $(x_1,x_2)$ plane. 
We identified two types of singular points, fold points and cusp points. In section~\ref{subsec:classification}, we obtained a more precise classification of these singular points, establishing that $\mathcal{X}_\phi$ vanishes at a fold point, while $\mathcal{X}_{\phi\phi}$ does not. At a cusp point, both $\mathcal{X}_{\phi}$ and $\mathcal{X}_{\phi\phi}$ vanish, while $\mathcal{X}_{\phi\phi\phi}$ does not. 
However, we did not consider the action function in that section.
In section~\ref{sec:maslov}, we discussed how the Lagrangian manifold $\Lambda^2$ can be generated by a phase function. In particular, we represented the asymptotic solution as an integral involving such a phase function. In section~\ref{subsec:maslovapproaches}, we considered the phase function $S+p_2(x_2-\mathcal{X}_2)$ and in section~\ref{subsec:maslovsingular}, we considered the function $\tau(x,\phi)$. We showed that the Lagrangian manifold is constituted by the points at which $\tau_\phi$ vanishes, while the singular points correspond to the points where $\tau_{\phi\phi}$ vanishes as well.

The theory of Lagrangian singularities, see e.g. Refs.~\cite{Arnold82,Arnold75,Arnold90,Poston78,Thom72,Berry80}, states that one can also study caustics by studying phase functions. By studying which derivatives of the phase function vanish, one obtains a classification of the different types of singular points. 
This classification is equivalent to our previous classification, which was based on the Jacobian and, for eikonal coordinates, on the derivates of $\mathcal{X}$ with respect to $\phi$.
In the process, one obtains a so-called normal form~\cite{Arnold82} for the action in the vicinity of a singular point. This normal form depends on the type of singular point under consideration and makes it possible to evaluate the integral over the phase function in terms of certain special functions~\cite{Maslov81}. Finally, the normal form does not depend on the phase function, only on the type of singular point that is being studied~\cite{Arnold82}. Of course, to obtain the right result, the phase function should generate the correct Lagrangian manifold.

In this section, we illustrate the correspondence between these two viewpoints for folds and cusps. In section~\ref{subsec:leadingorder}, we consider the Taylor expansion of the phase function at a singular point, based on Ref.~\cite{Dobrokhotov14}. Comparing it to the results from section~\ref{subsec:classification}, we show that the form of this Taylor expansion coincides with the normal forms that have been established in the literature~\cite{Arnold82,Thom72,Poston78}. Subsequently, we evaluate the integral~(\ref{eq:PsiSingular}) to obtain the leading-order term of the asymptotic solution near the fold and cusp points on the caustic. In section~\ref{subsec:uniform}, we first discuss why it is necessary to go beyond this leading-order approximation if we want to obtain a meaningful result in the vicinity of the cusp point. Subsequently, we discuss how to construct a uniform approximation~\cite{Ursell72,Connor81b} near the cusp. In this approximation, one does not consider the Taylor expansion of the action, but instead transforms the action to its normal form near a cusp using a change of variables. As a result, this approximation adequately captures the effect of the semiclassical phase on the focusing.

\subsection{Leading-order evaluations}  \label{subsec:leadingorder}

In section~\ref{subsec:maslovindex}, we evaluated the expression~(\ref{eq:precanonicalXsingularProperTime}) in regular points using the stationary phase approximation. We saw that, in regular points, the leading-order approximation~(\ref{eq:stationaryphase-regpoint}) coincides with the asymptotic solution~(\ref{eq:precanonicalX}) and we determined the Maslov index. However, we also saw that this asymptotic solution diverges near singular points. In particular, we noticed that $\tau_{\phi\phi}$ is proportional to $\mathcal{X}_\phi$ and hence vanishes at singular points. The main idea of the stationary phase approximation~\cite{Maslov81,Guillemin77,Dobrokhotov14} is that the leading-order approximation of the integral expression is determined by a small neighborhood of a stationary point. Since the second derivative $\tau_{\phi\phi}$ vanishes at the caustic, it is natural to look at higher-order derivatives. In this subsection, we discuss how we can obtain the leading-order approximation of the asymptotic solution by considering the first nonvanishing term in the Taylor expansion.
In section~\ref{subsec:classification}, we discussed that, when our problem is in general position, the only singularities that occur in the system are folds and cusps. We therefore confine our attention to these two types of singularities.

Let us consider a singular point $(\tau^*,\phi^*)$ on the Lagrangian manifold. To this point corresponds the point $\mathcal{X}(\tau^*,\phi^*)$ in the configuration space.
As before, we mark all quantities that are to be evaluated at the singular point with a star, e.g. $\mathcal{X}^*=\mathcal{X}(\tau^*,\phi^*)$.
Following the approach taken in Ref.~\cite{Dobrokhotov14}, we expand the phase function $\tau(x,\phi)$ in powers of $\phi$ around $\phi^*$ and in powers of $x$ around $\mathcal{X}^*$, that is,
\begin{equation}
  \tau(x,\phi) = \sum_j \frac{q_j(x)}{j!} (\phi-\phi^*)^j , \quad q_j(x) = a_i + \langle b_i , x - \mathcal{X}^* \rangle + \mathcal{O}\big( (x - \mathcal{X}^*)^2 \big) .
\end{equation}
Using the iteration method, values for these coefficients were obtained in Ref.~\cite{Dobrokhotov14}. Taking into account that $\mathcal{X}^*_\phi=0$, the values of the first few expansion coefficients are given by
\begin{align}
    a_0 &= \tau^* , \hspace*{1.8cm}  & \langle b_0 , x - \mathcal{X}^* \rangle &= \langle \mathcal{P}^* , x - \mathcal{X}^* \rangle \nonumber \\ 
    a_1 &= 0     ,  & \langle b_1 , x - \mathcal{X}^* \rangle &= \langle \mathcal{P}_\phi^* , x - \mathcal{X}^* \rangle \nonumber \\
    a_2 &= 0     ,  & \langle b_2 , x - \mathcal{X}^* \rangle &= \langle \mathcal{P}_{\phi\phi}^* , x - \mathcal{X}^* \rangle - \langle \mathcal{P}_\tau^* , \mathcal{X}_{\phi\phi}^* \rangle \langle \mathcal{P}^* , x - \mathcal{X}^* \rangle \nonumber \\
    a_3 &= - \langle \mathcal{P}_\phi^*, \mathcal{X}_{\phi\phi}^* \rangle \hspace*{-1.0cm}  && \nonumber \\
    a_4 &= - \langle \mathcal{P}_\phi^*, \mathcal{X}_{\phi\phi\phi}^* \rangle - 3 \langle \mathcal{P}_{\phi\phi}^*, \mathcal{X}_{\phi\phi}^* \rangle \hspace*{-5cm} && \label{eq:phi-expansion-coeffs}
\end{align}
As shown in appendix~\ref{app:conventionalCanonicalOp}, the same coefficients are obtained when we consider the Taylor expansion of the phase function of the conventional representation~(\ref{eq:precanonical-singular-conventional}) of the precanonical operator corresponding to singular charts.
In section~\ref{subsec:maslovindex}, we already saw that the coefficient $a_2 = \tau_{\phi\phi}^*$ vanishes, since $\mathcal{X}_{\phi}^* = 0$. Let us now look at the coefficients $a_3$ and $a_4$ and see whether they vanish at fold points and cusp points.

At a fold point, the second derivative $\mathcal{X}_{\phi\phi}^*$ is nonzero, as we established in section~\ref{subsec:classification}.
Furthermore, the determinant $\det(\mathcal{P}^*,\mathcal{P}_\phi^*) \neq 0$,
which means that the vectors $\mathcal{P}$ and $\mathcal{P}_\phi$ are both nonzero and are not parallel. In fact, we even showed that they are orthogonal, as $\langle \mathcal{P}^*,\mathcal{P}_\phi^* \rangle = 0$. Taking the derivative of the relation $\langle \mathcal{P} , \mathcal{X}_\phi \rangle = 0$, see equation~(\ref{eq:eikonalCoordinates}), with respect to $\phi$ and specializing to the singular point, we observe that $\langle \mathcal{P}^*, \mathcal{X}_{\phi\phi}^* \rangle = 0$. 
Together with the fact that $\det( \mathcal{P}^*,\mathcal{P}_\phi^* ) \neq 0$, this implies that $a_3 = - \langle \mathcal{P}_\phi^*, \mathcal{X}_{\phi\phi}^* \rangle \neq 0$. 
Actually, since both $\langle \mathcal{P}^*,\mathcal{P}_\phi^* \rangle = 0$ and $\langle \mathcal{P}^*, \mathcal{X}_{\phi\phi}^* \rangle = 0$, the vectors $\mathcal{X}_{\phi\phi}^*$ and $\mathcal{P}_\phi^*$ are either parallel or anti-parallel. Therefore, we even have $|a_3| = |\mathcal{P}_\phi^* | | \mathcal{X}_{\phi\phi}^* | \neq 0$.
Thus, at a fold point, the third derivative $\tau_{\phi\phi\phi}^*$ does not vanish.
This observation is in accordance with the theory of Lagrangian singularities~\cite{Arnold82}, which states that the phase function near a fold point can be expressed as a third-order polynomial in $\phi$. Within the general classification of singular points, the fold singularity is denoted by the symbol~$A_2$.

At a cusp point, the second derivative $\mathcal{X}_{\phi\phi}^*$ vanishes as well, whereas the third derivative $\mathcal{X}_{\phi\phi\phi}^*$ does not vanish, see section~\ref{subsec:classification}. Thus, at a cusp point, also the third derivative of the phase function vanishes, i.e. $a_3 = \tau_{\phi\phi\phi}^* = 0$. Let us therefore look at the coefficient $a_4$.
Taking the second derivative of the relation $\langle \mathcal{P} , \mathcal{X}_\phi \rangle = 0$ with respect to $\phi$ and specializing to the singular point, we now observe that $\langle \mathcal{P}^*, \mathcal{X}_{\phi\phi\phi}^* \rangle = 0$. Using the fact that $\det( \mathcal{P}^*, \mathcal{P}_{\phi}^* ) \neq 0$, we arrive at the conclusion that $a_4 = - \langle \mathcal{P}_\phi^*, \mathcal{X}_{\phi\phi\phi}^* \rangle \neq 0$. Since we also have $\langle \mathcal{P}^*,\mathcal{P}_\phi^* \rangle = 0$, we can even say that $|a_4| = |\mathcal{P}_\phi^* | | \mathcal{X}_{\phi\phi\phi}^* | \neq 0$.
The fact that $a_4$ is nonzero is in accordance with the theory of Lagrangian singularities~\cite{Arnold82}, which states that the phase function near a cusp point can be expressed as a fourth-order polynomial in $\phi$. Within the general classification of singular points, the cusp singularity is denoted by the symbol~$A_3$.

Using the above considerations, we can essentially simplify equation~(\ref{eq:PsiSingular}) for points near the caustic and rewrite its leading-order term in terms of well-established special functions.
Let us first consider a point $x_\text{fold}$ in the configuration space, to which corresponds a fold point $(\tau^*,\phi^*)$ on the Lagrangian manifold. As we discussed in the previous section, we need to add two contributions to obtain the wavefunction at points $x$ in the vicinity of $x_\text{fold}$. One of these contributions is given by equation~(\ref{eq:PsiRegular}) and corresponds to a regular point (on a regular chart) that is projected onto $x$. The other contribution comes from the singular chart and is given by equation~(\ref{eq:PsiSingular}). 
To obtain the leading-order term of the latter contribution near a fold point, we neglect~\cite{Dobrokhotov14} the terms in the Taylor expansion of the phase function~$\tau(x,\phi)$ that are of fourth or higher order in $\phi$. Furthermore, we expand the amplitude, that is, the collection of terms in front of the exponent containing the phase function, to zeroth order in $\phi$.
By a change of variables, one can subsequently express~\cite{Dobrokhotov14} the integral in equation~(\ref{eq:PsiSingular}) in terms of the Airy function, which is defined by
\begin{equation}  \label{eq:Airy-def}
  \text{Ai}(u) = \frac{1}{2\pi} \int_{-\infty}^\infty \exp\left( \frac{i}{3} t^3 + u t \right) \, \text{d} t .
\end{equation}
The next step of the procedure, which is described in detail in Ref.~\cite{Dobrokhotov14}, is to carefully establish to which order in $h$ the various terms in this expression correspond. 
To obtain the leading-order term, terms which give rise to higher-order contributions are subsequently neglected.
Finally, one obtains~\cite{Dobrokhotov14} that, for points $x$ in an $\mathcal{O}(h^{5/6})$-neighborhood of the point $x_\text{fold}$, the leading-order approximation to the expression~(\ref{eq:PsiSingular}) is given by
\begin{multline} \label{eq:Airy-final}
  \Psi(x) = \frac{e^{i\pi/4}}{\sqrt{2\pi h}} e^{-i \pi \mu_{\Omega_i^s}/2 } \Bigg( 
  2 \pi \sqrt[3]{\frac{2 h}{|a_3|}} \sqrt{| \det(\mathcal{P}^*,\mathcal{P}_\phi^*) |} \frac{A_0^0(\phi^*)e^{i \Phi_{sc}^*}}{\sqrt{R(\mathcal{X^*})}} \chi_0(\mathcal{X^*},\mathcal{P^*})  \\
  \times  \exp\left[ \frac{i}{h} (a_0 + \langle b_0, z \rangle) \right] \text{Ai}\left( \frac{2 \langle b_1, z \rangle}{2^{2/3} h^{2/3} a_3^{1/3}}  \right) + \mathcal{O}(h^{2/3})  \Bigg) .
\end{multline}
In this expression, $z = x - \mathcal{X}^*$ and the coefficients $a_i$ and $b_i$ are determined by equation~(\ref{eq:phi-expansion-coeffs}). Furthermore, as established in section~\ref{subsec:classification}, the determinant factorizes, that is, $|\det(\mathcal{P}^*,\mathcal{P}_\phi^*)|=|\mathcal{P}^*| |\mathcal{P}_\phi^*|$. For the graphene Hamiltonian~(\ref{eq:Dirac}), this expression becomes
\begin{multline} \label{eq:Airy-final-graphene}
  \Psi_\alpha(x) = \frac{2^{5/6} \pi^{1/2} e^{i\pi/4}}{h^{1/6}}
    \frac{\sqrt{|\mathcal{P}^*| |\mathcal{P}_\phi^*|}}{\sqrt[3]{| \langle \mathcal{P}_\phi^*, \mathcal{X}_{\phi\phi}^* \rangle |}}
    \frac{A_0^0(\phi^*) \exp(i \Phi_{sc,\alpha}^* -\frac{i \pi}{2} \mu_{\Omega_i^s})}{\sqrt{2 (E-U^*-m^*)^2 (E-U^*+m^*)}} 
    \\ \times    
    \left( \!\!\!\begin{array}{c} \mathcal{P}_1^* + i \alpha\mathcal{P}_2^* \\ E-U^*-m^* \end{array} \!\!\!\right)
    \exp\left[ \frac{i}{h} (\tau^* + \langle \mathcal{P}^* , x - \mathcal{X}^* \rangle) \right] 
    \\ \times
    \text{Ai}\left( -\frac{2^{1/3} \langle \mathcal{P}_\phi^* , x - \mathcal{X}^* \rangle}{h^{2/3}  \langle \mathcal{P}_\phi^*, \mathcal{X}_{\phi\phi}^* \rangle^{1/3}}  \right)
    + \mathcal{O}(h^{1/6}) ,
\end{multline}
where we have used the same manipulations that previously led to equation~(\ref{eq:PsiSingular-graphene}). Finally, let us briefly discuss the higher-order corrections to equation~(\ref{eq:Airy-final}). One of these corrections comes from the fourth-order term in the Taylor expansion of the phase function, see equation~(\ref{eq:phi-expansion-coeffs}), which was neglected in the derivation of equation~(\ref{eq:Airy-final}). By expanding the exponential function containing the higher-order terms, one can show~\cite{Dobrokhotov14} that this term gives rise to a contribution of~$\mathcal{O}(h^{1/6})$. Hence, it is part of the subleading term. A contribution of the same order is obtained when one expands the amplitude to first order in $\phi$ around $\phi^*$. Since we consider points $x$ in an $\mathcal{O}(h^{5/6})$-neighborhood of the point $x_\text{fold}$, the expansion of the amplitude with respect to $x$ gives rise to contributions that are of higher order than $\mathcal{O}(h^{1/6})$.

Let us now consider a point $x_\text{cusp}$ in the configuration space, to which corresponds a cusp point $(\tau^*,\phi^*)$ on the Lagrangian manifold. For points $x$ in its vicinity, the wavefunction is given by equation~(\ref{eq:PsiSingular}), as discussed in section~\ref{sec:maslov}. In the expansion~(\ref{eq:phi-expansion-coeffs}) of the phase function, the first nonvanishing coefficient comes from the fourth derivative. Neglecting higher-order terms in the Taylor expansion, expanding the amplitude to zeroth order in $\phi$ and performing a change of variables, one can subsequently express~\cite{Dobrokhotov14} the integral in equation~(\ref{eq:PsiSingular}) in terms of the Pearcey function~\cite{Pearcey46}, which is defined by
\begin{equation} \label{eq:Pearcey-def}
  \text{P}^{\pm}(u,v) = \int_{-\infty}^\infty \exp\left( \pm i t^4 + i u t^2 + i v t \right) \, \text{d} t .
\end{equation}
Note that the superscript on the Pearcey function corresponds to the sign in front of the coefficient $t^4$.
As before, one then assesses to which order in $h$ the various terms in the expression correspond and retains only the leading-order terms.
Finally, one obtains~\cite{Dobrokhotov14} that, for points $x$ in an $\mathcal{O}(h^{7/8})$-neighborhood of the point $x_\text{cusp}$, the leading-order approximation to the asymptotic solution~(\ref{eq:PsiSingular}) for the wavefunction is given by
\begin{multline} \label{eq:Pearcey-final}
  \Psi(x) = \frac{e^{i\pi/4}}{\sqrt{2\pi h}} e^{-i \pi \mu_{\Omega_i^s}/2 } \Bigg( 
  \sqrt[4]{\frac{24 h}{|a_4|}} \sqrt{| \det(\mathcal{P}^*,\mathcal{P}_\phi^*) |} \frac{A_0^0(\phi^*)e^{i \Phi_{sc}^*}}{\sqrt{R(\mathcal{X^*})}} \chi_0(\mathcal{X^*},\mathcal{P^*})
  \\ \hspace*{1cm} \times \exp\left[ \frac{i}{h} (a_0 + \langle b_0, z \rangle) \right] \text{P}^\pm \left( \sqrt{\frac{6}{h |a_4|}}\langle b_2, z \rangle, \sqrt[4]{\frac{24}{h^3 |a_4|}}\langle b_1, z \rangle  \right) 
  \\ + \mathcal{O}(h^{1/2})  \Bigg) .
\end{multline}
In this expression, one uses $P^+$ when $a_4 > 0$ and $P^-$ when $a_4 < 0$. Thus, the superscript on the Pearcey function corresponds to the sign of $a_4$.
As before, the coefficients $a_i$ and $b_i$ are given by equation~(\ref{eq:phi-expansion-coeffs}) and $z = x - \mathcal{X}^*$. Note in particular that, since $\mathcal{X}_{\phi\phi}^*$ is zero at a cusp point, the second term in both $a_4$ and $b_2$ vanishes, leaving only one term. Using the fact that the determinant factorizes at the singular point, we obtain that for the graphene Hamiltonian~(\ref{eq:Dirac}) this expression becomes
\begin{multline} \label{eq:Pearcey-final-graphene}
  \Psi_\alpha(x) = \frac{6^{1/4} e^{i\pi/4}}{\pi^{1/2} h^{1/4}} 
    \frac{\sqrt{|\mathcal{P}^*| |\mathcal{P}_\phi^*|}}{\sqrt[4]{| \langle \mathcal{P}_\phi^*, \mathcal{X}_{\phi\phi\phi}^* \rangle |}}
    \frac{A_0^0(\phi^*) \exp(i \Phi_{sc,\alpha}^* -\frac{i \pi}{2} \mu_{\Omega_i^s})}{\sqrt{2 (E-U^*-m^*)^2 (E-U^*+m^*)}}
    \\ \hspace*{1.2cm}  \times 
    \text{P}^\pm \left( \sqrt{\frac{6}{h |\langle \mathcal{P}_\phi^*, \mathcal{X}_{\phi\phi\phi}^* \rangle|}} \langle \mathcal{P}_{\phi\phi}^* , x - \mathcal{X}^* \rangle, \sqrt[4]{\frac{24}{h^3 |\langle \mathcal{P}_\phi^*, \mathcal{X}_{\phi\phi\phi}^* \rangle|}} \langle \mathcal{P}_\phi^* , x - \mathcal{X}^* \rangle \right)
    \\ \times    
    \left( \!\!\!\begin{array}{c} \mathcal{P}_1^* + i \alpha\mathcal{P}_2^* \\ E-U^*-m^* \end{array} \!\!\!\right)
    \exp\left[ \frac{i}{h} (\tau^* + \langle \mathcal{P}^* , x - \mathcal{X}^* \rangle) \right] 
    + \mathcal{O}(h^0) .
\end{multline}
The terms of $\mathcal{O}(h^0)$ in this expression arise from the fifth-order term in the Taylor expansion of the phase function and from the first-order term in the expansion of the amplitude with respect to $\phi$.

\subsection{Uniform approximation near the cusp}  \label{subsec:uniform}

In the previous subsection, we derived that the leading-order approximation near the cusp is given by equations~(\ref{eq:Pearcey-final}) and~(\ref{eq:Pearcey-final-graphene}). Within this approximation, the semiclassical phase does not influence the wavefunction. Indeed, when we compute $\lVert \Psi \rVert^2 = \Psi^\dagger \Psi$, we observe that the term containing $\Phi_{sc}$ drops out. Since the main focus lies in the vicinity of the cusp, this approximation predicts that the focusing does not depend on the semiclassical phase. However, since the semiclassical phase modifies the phase of each of the trajectories, it should affect the way they interfere. Therefore, the semiclassical phase should also affect the position of the focus and its intensity. To capture this effect, we need to go beyond the leading-order approximation.
The first way to go beyond the leading-order approximation is to include higher-order terms in the Taylor expansions that we made in the previous subsection. This method was employed in Ref.~\cite{Reijnders17a}, where it gave good results for the position of the intensity maximum. However, the approximation overestimated the height of the maximum. In the same paper, better results were obtained with the uniform approximation~\cite{Ursell72,Connor81b}. In this paper, we therefore only consider the latter method and its implementation. We review it in this subsection, based on Refs.~\cite{Ursell72,Connor81b}.

As we already briefly discussed, the theory of Lagrangian singularities~\cite{Arnold82} states that phase functions have a normal form in the vicinity of a singular point. To show what this means, let us consider the phase function $\tau(x,\phi)$ in the vicinity of a cusp point $x_\text{cusp}$.
In the previous subsection, we established that the fourth derivative $\tau_{\phi\phi\phi\phi}$ does not vanish at this point.
The theory of Lagrangian singularities then dictates that, for points $x$ in the vicinity of $x_\text{cusp}$, there is a smooth change of variables $\zeta = \zeta(x,\phi)$ such that
\begin{equation}  \label{eq:tau-normal-form}
  \tau = \pm \zeta^4 + w_2(x) \zeta^2 + w_1(x) \zeta + w_0(x) .
\end{equation}
This transformation is invertible, with inverse $\phi = \phi(x, \zeta)$. The sign in front of $\zeta^4$ equals the sign of $a_4$, defined in equation~(\ref{eq:phi-expansion-coeffs}). Furthermore, at the point $x_\text{cusp}$, the coefficients $w_2$ and $w_1$ vanish.
It is important to note that this change of variables is exact and that we did not neglect any higher-order terms in equation~(\ref{eq:tau-normal-form}). This is in contrast to the previous subsection, where we neglected the higher-order terms in the Taylor expansion to obtain a fourth-order polynomial. Instead, in equation~(\ref{eq:tau-normal-form}) we have eliminated the higher-order terms by the change of variables.

Although one can prove that the phase function has the normal form~(\ref{eq:tau-normal-form}) in the vicinity of a cusp point, this proof is not constructive. In particular, it requires the Malgrange preparation theorem~\cite{Arnold82,Poston78}.
Thus, the proof of the normal form does not provide us with a way to determine the transformation $\zeta = \zeta(x,\phi)$. In Ref.~\cite{Connor81b}, two methods are discussed to determine the transformation and the functions $w_i(x)$: an algebraic method and an iterative method. We discuss their iterative method below, referring the reader interested in the algebraic method to Ref.~\cite{Connor81b}.

In what follows, we consider a fixed point $x$. Since we require that the mapping $\zeta = \zeta(x,\phi)$ is one-to-one, the stationary points of $\tau(x,\phi)$ should be mapped to the stationary points of the fourth-order polynomial. Since the derivative of the phase function generates the Lagrangian manifold, we may also think of this step as making sure that the Lagrangian manifolds coincide, at least locally. Since $\zeta(x,\phi)$ is invertible, the derivative $\partial\zeta/\partial\phi$ is nonzero. Therefore, the stationary points of the right-hand side of equation~(\ref{eq:tau-normal-form}) are defined by
\begin{equation} \label{eq:def-roots-comp}
  0 = \pm 4 \zeta^3 + 2 w_2 \zeta + w_1 .
\end{equation}
We can classify the roots of this equation using the discriminant
\begin{equation} \label{eq:discriminant}
  \Delta = \mp 2^7 w_2^3 - 2^4 3^3 w_1^2 .
\end{equation}
There are three distinct cases that should be considered. When $\Delta$ is positive, all three roots are real and distinct. When $\Delta$ is zero, the roots are still real, but there is a multiple root. Finally, we have one real root and two complex conjugate roots when $\Delta$ is negative. Since a real extremum of the phase function corresponds to a trajectory on the configuration space, these three regimes correspond to three clearly identifiable regions of the configuration space. The regime $\Delta>0$ corresponds to the region inside the caustic, where each point lies on three trajectories. The caustic corresponds to $\Delta=0$. The regime $\Delta<0$ corresponds to the region outside the caustic, where each point lies on a single trajectory. One could say that the two complex roots correspond to ``complex trajectories''.

Let us first consider a point $x$ inside the caustic, i.e. in the interference region. This point lies on three trajectories, each of which corresponds to a stationary point $(\tau_i, \phi_i)$. When we label the three real roots of equation~(\ref{eq:def-roots-comp}) by $\zeta_i$, we therefore obtain the following set of three equations
\begin{equation} \label{eq:tau-stationary-points}
  \begin{aligned}
    \tau_1 &= \pm \zeta_1^4 + w_2 \zeta_1^2 + w_1 \zeta_1 + w_0, \\
    \tau_2 &= \pm \zeta_2^4 + w_2 \zeta_2^2 + w_1 \zeta_2 + w_0, \\
    \tau_3 &= \pm \zeta_3^4 + w_2 \zeta_3^2 + w_1 \zeta_3 + w_0. 
  \end{aligned}
\end{equation}
Since all stationary points are distinct, we can subtract both the second and the third equation from the first, giving rise to the following pair of equations:
\begin{equation} \label{eq:w-self-consistency}
  \begin{aligned}
    \tau_1 - \tau_2 &= \pm (\zeta_1^4-\zeta_2^4) + w_2 (\zeta_1^2-\zeta_2^2) + w_1 (\zeta_1-\zeta_2), \\
    \tau_1 - \tau_3 &= \pm (\zeta_1^4-\zeta_3^4) + w_2 (\zeta_1^2-\zeta_3^2) + w_1 (\zeta_1-\zeta_3). 
  \end{aligned}
\end{equation}
We can then determine the roots $\zeta_i$ and the parameters $w_1$ and $w_2$ using an iterative procedure. Before we can start this procedure, we require initial guesses $w_{i,0}$ for the parameters $w_i$. Since we are in the vicinity of the cusp point, a reasonable first estimate is provided by the result of the Taylor expansion, i.e.
\begin{equation}  \label{eq:w-initial-guesses}
  w_{2,0} = \sqrt{\frac{6}{|\langle \mathcal{P}_\phi^*, \mathcal{X}_{\phi\phi\phi}^* \rangle|}} \langle \mathcal{P}_{\phi\phi}^* , x - \mathcal{X}^* \rangle, 
  \quad 
  w_{1,0} = \sqrt[4]{\frac{24}{|\langle \mathcal{P}_\phi^*, \mathcal{X}_{\phi\phi\phi}^* \rangle|}} \langle \mathcal{P}_\phi^* , x - \mathcal{X}^* \rangle .
\end{equation}
Inserting these into equation~(\ref{eq:def-roots-comp}), we obtain three roots $\zeta_i$. These can subsequently be inserted into equation~(\ref{eq:w-self-consistency}) to obtain new guesses $w_{i,1}$ for the parameters $w_i$. This procedure is repeated until self-consistency in all parameters is reached. Finally, one determines $w_0$ using one of the three equations~(\ref{eq:tau-stationary-points}). Note that, throughout this procedure, one should use the sign in front of $\zeta^4$ that corresponds to the sign of $a_4$, see equation~(\ref{eq:phi-expansion-coeffs}).

In the procedure described above, we used all three stationary points to obtain values for the parameters $w_i$. When we consider a point $x$ outside the caustic, i.e. a point that only lies on a single trajectory, we only know one stationary point. We do not have any information on the two complex stationary points of the phase function. In principle, one could think about extending the phase function~(\ref{eq:defTauSingular}) into the complex plane in order to see if an approximation can be constructed in this way. However, since the solutions $\big( \mathcal{X}(\tau,\phi), \mathcal{P}(\tau,\phi) \big)$ of the Hamiltonian system are typically determined numerically, we only have an interpolating function that describes these solutions. Since we do not know their functional form, we also do not know the functional form of $\tau$, which makes this procedure even more complicated. We therefore do not consider this case in this paper, and limit our application of the uniform approximation to the interference region.

When we are on the caustic, the discriminant $\Delta=0$. One then obtains a relation between $w_1$ and $w_2$ from equation~(\ref{eq:discriminant}), which can subsequently be used to obtain values for the roots $\zeta_i$ and the parameters $w_i$. Since arbitrarily close to a point on the caustic is a point inside the interference region, we do not discuss this case here. Instead, we refer the interested reader to Ref.~\cite{Connor81b}.

Now that we have seen how we can obtain the stationary points $\zeta_i$ and the parameters $w_i$ for a point inside the interference region, let us consider the integral~(\ref{eq:PsiSingular}). 
To simplify the notation, we define a new function $g(x,\phi)$ which contains all factors in the amplitude of the integrand, except for the determinant and the $h$-dependence.
In particular, this function $g(x,\phi)$ contains the semiclassical phase.
Subsequently, we change the integration variable from $\phi$ to $\zeta$. Using equation~(\ref{eq:tau-normal-form}), we obtain
\begin{multline}  \label{eq:Psi-uniform-change-vars}
  \Psi(x) = \frac{1}{\sqrt{h}} \int_{-\infty}^\infty \text{d} \phi \sqrt{|\det(\mathcal{P},\mathcal{P}_\phi)|} \, g(x,\phi) e^{\frac{i}{h}\tau(x,\phi)} \big( 1 + \mathcal{O}(h) \big)  \\ 
    = \frac{e^{\frac{i}{h} w_0(x)}}{\sqrt{h}} \int_{-\infty}^\infty \text{d} \zeta \, G(x,\zeta) e^{\frac{i}{h}\left(\pm \zeta^4 + w_2(x) \zeta^2 + w_1(x) \zeta\right)} \big( 1 + \mathcal{O}(h) \big) ,
\end{multline}
where we have defined the new amplitude function
\begin{equation}  \label{eq:uniform-change-vars}
  G(x,\zeta) = \left| \frac{\text{d} \phi}{\text{d} \zeta} \right| \sqrt{|\det(\mathcal{P},\mathcal{P}_\phi)|} \, g(x, \phi(x, \zeta)) .
\end{equation}
The next step in the procedure~\cite{Connor81b,Ursell72} is to expand this new amplitude function in powers of $\zeta$, i.e.
\begin{equation}  \label{eq:uniform-amp-expansion}
  G(x,\zeta) = D_0(x) + D_1(x) \zeta + D_2(x) \zeta^2 + \mathcal{O}(\zeta^3) .
\end{equation}
When we insert this expansion into the integral~(\ref{eq:Psi-uniform-change-vars}), we obtain the uniform approximation for the wavefunction, as we will see shortly. However, let us first consider how we can determine the constants $D_i$. To this end, we neglect the higher-order terms in the Taylor expansion~(\ref{eq:uniform-amp-expansion}) and combine this equation with equation~(\ref{eq:uniform-change-vars}). Subsequently, we specialize to the three stationary points. This gives us a system of three linear equations in the three variables $D_0$, $D_1$ and $D_2$. We can solve these equations when we know the value of the derivative $\text{d} \phi/\text{d} \zeta$ at each of the stationary points. We determine this derivative by considering the second derivative of relation~(\ref{eq:tau-normal-form}) with respect to $\zeta$. This gives
\begin{equation}
  \frac{\partial^2 \tau}{\partial \phi^2} \left( \frac{\text{d} \phi}{\text{d} \zeta} \right)^2 + \frac{\partial \tau}{\partial \phi} \frac{\text{d}^2 \phi}{\text{d} \zeta^2} 
    = \pm 12 \zeta^2 + 2 w_2 .
\end{equation}
When we are at a stationary point, the second term on the left-hand side vanishes, since $\tau_\phi$ vanishes. Furthermore, at a stationary point the second derivative of the action is given by $\tau_{\phi\phi} = - \langle \mathcal{P}_\phi, \mathcal{X}_\phi \rangle$, see equation~(\ref{eq:tauphiphi-stationary}). Since the terms on the right-hand side are known as well, we have thus obtained an expression for the derivative $\text{d} \phi/\text{d} \zeta$ at a stationary point. Note that we can generally choose the mapping $\zeta(x,\phi)$ to be orientation preserving, rendering the derivative positive.
Hence, the constants $D_0$, $D_1$ and $D_2$ are determined by the following system of three linear equations, where the index $i$ labels the stationary points.
\begin{equation}  \label{eq:uniform-amplitude-linear-eqs}
  \left( \frac{\pm 12 \zeta_i^2 + 2 w_2}{\partial^2 \tau/\partial \phi^2|_{\phi=\phi_i}} \, |\det(\mathcal{P},\mathcal{P}_\phi)| \right)^{1/2}  g(x, \phi_i) = D_0 + D_1 \zeta_i + D_2 \zeta_i^2
\end{equation}
With these constants $D_i$, we can subsequently obtain the uniform approximation to the wavefunction.

Before we evaluate the integral~(\ref{eq:Psi-uniform-change-vars}), we define the derivatives of the Pearcey function~(\ref{eq:Pearcey-def}) with respect to $u$ and $v$. They are given by
\begin{align}
  \text{P}_v^{\pm}(u,v) &= i \int_{-\infty}^\infty t \exp\left( \pm i t^4 + i u t^2 + i v t \right) \, \text{d} t .  \label{eq:def-Pearcey-diff-v} \\
  \text{P}_u^{\pm}(u,v) &= i \int_{-\infty}^\infty t^2 \exp\left( \pm i t^4 + i u t^2 + i v t \right) \, \text{d} t . \label{eq:def-Pearcey-diff-u}
\end{align}
When one inserts the expansion~(\ref{eq:uniform-amp-expansion}) in the integral~(\ref{eq:Psi-uniform-change-vars}), one can show that~\cite{Ursell72}
\begin{multline} \label{eq:uniform-final}
  \Psi(x) = e^{\frac{i}{h} w_0(x)} \left[    
    h^{-1/4} D_0 \text{P}^{\pm} \left( \frac{w_2}{h^{1/2}} , \frac{w_1}{h^{3/4}} \right)
    -i h^{0} D_1 \text{P}_v^{\pm} \left( \frac{w_2}{h^{1/2}} , \frac{w_1}{h^{3/4}} \right) \right. \\
    \left. -i h^{1/4} D_2 \text{P}_u^{\pm} \left( \frac{w_2}{h^{1/2}} , \frac{w_1}{h^{3/4}} \right)
  \right] + \mathcal{O}(h^{3/4}) .
\end{multline}
This expression is the uniform approximation to the asymptotic solution~(\ref{eq:PsiSingular}). Comparing it to the leading-order approximation~(\ref{eq:Pearcey-final}), we observe that it consists of three terms instead of just one. These two additional terms are of higher order in $h$, making it a more accurate approximation. Furthermore, because we now have three terms, complex phases in the amplitude no longer cancel when computing the intensity $\lVert \Psi \rVert^2$. Thus, within the uniform approximation the semiclassical phase, which is encoded in the amplitude expansion coefficients $D_i$, has a clear influence on the intensity. Unfortunately, this dependence is less explicit than in our previous approximations.

The main limitation of the uniform approximation is that we can only use it within the region where interference takes places, i.e. inside the caustic. As we have seen, this limitation is not easy to circumvent, as this would require knowledge of the complex stationary points of the action. On the other hand, the interference region is the most important region, as the maximum of the intensity lies in this region. Finally, we remark that one can also construct a uniform approximation near a fold point~\cite{Chester57} see also e.g. Ref.~\cite{Connor81b}. However, we do not consider this uniform Airy approximation in this paper, as we are mainly interested in the intensity maximum near the cusp.

\section{Numerical implementation and comparison to tight-binding calculations}  \label{sec:examples}

In this section, we discuss how one can implement the results of the previous two sections to obtain numerical values. We also discuss the results of our implementation in Wolfram Mathematica~\cite{Mathematica}.
In section~\ref{subsec:classical-trajectories}, we discuss the classical trajectories, the variational system and the inverse functions $\tau_i(x),\phi_i(x)$. In the next subsection, we implement the leading-order approximations~(\ref{eq:Pearcey-final-graphene}) and~(\ref{eq:Airy-final-graphene}) together with the WKB approximation~(\ref{eq:PsiRegular-graphene}) for small $h$. In particular, we take a close look at their region of applicability. In section~\ref{subsec:numerics-large-h}, we consider larger values of $h$, which correspond to physically more realistic situations. Since we need the uniform approximation to obtain good results for this case, we discuss its implementation and limitations. The results of the uniform approximation are compared with the results of tight-binding calculations, performed using the Kwant package~\cite{Groth14}, in section~\ref{subsec:numerics-large-h-comparison}. We discuss different sample setups and the origin of the observed deviations. Finally, in section~\ref{subsec:numerics-large-h-semiclassical-phase}, we discuss how the semiclassical phase influences the maximum of the intensity. We compare the uniform approximation with tight-binding results and also consider what happens when one incorporates the semiclassical phase into the trajectories.

\subsection{Classical trajectories}  \label{subsec:classical-trajectories}

\begin{figure}[tb]
  \begin{center}
    \includegraphics{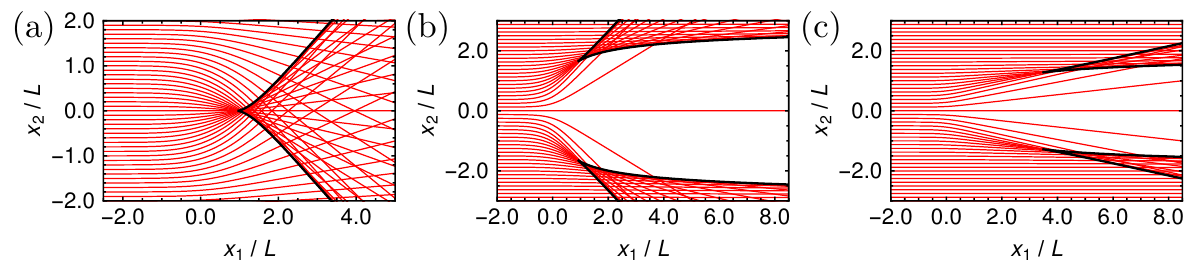}
    \caption{Trajectories obtained by integrating Hamilton's equations~(\ref{eq:Ham-Dirac-2}) for different potentials $\tilde{U}(\tilde{x})$ and masses $\tilde{m}(\tilde{x})$. The black lines indicate the caustics. (a) Gaussian potential well~(\ref{eq:pot-gaussian-well}) with $\tilde{U}_0=\tfrac{1}{2}$ and $\tilde{m}(\tilde{x})=0$. (b) Gaussian potential barrier~(\ref{eq:pot-gaussian-well}) with $\tilde{U}_0=-\tfrac{1}{2}$ and $\tilde{m}(\tilde{x})=0$. (c) Gaussian mass $\tilde{m}(\tilde{x}) = \tilde{m}_0 \exp(-\tilde{x}^2)$ with $|\tilde{m}_0| = \tfrac{1}{2}$ and $\tilde{U}(\tilde{x})=0$.}
    \label{fig:trajectories}
  \end{center}
\end{figure}

The first step in the numerical implementation of the semiclassical approximation consists of obtaining the classical trajectories. To this end, we numerically integrate Hamilton's equations~(\ref{eq:Ham-Dirac-2}) with the initial condition~(\ref{eq:Lambda1}). In this section, we mostly consider a Gaussian potential, given by, in units with dimensions,
\begin{equation} \label{eq:pot-gaussian-well}
  U(x) = - U_0 \exp(-x^2/L^2) , \quad \text{whence} \quad \tilde{U}(\tilde{x}) = - \tilde{U}_0 \exp(-\tilde{x}^2) 
\end{equation}
in dimensionless units. The starting point $x_1^0$ of the integration is always chosen in such a way that the potential $U(x_1^0,0)$ is very small. Typically, we set $x_1^0 \leq -5L$. Therefore, we can take $p_1^0(\phi)$ to be the constant $E$ without making noticeable errors. From the numerical integration we obtain the solutions $\big\{ \mathcal{X}(\tau,\phi),\mathcal{P}(\tau,\phi)\big\}$, which form the Lagrangian manifold $\Lambda^2$ parametrized by eikonal coordinates $(\tau,\phi)$.

In this section, our main example is a potential well, for which $U_0>0$. When we set $\tilde{U}_0 = \tfrac{1}{2}$ and set the mass $\tilde{m}(\tilde{x})$ to zero, we obtain the trajectories shown in figure~\ref{fig:trajectories}(a). We observe that the trajectories are focused and that a cusp caustic is formed. Since the potential is symmetric in $x_2$, the trajectories have the same symmetry. In particular, the cusp point lies on the $x_1$-axis and we have $x_\text{cusp} = (x_{1,\text{cusp}},0)$.

When we instead consider a Gaussian potential barrier, obtained by setting $\tilde{U}_0=-\tfrac{1}{2}$, we obtain the trajectories shown in figure~\ref{fig:trajectories}(b). This time, the potential bends the trajectories outwards, and two cusp caustics are formed, which lie symmetrically around the $x_1$-axis. A third situation one can consider is a Gaussian mass, $\tilde{m}(\tilde{x}) = \tilde{m}_0 \exp(-\tilde{x}^2)$, with the potential $\tilde{U}(\tilde{x})$ set to zero. Since Hamilton's equations~(\ref{eq:Ham-Dirac-2}) only depend on the mass through $\tilde{m}^2$, the sign of the mass does not influence the classical trajectories. Figure~\ref{fig:trajectories}(c) shows the trajectories obtained for $|\tilde{m}_0| = \tfrac{1}{2}$. We see that the classical trajectories are again bent outwards, although much slower than by a potential barrier of equal height.

We obtain the inverse functions $\tau_i(x), \phi_i(x)$ on each leaf in three steps. First, we compute the positions $x$ for a large number of points $\phi$ and $\tau$ on a grid and store this data in a table. Second, we split the data into the different leaves. Third, we interpolate the data to obtain the inverse functions. However, the density of trajectories is far from constant in our problem, as can for instance be seen in figure~\ref{fig:trajectories}(b). In order to obtain a good interpolating function, the density of trajectories should not be too low. Therefore, we use an adaptive step size algorithm to ensure that the separation between neighboring trajectories does not exceed a certain threshold at their initial and final interpolation points. This algorithm, very similar to algorithms typically used in solvers for ordinary differential equations, see e.g. Ref.~\cite{DeVries94}, leads to a varying step size in $\phi$.

We compute the (higher-order) derivatives of $\mathcal{X}$ and $\mathcal{P}$ with respect to $\phi$ using the variational system~(\ref{eq:varsys}). Furthermore, by taking the derivative of equation~(\ref{eq:BerryPhaseEikonal}) with respect to $\tau$, we obtain a differential equation for the Berry phase. By adding this equation to the dynamical system, we can simultaneously obtain the Berry phase along the trajectories.

\subsection{Implementation of the semiclassical approximation: small $h$}  \label{subsec:numerics-small-h}

Now that we have computed the classical trajectories, we want to obtain numerical values for the asymptotic solution for the wavefunction. Since all our results are expressed as asymptotic series in $h$, we start our discussion of the numerical implementation by examining the deep semiclassical limit, in which $h$ is small. In the previous section, we established an approximate region of validity for the leading-order approximations~(\ref{eq:Pearcey-final-graphene}) and~(\ref{eq:Airy-final-graphene}) in terms of powers of $h$. However, we did not write down an exact region of validity. In fact, this region can vary slightly depening on the problem that one considers and needs to be determined by inspecting the numerical results. For instance, to establish the region of validity of the leading-order Pearcey approximation~(\ref{eq:Pearcey-final-graphene}), one plots both this approximation and the WKB approximation~(\ref{eq:PsiRegular-graphene}) in a region around the cusp.
Since both approximations represent the asymptotic solution in a certain region of configuration space, this plot should show a small transition region, in which both approximations give the same result. This region then marks the end of the region of validity of the leading-order Pearcey approximation, and the start of the region of validity of the WKB approximation. We emphasize that, throughout the whole procedure, we do not match the different results by adjusting their coefficients. Instead, both approximations~(\ref{eq:Pearcey-final-graphene}) and~(\ref{eq:PsiRegular-graphene}) represent the asymptotic solution in a certain region and all that has to be done is to determine the boundary of this region.

In this subsection, we consider an electron with energy $E=200$~meV incident on a Gaussian potential well~(\ref{eq:pot-gaussian-well}) with $U_0=100$~meV and $L=10^4$~nm. We set the mass $m(x)$ to zero. Hence, the semiclassical parameter $h = \hbar v_F/(E L)$ equals $3.2\cdot 10^{-4}$, which shows that we are indeed in the deep semiclassical regime. We take the initial amplitude $A_0^0(\phi)$ to be constant, and set $A_0^0(\phi) = 1$. Since the potential is symmetric in $x_2$, the trajectories have this same symmetry. In particular, the trajectory with $\phi=0$ coincides with the $x_1$-axis and the cusp point $x_\text{cusp} = (x_{1,\text{cusp}},0)$ lies on this trajectory.

Since the mass $m(x)$ vanishes and our potential is symmetric, we have $\Psi_\alpha(x,-y) = \sigma_x \Psi_\alpha(x,y)$, see equation~(\ref{eq:symm-zero-mass}). Therefore, the intensity $\lVert \Psi_\alpha \rVert = (\Psi_\alpha^\dagger \Psi_\alpha)^{1/2}$ is symmetric about the $x_1$-axis. Furthermore, we have $\lVert \Psi_{K} \rVert = \lVert \Psi_{K'} \rVert$ by equation~(\ref{eq:symm-between-valleys-zero-mass}). Hence, the intensities in both valleys are equal. In this subsection, we therefore omit the valley index when we consider the norm of the wavefunction, i.e. we write $\lVert \Psi \rVert$.

We determine the region of validity for the Pearcey approximation~(\ref{eq:Pearcey-final-graphene}) by comparing it to the other approximations along the $x_1$-axis and along a line perpendicular to the $x_1$-axis. Unfortunately, the Pearcey function is not implemented in most computer algebra systems, including Wolfram Mathematica~\cite{Mathematica}. Therefore, we have implemented it using the contour integral method described in Ref.~\cite{Connor82}. Other implementation schemes can be found in Ref.~\cite{Connor81a}. The (higher-order) derivatives of $\mathcal{X}$ and $\mathcal{P}$ with respect to $\phi$ are obtained by numerically integrating the variational system. Evaluating them at the cusp point, we obtain that the coefficient $a_4$ is positive for our example. Subsequently, we can easily implement the Pearcey approximation~(\ref{eq:Pearcey-final-graphene}).
The implementation of the WKB approximation in the interference region is simplified by the fact that we probe along the trajectory through the cusp point, which means that we already know $\phi$ and $\tau$ on the middle leaf of the Lagrangian manifold. To implement the Airy approximation, we compute the point on the fold caustic that is closest to the point under consideration. 
We subsequently use the parameters of this point in equation~(\ref{eq:Airy-final-graphene}), and add the WKB approximation on the third leaf of the Lagrangian manifold to the result to obtain the Airy approximation. Finally, we construct the uniform approximation~(\ref{eq:uniform-final}) in the region where the Lagrangian manifold has three leaves using the procedure described in section~\ref{subsec:uniform}.

\begin{figure}[tb]
  \begin{center}
    \includegraphics{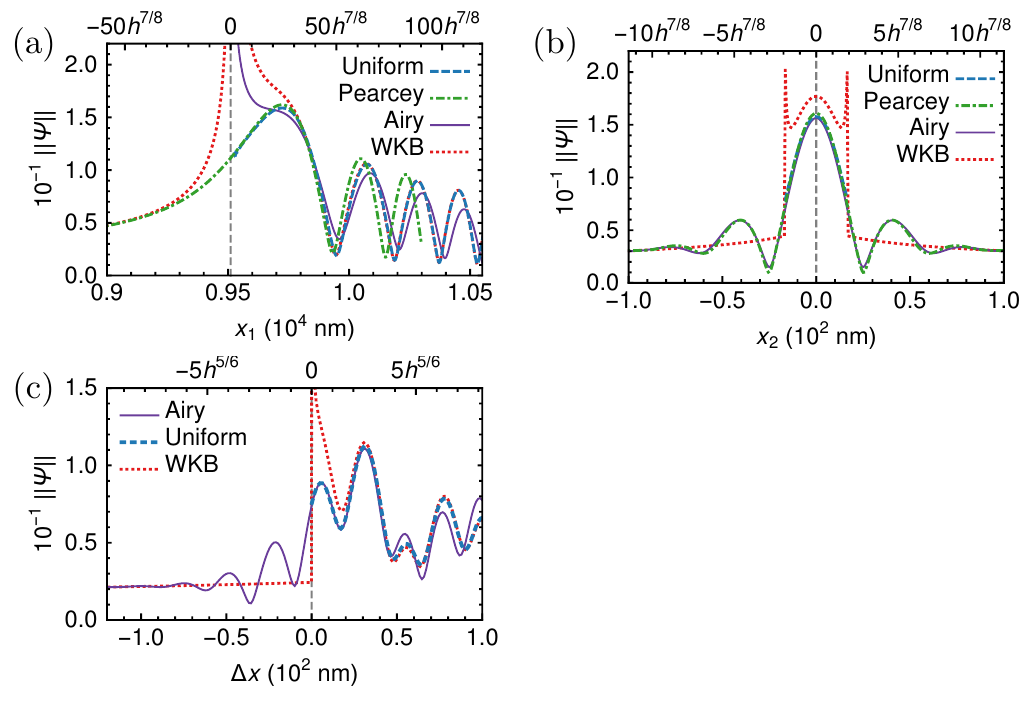}
    \caption{Comparison of the different semiclassical approximations near the caustic for $E=200$~meV, $U_0=100$~meV and $L=10^4$~nm. The position of the caustic is indicated by a vertical, dashed grey line. The bottom scale shows the position in nm, with the origin of the potential lying at $(0,0)$. The top scale shows the dimensionless distance to the caustic in the relevant power of the semiclassical parameter, which is $h^{7/8}$ for the cusp and $h^{5/6}$ for the fold. (a) Comparison along the $x_1$-axis. At a distance of about $40 h^{7/8}$, the Pearcey approximation~(\ref{eq:Pearcey-final-graphene}) smoothly joins the WKB approximation~(\ref{eq:PsiRegular-graphene}), which diverges near the cusp. The uniform approximation~(\ref{eq:uniform-final}) interpolates between these two approximations. (b) Comparison along the line $x_1=9.7\cdot10^3$~nm, close to the maximum. The Airy, Pearcey and uniform approximation give similar results, whereas the WKB approximation is very different. (c) Comparison along a line perpendicular to the caustic at the point $x_\text{fold}=(10117, 98)$~nm. The Airy approximation, consisting of the sum of the result~(\ref{eq:Airy-final-graphene}) and the WKB approximation on the third leaf, smoothly joins the WKB approximation at $\Delta \tilde{x}=|x-x_\text{fold}|/L=-8h^{5/6}$ and at $\Delta\tilde{x}=3.5h^{5/6}$.}
    \label{fig:compL10000}
  \end{center}
  \end{figure}

Two comparisons of these different approximations are shown in figures~\ref{fig:compL10000}(a) and (b). We observe that both the Airy approximation and the WKB approximation diverge near the cusp point, as shown theoretically in the previous sections. Along the $x_1$-axis, we observe that the values of the Pearcey approximation and the WKB approximation are very close to each other when we are at a distance of approximately $40 h^{7/8}$ from the cusp. This indicates that we should use the Pearcey approximation until this point, and the WKB approximation after. Near the cusp, the Pearcey approximation almost coincides with the uniform approximation. Thus, for the small value of $h$ that we are considering, the leading-order term is sufficient to approximate the behavior near the cusp and we do not require the higher-order corrections that are comprised within the uniform approximation. Further away from the cusp, the uniform approximation smoothly coincides with the WKB approximation.
On the line parallel to the $x_2$-axis that goes through the maximum, the situation is somewhat different. Here we observe that the Pearcey approximation and the Airy approximation already coincide at a rather small distance from the $x_1$-axis.
In a similar way, we compare the various approximations along a line perpendicular to a point on the fold caustic, see figure~\ref{fig:compL10000}(c). On the inside of the caustic, where we have interference, we observe that the WKB approximation and the Airy approximation have similar values between $3 h^{5/6}$ and $4 h^{5/6}$ from the fold. On the outside of the caustic, the Airy approximation shows additional oscillations that are not exhibited by the WKB approximation and that are significant until a distance of approximately $8 h^{5/6}$ from the fold.

  \begin{figure}[tb]
  \begin{center}
    \includegraphics{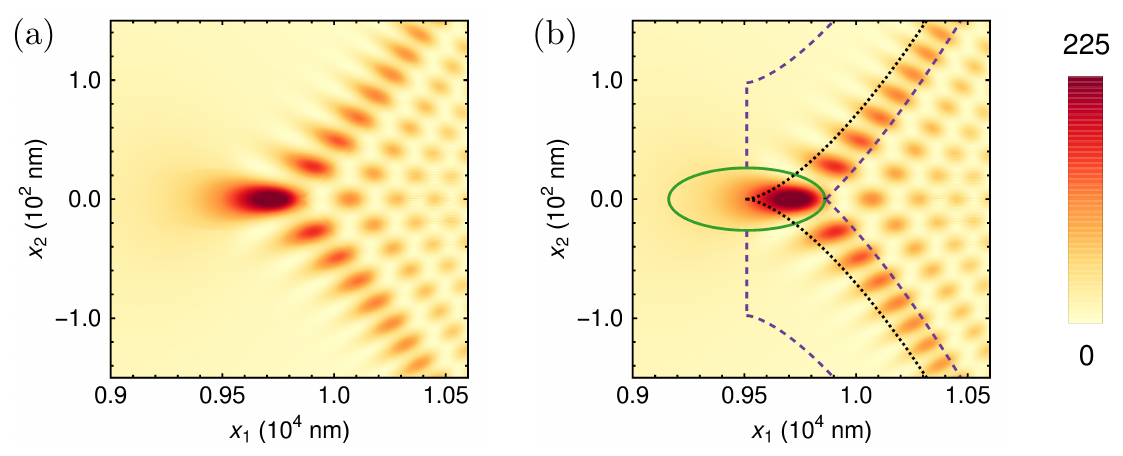}
    \caption{(a) Intensity $\lVert \Psi \rVert^2 = \Psi^\dagger \Psi$ obtained by using the various semiclassical approximations in the appropriate regions. (b) The intensity $\lVert \Psi \rVert^2$ together with the regions in which the various approximations were used. The Pearcey approximation~(\ref{eq:Pearcey-final-graphene}) was used inside the green ellipse, while the Airy approximation~(\ref{eq:Airy-final-graphene}) was used between the dashed purple lines. The WKB approximation~(\ref{eq:PsiRegular-graphene}) was used to create the rest of the figure. The dotted black line represents the caustic.}
    \label{fig:densityL10000}
  \end{center}
  \end{figure}

Now that we have determined their regions of validity, we can combine the Pearcey approximation, the Airy approximation and the WKB approximation to obtain the intensity $\lVert \Psi \rVert^2$ in a large region. We first compute the Pearcey approximation~(\ref{eq:Pearcey-final-graphene}) on a rectangular grid around the cusp. Subsequently, we consider only the points within an ellipse with semimajor axis $40 h^{7/8}$ and semiminor axis $3 h^{7/8}$.
On the fold caustic, we use an adaptive step size algorithm to construct a grid of points that are roughly equally spaced. At each of those points, we construct the line perpendicular to the fold. On this line, we create a regular grid of points that extends $8 h^{5/6}$ in the outward direction and $3.5 h^{5/6}$ in the inward (interference) direction and compute the Airy approximation on this grid.
Outside of these regions, we use the WKB approximation. We use the adaptive step size algorithm, see section~\ref{subsec:classical-trajectories}, to control the maximal spacing between the trajectories. Outside of the caustic, where no interference takes place, we use a rather large stepsize, as the variations in the wavefunction are much smaller than inside the caustic.
Figure~\ref{fig:densityL10000} shows the result of combining the various approximations in this way. We note that it is very smooth, which indicates that we have correctly determined the regions of validity for each of the approximations.

\subsection{Physically realistic situations: large $h$}  \label{subsec:numerics-large-h}

\begin{figure}[tb]
  \begin{center}
    \includegraphics{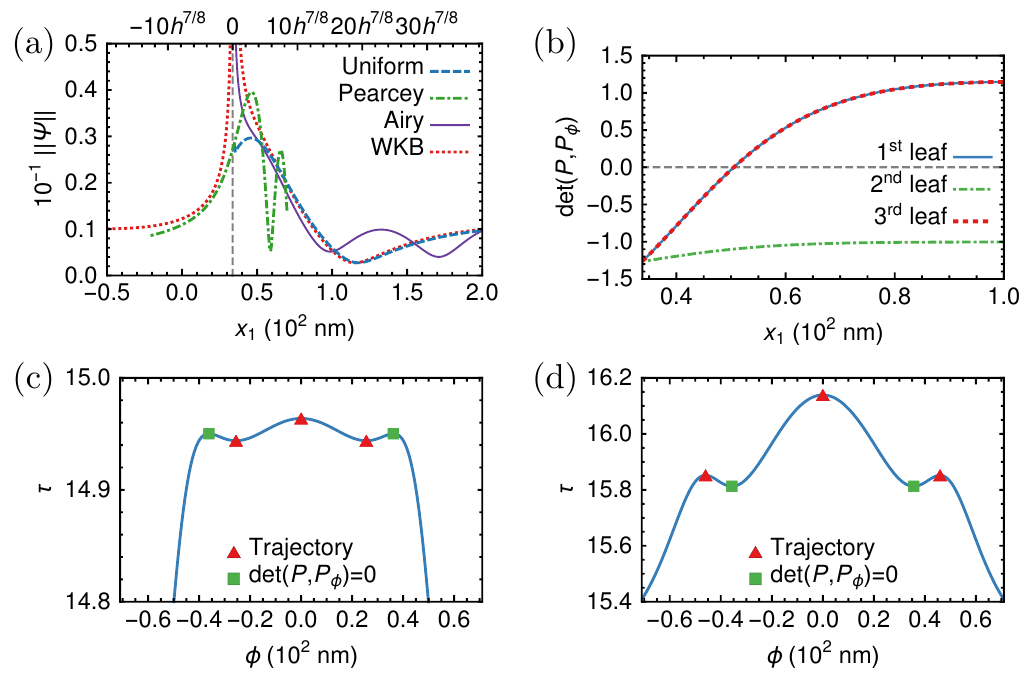}
    \caption{(a) Comparison of the different semiclassical approximations along the $x_1$-axis for $E=200$~meV, $U_0=100$~meV and $L=35.5$~nm. The position of the caustic is indicated by a vertical, dashed grey line. The bottom scale shows the position in nm, with the origin of the potential lying at $(0,0)$. The top scale shows the dimensionless distance to the caustic in units of $h^{7/8}$. (b) The value of the determinant $\det(\mathcal{P},\mathcal{P}_\phi)$ for each of the three points on the Lagrangian manifold that are projected onto $(x_1,0)$. (c), (d) The action $\tau(x,\phi)$ as a function of $\phi$ for (c) $x=(40,0)$~nm, (d) $x=(80,0)$~nm. There are two types of extrema, corresponding to trajectories and to points where the determinant vanishes.}
    \label{fig:compL35_5uniform}
  \end{center}
  \end{figure}

In the previous subsection, we considered the deep semiclassical limit. We saw that, within this limit, each of the approximations has a well-defined regime of validity and that their combination gives rise to a smooth intensity.
However, the length scale that we considered is much larger than the length scales that are typically considered in graphene devices. Therefore, we now decrease the length scale to $L=35.5$~nm, keeping the other parameters the same as in the previous subsection. The semiclassical parameter then becomes $h = \hbar v_F/(E L) = 0.09$, meaning that we are outside the deep semiclassical regime. 
In figure~\ref{fig:compL35_5uniform}(a), we show a comparison of the various approximations along the $x_1$-axis for this reduced length scale.
An important observation is that the region in which the Pearcey approximation and the WKB approximation coincide has now disappeared. On the contrary, the uniform approximation still coincides with the WKB approximation far away from the cusp.
The Pearcey approximation also predicts a much larger maximal value than the uniform approximation, although the positions of their maxima roughly coincide. These observations indicate that the leading-order term in the asymptotic expansion is no longer sufficient near the cusp. Instead, the higher-order corrections that are included in the uniform approximation prove to be essential.
Similar behavior was observed in Ref.~\cite{Reijnders17a}, where the Pearcey approximation was compared with the uniform approximation and an exact result. In that example, the uniform approximation coincided with the exact result, whereas the Pearcey approximation did not.

At this point, we would like to discuss a technical point regarding the implementation of the uniform approximation.
In the previous sections, we extensively discussed the region of validity of the representation~(\ref{eq:PsiSingular}) of the asymptotic solution corresponding to singular charts. In particular, we established that it is only valid when $\det(\mathcal{P},\mathcal{P}_\phi)$ does not vanish, which gives a boundary on the maximal size of the singular chart. However, for our current parameters, this region turns out to be quite small. When $x_1>x_{1,\text{cusp}}$, the point $(x_1,0)$ corresponds to three points on the Lagrangian manifold, with each of these points lying on a distinct leaf. If the determinant $\det(\mathcal{P},\mathcal{P}_\phi)$ vanishes on one of these leaves for a certain point $(x_{1,\text{crit}},0)$, we cannot use the uniform approximation past this point. Looking at figure~\ref{fig:compL35_5uniform}(b), we conclude that, for our current set of parameters, we cannot use the uniform approximation past $x_{1,\text{crit}} \approx 50$~nm.

We can get a better understanding of what is going on by looking at figures~\ref{fig:compL35_5uniform}(c) and (d). They show the action $\tau(x,\phi)$, defined by equation~(\ref{eq:defTauSingular}), as a function of $\phi$ for two points $(x_1,0)$: a point with $x_{1,\text{cusp}} < x_1 < x_{1,\text{crit}}$ in panel (c) and a point with $x_1 > x_{1,\text{crit}}$ in panel (d). In both cases, the action has five extrema. Three of these extrema correspond to a trajectory: for this value of $\phi$ we have $x=\mathcal{X}(\tau(x,\phi),\phi)$. The other two extrema do not correspond to a trajectory. Instead, the determinant $\det(\mathcal{P},\mathcal{P}_\phi)$ vanishes at these points. Hence, for these values of $\phi$, the vectors $\mathcal{P}$ and $\mathcal{P}_\phi$ are parallel. The vector $x-\mathcal{X}$ is nonzero and is orthogonal to both $\mathcal{P}$ and $\mathcal{P}_\phi$, see equations~(\ref{eq:defTauSingular}) and~(\ref{eq:tauStationaryLagMan}).

When $x_{1,\text{cusp}} < x_1 < x_{1,\text{crit}}$, the three extrema corresponding to a trajectory lie between those corresponding to the vanishing determinant. This is in agreement with the observation that $\det(\mathcal{P},\mathcal{P}_\phi)$ does not vanish at the cusp point. In this case, the second derivatives at the extrema corresponding to a trajectory have alternating signs: negative for the extremum corresponding to the second leaf, and positive for the extrema corresponding to the first and third leaf of the Lagrangian manifold. When $x_1 > x_{1,\text{crit}}$, the situation is very different. This time, the extrema are of alternating type, as the extrema corresponding to a vanishing determinant lie in between the extrema corresponding to a trajectory. In particular, the extrema corresponding to a trajectory all have a negative second derivative. 

The additional extrema of $\tau(x,\phi)$, which are caused by the vanishing of the determinant $\det(\mathcal{P},\mathcal{P}_\phi)$, complicate our analysis. Since these points are stationary points of the phase function, they contribute to the stationary phase approximation of the integral~(\ref{eq:PsiSingular-graphene}). However, their contributions do not have any physical significance, since the extrema do not correspond to a point on the Lagrangian manifold, or, equivalently, to a trajectory in phase space. As discussed in the previous sections, see also Refs.~\cite{Dobrokhotov14,Dobrokhotov14b,Guillemin77}, the physically relevant contributions to the integral only come from points on the Lagrangian manifold. Therefore, we should not take these stationary points into account when we compute the integral~(\ref{eq:PsiSingular-graphene}). Formally, this can be done by excluding a small area around the stationary point from the integration interval, since, within the stationary phase approximation, only a small region around a stationary point contributes to the result~\cite{Maslov81,Guillemin77,Dobrokhotov14}. In practice, however, one hardly ever computes the integral explicitly. Instead, one typically constructs either the leading-order approximation, see section~\ref{subsec:leadingorder}, or the uniform approximation, see section~\ref{subsec:uniform}. We remark that the additional extrema do not contribute to the leading-order approximation: since the amplitude is proportional to the determinant, it vanishes at these points. However, these points can still give rise to contributions of $\mathcal{O}(h^{0})$ or higher through the derivatives of the amplitude. An example where such contributions were explicitly taken into account can be found in Ref.~\cite{Reijnders17a}.

When we constructed the uniform approximation in section~\ref{subsec:uniform}, we did not take the additional extrema into account. From the above considerations, we conclude that this is correct, which justifies our use of the result~(\ref{eq:uniform-final}). However, in the construction we implicitly assumed that the second derivatives at the extrema have alternating signs. This can be seen from equation~(\ref{eq:uniform-amplitude-linear-eqs}), where we divide the second derivative of the normal form~(\ref{eq:tau-normal-form}) by the second derivative of the phase function. As we discussed, this assumption corresponds to the situation $x_{1,\text{cusp}} < x_1 < x_{1,\text{crit}}$. Since we would also like to use the uniform approximation for larger values of $x_1$, let us see if there is a way to smoothly continue the approximation past the point $x_{1,\text{crit}}$.

A key observation is that the sign of $\det(\mathcal{P},\mathcal{P}_\phi)/\tau_{\phi\phi}$ does not change when we pass the point $x_{1,\text{crit}}$, since both the sign of the determinant and the sign of the second derivative change on the first and third leaf, see figures~\ref{fig:compL35_5uniform}(b)--(d). Therefore, we obtain smooth amplitude functions $D_i(x)$ when we replace $| \det(\mathcal{P},\mathcal{P}_\phi) |$ by $- \det(\mathcal{P},\mathcal{P}_\phi) $ in equation~(\ref{eq:uniform-amplitude-linear-eqs}). In this expression, one chooses a minus sign since the determinant is negative on all leaves for $x_{1,\text{cusp}} < x_1 < x_{1,\text{crit}}$. The uniform approximation that results from this replacement is smooth across the point $x_{1,\text{crit}}$ and coincides with our previous approximation in the region $x_{1,\text{cusp}} < x_1 < x_{1,\text{crit}}$. From a more theoretical point of view, one can view the replacement of the absolute value as a way to obtain the correct value of the Maslov index. This index naturally changes when one makes a transition from one chart to another, as discussed in section~\ref{subsec:maslovindex}. The uniform approximation for $x_1 > x_{1,\text{crit}}$ can then be seen as an implementation of the representation~(\ref{eq:precanonicalXsingularProperTime}) in regular points, as discussed in section~\ref{subsec:maslovsingular}. 

The uniform approximation that is plotted in figure~\ref{fig:compL35_5uniform}(a) was constructed using the method explained above. We observe that both its amplitude and its phase smoothly coincide with the WKB approximation for large values of $x_1$, which indicates that our continuation is correct.
Note that this continuation is only possible because nothing essentially changes at the point $x_{1,\text{crit}}$, in the sense that there is no physical singularity at this point. The difficulties that we experience at this point are rather a consequence of the representation~(\ref{eq:PsiSingular}), just as the additional extrema of the phase function are.

\subsection{Comparison with tight-binding calculations}   \label{subsec:numerics-large-h-comparison}

We can assess the quality of our various semiclassical approximations, and of the uniform approximation in particular, by comparing their outcomes with numerical results. Using the Kwant package~\cite{Groth14}, we numerically compute the wavefunction within the tight-binding approximation. In this approximation, we only take nearest-neighbor interactions between the carbon atoms into account. The continuum Dirac Hamiltonian is obtained from this approximation within the limit of small $k$, see e.g. Ref.~\cite{Katsnelson13}.
We construct a graphene lattice using the lattice vectors $a_{CC}(\sqrt{3}, 0)$ and $a_{CC}(\sqrt{3},1)/2$, placing atoms at the positions $(0,0)$ and $(0,1)$ within each unit cell. We subsequently set the nearest-neighbor interaction to $t=-3.0$~eV, thus creating a honeycomb lattice. We consider a sample of both large width and large length and fix the potential on each site. At both ends of this sample, we attach a lead whose width is equal to the width of the sample. Electrons are injected through one of these leads, and are collected at the other lead.

In order to probe the effect of different setups, we consider three types of samples in our numerical calculations.
In the first type of sample, the electrons propagate along the $x_1$-direction of our graphene lattice, which means that the sample has so-called zigzag edges. The behavior at this type of edges can be captured within a continuum approximation by requiring the wavefunction of sublattice $A$ to be zero on one side of the sample and the wavefunction of sublattice $B$ to be zero on the other side~\cite{Brey06}. In reciprocal space, the two cones corresponding to the $K$ and $K'$ valleys are well separated in this case, and one can select modes within each of the cones.
In the second type of sample, the electrons also propagate along the $x_1$-direction. However, this time, we attach leads to the sample at both sides, through which electrons can exit. These drain leads effectively reduce reflections at the edges. In the third type of sample, the electrons propagate along the $x_2$-direction of our graphene lattice, which means that the sample has so-called armchair edges. This leads to a more complicated boundary condition, which couples the valleys~\cite{Brey06}. In this case, the cones become degenerate and we cannot select modes belonging to a particular valley.

For all of our samples, the incoming wave is an eigenstate of the lead. For the zigzag sample, we use the mode with the lowest transversal momentum, which therefore has the largest longitudinal momentum.  Because of the zigzag boundary conditions, there is a small difference between the values of the wavefunction on the $A$ and $B$ sublattices. On each sublattice, the wavefunction has the form of a cosine, with its maximum near $x_2=0$. Although this means that the transversal momentum is not really zero and the initial amplitude is not quite constant in our numerical calculations, we expect the consequences of these effects to become less relevant as the sample width increases. In order to make an adequate comparison between the semiclassical result and our numerical results, we average the latter over the two sublattices. Specifically, we replace $|\Psi_i|^2$, the square of the absolute value of the wavefunction on site $i$, by $|\Psi_{i,\text{av}}|^2 = \tfrac{1}{2} |\Psi_i|^2 + \tfrac{1}{6} \sum |\Psi_j|^2$, where we sum over the three neighboring sites $j$.
We observe that, in the absence of a mass term, this averaging procedure does not change the results significantly. Furthermore, the results do not change significantly when we change the vertical position of our array of sites by a small amount.
Finally, since the semiclassical results are computed using an initial amplitude of unity, we normalize the initial numerical wavefunction using the average of $|\Psi_i|^2$ over 30 sites that lie on the $x_1$-axis and are located at the beginning of the sample.

For the second type of sample, where we have drain leads at the sides, we consider the same incoming mode as for the first type of sample. We observe that the results are influenced by the Bloch phase, since the results along the $x_1$-axis exhibit fairly large oscillations that disappear when we plot only every third point. All three lines that can be generated in this way show the same qualitative behavior, although there are some small differences in values. We therefore only consider every third point of the averaged results, and subsequently normalize this subset using points from the beginning of the sample.

For the third type of sample, which has armchair edges, the Bloch phase is clearly visible in the wavefunction of the transversal eigenmodes of the lead. Plotting the three subsets that are obtained by selecting every third site in a plot of $|\Psi_i|^2$, we observe that each of these subsets is well described by the absolute value of a trigonometric function. Giving the trigonometric function that describes one of these subsets a phase shift of $\pm \pi/3$, we obtain the other two subsets. As incoming mode we choose the mode with the largest longitudinal momentum among the modes for which each of these subsets has a single maximum. Since averaging this mode leads to a result that no longer resembles the original mode, we do not average the wavefunction for this sample. We believe that this choice is justifiable since, in the absence of a mass, the two components of the eigenfunction $\chi_0$ only differ by a phase factor, see equation~(\ref{eq:chi0Dirac}). When considering the results along the propagation direction, we observe a small difference between the two sublattices $A$ and $B$, leading to additional oscillations. Splitting the results into two sublattices and normalizing them separately, we obtain results that are roughly identical for a sample width of 2000 $a_{CC}$, but clearly differ in peak height for a sample width of 3000 $a_{CC}$. Finally, going to a neighboring line of sites, which has a different initial wavefunction because of the Bloch phase, does not essentially change the position of the maximum or its height after normalizing to the initial value of the wavefunction.

  \begin{figure}[tb]
  \begin{center}
    \includegraphics{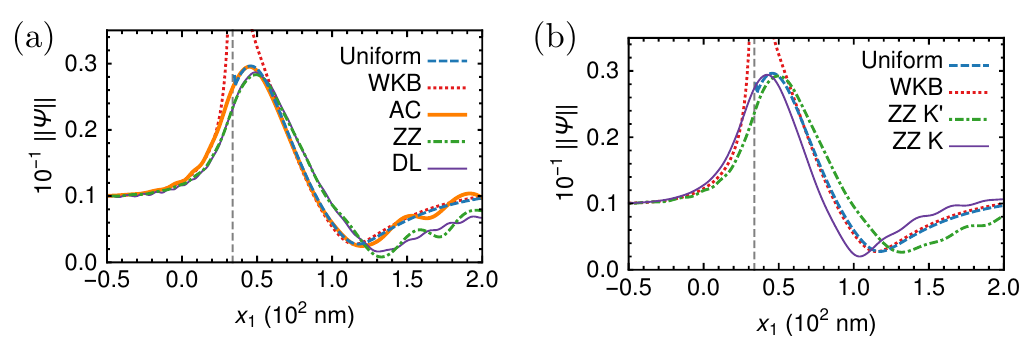}
    \caption{Comparison of the results of tight-binding calculations with the uniform approximation and the WKB approximation along the $x_1$-axis for $E=200$~meV, $U_0=100$~meV and $L=35.5$~nm. The position of the caustic is indicated by a vertical, dashed grey line. (a) Results of tight-binding calculations for a zigzag (ZZ) sample, a sample with drain leads (DL) and for a sample with armchair edges (AC). All simulated samples had a width of $2000~a_{CC}=284$~nm and extended from $-3250 a_{CC}\approx -462$~nm to $1750 a_{CC}\approx 249$~nm along the propagation direction. For the ZZ and DL sample, we selected the incoming mode within the $K'$-valley. (b) Results of tight-binding calculations for a zigzag sample of width $4000~a_{CC}=568$~nm, with the same length as the sample considered in (a). We show the result for both the $K'$-valley and the $K$-valley.}
    \label{fig:compL35_5Kwant}
  \end{center}
  \end{figure}

The results for the different types of samples are shown in figure~\ref{fig:compL35_5Kwant}, in which we compare the wavefunction along the direction of propagation with the uniform approximation and the WKB approximation. We observe that the results of the armchair sample match the uniform approximation very well. The results of the zigzag and drain lead samples compare well to the uniform approximation, but show some important deviations, which we elucidate in the next paragraphs.

We first observe that all three types of samples show additional oscillations at larger values of $x_1$, which are especially visible beyond the minimum. Comparing the zigzag samples in figures~\ref{fig:compL35_5Kwant}(a) and (b), we observe that the amplitude of the oscillations decreases when we increase the sample width. Furthermore, the oscillations are larger for a sample with zigzag edges than for a sample with drain leads on the sides, through which the electrons can exit. We therefore ascribe these additional oscillations to reflections from the sides of the sample. This conclusion is supported by the fact that the oscillations did not change when we increased the length of one of the zigzag samples. We also observe that the peak height increases slightly when we increase the width of the sample from 284~nm to 562~nm. However, the position of the maximum does not change significantly. When we further increased the width of the sample to 710~nm, the peak height only marginally changed, indicating that we do not need to consider samples with a width larger than 562~nm. For the armchair sample, the position of the maximum also seemed to be practically independent of the sample width.

Looking at figure~\ref{fig:compL35_5Kwant}(b), we see that the position of the maximum is different for the valleys $K'$ and $K$. This is not only true for the zigzag sample, but also for the sample with drain leads, which exhibits similar peak positions. This indicates that this effect is probably not caused by the type of boundary conditions at the edges of the sample. Instead, we believe it to be due to so-called trigonal warping~\cite{Ajiki96,Ando98,Katsnelson13}. Within the Dirac approximation, the dispersion relation~(\ref{eq:L0Dirac}) is rotationally invariant. However, the trigonal warping term, the second-order term in the expansion of the tight-binding Hamiltonian near the $K$ and $K'$ points, breaks this rotational symmetry, replacing it by a trigonal symmetry. 
Trigonal warping most strongly affects samples in which the waves propagate along the zigzag direction, modifying the longitudinal momenta in the two valleys in opposite ways. An earlier study~\cite{Reijnders17b}, which was concerned with a different lensing problem, already showed that the trigonal warping term can significantly affect the position of the maximum. This influence was found to be important for low energies as well. Furthermore, it was found to be dependent on the sample orientation, being maximal for zigzag samples and minimal for armchair samples.

We can estimate how strongly our system is affected by trigonal warping by computing the position of the cusp point for the Hamiltonian including trigonal warping. For the zigzag sample, this Hamiltonian is symmetric with respect to $p_y$. Hence, the trajectory with $\phi=0$ still coincides with the $x_1$-axis. Thus, the cusp point remains on the $x_1$-axis and we can find its position by finding the root of the Jacobian $J=\det(X_t, X_\phi)$ on the central trajectory. For the energy and potential that we consider in this section, we find that the cusp point lies at $\tilde{x}_{1,\text{cusp},K}=0.818$ for the $K$-valley and at $\tilde{x}_{1,\text{cusp},K'}=1.095$ for the $K'$-valley. For the Dirac Hamiltonian, the cusp point lies at $\tilde{x}_{1,\text{cusp}}=0.951$. For $L=35.5$~nm, this translates to $x_{1,\text{cusp},K}=29.0$~nm for the $K$-valley, $x_{1,\text{cusp},K'}=38.9$~nm for the $K'$-valley and $x_{1,\text{cusp}}=33.8$~nm for the Dirac Hamiltonian.
Although the actual peaks lie at somewhat larger values of $x_1$, the distances between these peaks are in very good agreement with the distances between the calculated cusp points. This confirms out hypothesis that the observed deviation from the uniform approximation can be ascribed to trigonal warping. 
By the same logic, samples in which the waves propagate along the armchair direction should be less affected by trigonal warping, since this effect does not alter the longitudinal momentum of the mode with zero transversal momentum. This behavior is indeed observed in figure~\ref{fig:compL35_5Kwant}(a). For samples with intermediate orientations, we expect a deviation from the uniform approximation that lies between these two cases, see also Ref.~\cite{Reijnders17b}.

\begin{figure}[tb]
  \begin{center}
    \includegraphics{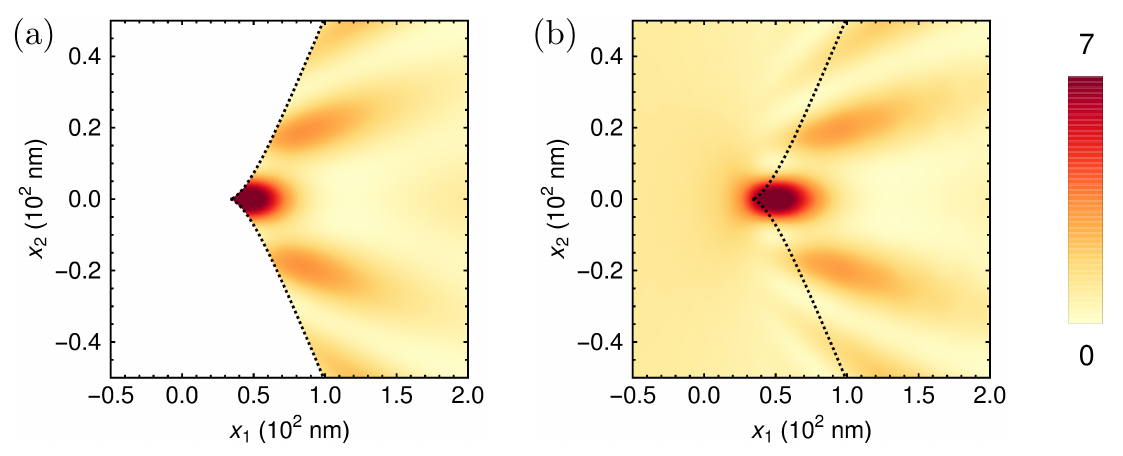}
    \caption{Comparison of the intensity $\lVert \Psi \rVert^2$ for $E=200$~meV, $U_0=100$~meV and $L=35.5$~nm. (a) The intensity $\lVert \Psi \rVert^2$ predicted by the uniform approximation (b) The intensity $\lVert \Psi_{K'} \rVert^2$ for an electron in the $K'$-valley, obtained from a tight-binding calculation for a zigzag sample with a width of $4000~a_{CC} \approx 568$~nm. The length of the sample corresponds to the length of the samples considered in figure~\ref{fig:compL35_5Kwant}.}
    \label{fig:densityL35_5Kwant}
  \end{center}
  \end{figure}

In figure~\ref{fig:densityL35_5Kwant}(a), we show the intensity $\lVert \Psi \rVert^2$ obtained from the uniform approximation, which is the same for both valleys. Remember that we can only compute the uniform approximation in the region where the Lagrangian manifold has three leaves, i.e. in the interference region inside the caustic. Figure~\ref{fig:densityL35_5Kwant}(b) shows the intensity $\lVert \Psi_{K'} \rVert^2$ obtained from tight-binding calculations for a zigzag sample with a width of 568~nm. We observe that the general agreement between the two figures is very good. The tight-binding result for the $K$-valley looks very similar to the result for the $K'$-valley, although the maxima are slightly shifted to the left, as we already observed in figure~\ref{fig:compL35_5Kwant}.

\subsection{Engineering the semiclassical phase}  \label{subsec:numerics-large-h-semiclassical-phase}

The previous example showed that the uniform approximation successfully describes focusing of electrons by a Gaussian potential well. While studying this example, we did not observe any effects that we could specifically connect to the presence of the semiclassical phase. 
We observed that the position of the maximum predicted by the Pearcey approximation is roughly equal to the position predicted by the uniform approximation and observed in the results of tight-binding calculations. This indicates that the semiclassical phase did not play a large role in our example, since the position of the maximum does not depend on the semiclassical phase within the Pearcey approximation. The goal of this subsection is to construct an example in which the semiclassical phase has a substantial influence on the maximum.

In section~\ref{sec:semiclassics}, we showed that the semiclassical phase equals the Berry phase in the absence of a mass term. In particular, we noted that the semiclassical phase equals half the difference between the initial and final angles in momentum space, i.e. $\Phi_{sc,\alpha} = -\tfrac{\alpha}{2} \Delta \phi_p$. Thus, we could say that the semiclassical phase is a local quantity in this case, as it only depends on the point in phase space, instead of on the entire path in momentum space.
The effect of the semiclassical phase is also constrained by the symmetries that we discussed in section~\ref{sec:assumptions}. When the initial amplitude and the potential are symmetric in $x_2$, these symmetries dictate that $\Psi_\alpha(x_1,x_2) = \sigma_x \Psi_\alpha(x_1,-x_2)$. Thus, $\lVert \Psi_\alpha(x_1,x_2) \rVert = \lVert \Psi_\alpha(x_1,-x_2) \rVert$ and the intensity is symmetric in $x_2$. For a symmetric potential, the semiclassical phase therefore cannot cause a transveral shift of the maximum, as this would break the aforementioned symmetry. Furthermore, we have symmetry between the two valleys, since $\lVert \Psi_K (x) \rVert = \lVert \Psi_{K'}(x) \rVert$.

We can break both of these symmetries by considering a nonzero mass, as we discussed in section~\ref{sec:assumptions}. When we consider a mass and a potential that are both symmetric, we obtain two new symmetries, which are however much less restrictive. They state that $\lVert \Psi_{\alpha,m}(x_1,x_2) \rVert = \lVert \Psi_{\alpha,-m}(x_1,-x_2) \rVert$ and that $\lVert \Psi_{\alpha}(x_1,x_2) \rVert = \lVert \Psi_{-\alpha}(x_1,-x_2) \rVert$, meaning that changing valley and reversing the sign of the mass both lead to a mirror reflection in the $x_1$-axis. In particular, these symmetries do not exclude a transversal shift of the maximum: the maximum no longer necessarily lies on the $x_1$-axis.
Looking at the semiclassical phase, we observe that both components $\Phi_{B,\alpha}$ and $\Phi_{A,\alpha}$ are nonzero in the presence of a mass term, see equation~(\ref{eq:phases-B-A}). Furthermore, the Berry phase is no longer given by the angle in momentum space. Because of these two differences, the semiclassical phase now depends on the path in phase space, instead of only on a point. Thus, one may say that it is a nonlocal quantity.

In order to accurately determine the influence of the semiclassical phase, we would like to compare two situations in which only the semiclassical phase of the trajectories differs. 
To be able to do this, we have to construct a specific combination of the potential $U(x)$ and the mass $m(x)$ which does not alter the trajectories, but modulates the semiclassical phase. Since a negative potential bends the trajectories inwards and a mass term bends them outwards, one should be able to obtain straight trajectories by delicately balancing the two.
With such a ``semiclassical phase modulator'', we can spatially decouple the modulation of the semiclassical phase from the focusing itself. We therefore identify two different regions: the semiclassical phase is modulated in the first region and focusing takes place in the second region.
In this case, the semiclassical phase can be written as a sum of two terms: one term coming from the semiclassical phase modulator and one term coming from the focusing. When the electrons are focused by a potential only, i.e. when the mass term vanishes in the focusing region, this second contribution is simply given by the angle in momentum space. However, the total semiclassical phase is still a nonlocal quantity in this case, since the contribution from the modulator is nonlocal in nature.

Let us now consider how we can construct such a semiclassical phase modulator. We demand that the only difference between the asymptotic solution in this region and the asymptotic solution for a region without potential and mass lies in the semiclassical phase.
From the discussion in section~\ref{sec:maslov}, it is clear that this requires that the Lagrangian manifolds of the two problems coincide. In particular, this means that that the classical actions for the two problems should be equal. Since the action equals $\tau$ in eikonal coordinates, it also means that the parametrizations of the two Lagrangian manifolds in eikonal coordinates have to coincide. Looking at Hamilton's equations~(\ref{eq:Ham-Dirac-2}) in eikonal coordinates, we observe that all of the above requirements are satisfied when $C(x)$ coincides for both situations. When we set
\begin{equation} \label{eq:BPh-pot-mass-combi}
  U(x) = E - \sqrt{E^2+m^2(x)} ,
\end{equation}
we observe that
\begin{equation}
  C(x) = \frac{1}{\sqrt{(U(x)-E)^2-m^2(x)}} = \frac{1}{E} ,
\end{equation}
which is the same as the value it assumes in the absence of a potential and a mass. 
Note that the potential~(\ref{eq:BPh-pot-mass-combi}) is negative, as we already anticipated. It therefore bends the trajectories inwards, whereas the mass term bends the trajectories outwards. Together, these two effects lead to trajectories that are straight lines.
Hence, when the potential and mass are related by equation~(\ref{eq:BPh-pot-mass-combi}), we obtain the same Lagrangian manifold as we obtain when both $U(x)$ and $m(x)$ vanish.

Subsequently, let us consider how we can choose the mass $m(x)$ in such a way that our modulator has a large effect on the semiclassical phase and thereby on the focusing.
Since we consider particles that come in from the left, the transversal momentum $p_2$ vanishes in the region where the modulator acts.
The semiclassical phase is therefore proportional to the integral of $p_1 \partial(U+m)/\partial x_2$ along the trajectories, see equations~(\ref{eq:BerryPhaseEikonal}) and~(\ref{eq:L1Dirac}).
In particular, we observe that a mass that is even in $x_2$ leads to a semiclassical phase that is odd in $x_2$. Such a mass is therefore able to create a large difference in semiclassical phase between trajectories with positive and negative $\phi$.
We remark that, in order to remain within the regime of applicability of the semiclassical approximation, we should make sure that $m(x)$ increases and decreases smoothly in the coordinate $x_1$.
Taking these requirements into account, we arrive at the following choice for $m(x)$, in dimensionless units:
\begin{equation} \label{eq:BPh-mass}
  \tilde{m}(\tilde{x}) = \frac{3}{8} \left( 1 - \frac{1}{\cosh\left[\tfrac{5}{4} \tilde{x}_2\right]} \right)
    \left( \tanh\left[ \tfrac{9}{5} ( \tilde{x}_1 - \tilde{x}_{1,\text{b}1} ) \right] - 
      \tanh\left[ \tfrac{9}{5} (\tilde{x}_1 - \tilde{x}_{1,\text{b}2} ) \right] \right)
\end{equation}
The derivative of this expression with respect to $\tilde{x}_2$ is of order unity, which is large enough to create a sizeable effect, yet small enough to remain within the semiclassical regime. 

\begin{figure}[tb]
  \begin{center}
    \includegraphics[width=4.8cm]{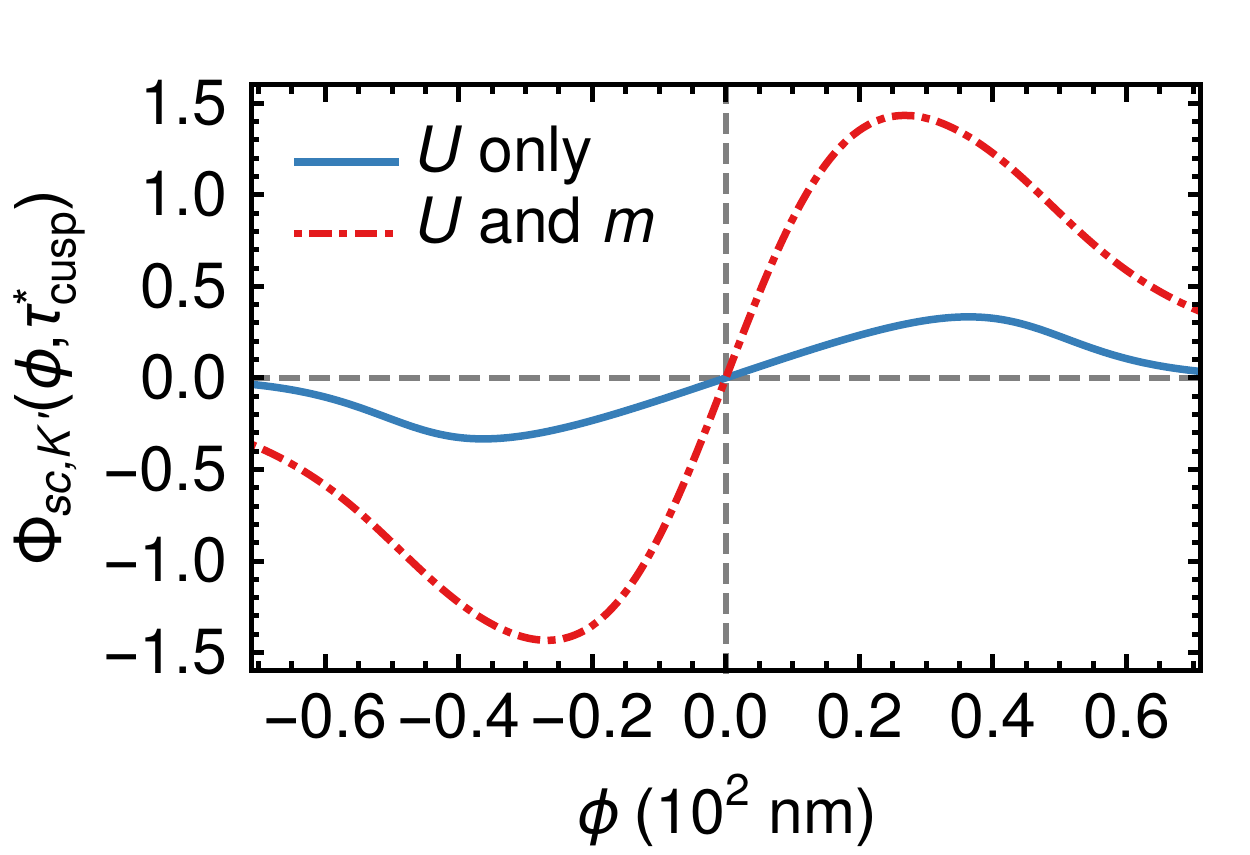}
    \caption{The semiclassical phase $\Phi_{sc,K'}$ for an electron in the $K'$-valley as a function of $\phi$ at time $\tau^*_\text{cusp}$, i.e. along the wavefront whose projection passes through $x_\text{cusp}$. The blue (solid) line shows the result for focusing by the Gaussian potential well~(\ref{eq:pot-gaussian-well}). The red (dashed-dotted) line shows the semiclassical phase that is acquired when this potential is preceded by our semiclassical phase modulator. In this region, the mass $m(x)$ is given by equation~(\ref{eq:BPh-mass}) and the potential by equation~(\ref{eq:BPh-pot-mass-combi}). This combination does not alter the classical trajectories, but significantly increases the semiclassical phase.}
    \label{fig:BPhcomp}
  \end{center}
  \end{figure}

Finally, we combine our semiclassical phase modulator with the Gaussian potential~(\ref{eq:pot-gaussian-well}). As in sections~\ref{subsec:numerics-large-h} and~\ref{subsec:numerics-large-h-comparison}, we set $E=200$~meV, $U_0=100$~meV and $L=35.5$~nm. 
To make sure that the modification of the semiclassical phase happens well before the Gaussian potential focusses the trajectories, we choose $\tilde{x}_{1,\text{b}1}=x_{1,\text{b}1}/L=-10$ and $\tilde{x}_{1,\text{b}2}=-5$. To probe the effect of our modulator on the semiclassical phase, we plot this phase as a function of $\phi$ at time $\tau^*_\text{cusp}$.
This means that we plot the semiclassical phase along the wavefront whose projection passes through $x_\text{cusp}$.
Figure~\ref{fig:BPhcomp} shows that our modulator leads to a semiclassical phase that is indeed much larger than in the example discussed in the previous subsection.

  \begin{figure}[tb]
  \begin{center}
    \includegraphics{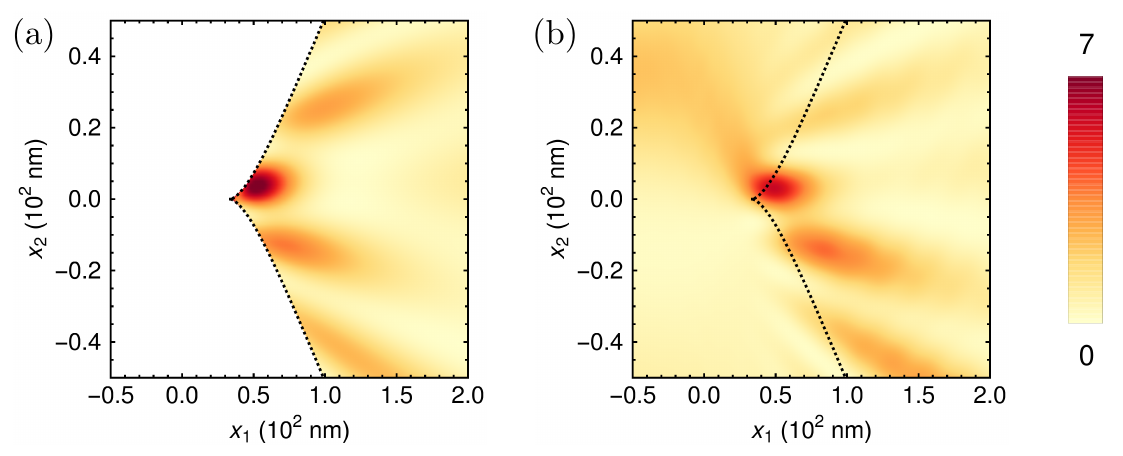}
    \caption{Comparison of the intensity $\lVert \Psi_{K'} \rVert^2$ for electrons in the $K'$-valley with $E=200$~meV. The incoming electrons are focused by the Gaussian potential~(\ref{eq:pot-gaussian-well}) with $U_0=100$~meV and $L=35.5$~nm. In front of this potential, there is a region in which the mass is given by equation~(\ref{eq:BPh-mass}) with $\tilde{x}_{1,\text{b}1}=-10$ and $\tilde{x}_{1,\text{b}2}=-5$ and the potential is given by equation~(\ref{eq:BPh-pot-mass-combi}). (a) Result of the uniform approximation (b) Result of a tight-binding calculation for a zigzag sample with a width of $4000~a_{CC} \approx 568$~nm.}
    \label{fig:densityMassL35_5Kwant}
  \end{center}
  \end{figure}

We can subsequently compute the intensity $\lVert \Psi_{\alpha} \rVert^2$ for an electron incident on our setup with the uniform approximation. For an electron in the $K'$-valley, we obtain the result shown in figure~\ref{fig:densityMassL35_5Kwant}(a). When we compare it to the result shown in figure~\ref{fig:densityL35_5Kwant}(a), we observe that the intensity maximum is no longer on the $x_1$-axis. Instead, it occurs at positive $x_2$. Since the wavefunctions in the two valleys are related by $\Psi_K (x_1,x_2) = \Psi_{K'}(x_1,-x_2)$, the maximum for an electron in the $K$-valley occurs at negative $x_2$.
Taking a closer look at figure~\ref{fig:densityMassL35_5Kwant}(a), we observe that the uniform approximation diverges near the cusp point. Although a small divergence is also visible in the results of the previous subsection, where we only consider a Gaussian potential well, the divergence observed here is much stronger. Since both situations give rise to the same Lagrangian manifold, the quantities in the uniform approximation that are derived from it coincide as well. Indeed, we observe that these quantities are well-behaved as we approach the cusp point.
However, in the setup with the semiclassical phase modulator, the amplitudes $D_1$ and $D_2$ strongly diverge when we approach the cusp point. More reseach is needed to decide whether this is a fundamental limitation of the method, or rather the result of errors that arise from the various interpolation procedures that are used to obtain the final result.

We would like to compare the uniform approximation to the result of a tight-binding calculation. Based on the results discussed in the previous subsection, we chose to perform these calculations for a zigzag sample. Although an armchair sample would probably be less affected by trigonal warping, one cannot separate the two valleys in such a sample, making it impossible to study the deflection of the focus for the different valleys. Furthermore, the wavefunction for a zigzag sample does not show the additional oscillations that are present in the wavefunction for a sample with drain leads. Although this comes at the price of additional reflections by the sides of the sample, their effect can be reduced by considering a sufficiently wide sample.
In figure~\ref{fig:densityMassL35_5Kwant}(b), we show the intensity $\lVert \Psi_{K'} \rVert^2$ obtained from a tight-binding calculation of a zigzag sample with a width of $568$~nm.
This figures clearly confirms the lateral shift of the maximum. However, the shape of the maximum is slightly different than in the uniform approximation, and the maximal value somewhat lower.
Furthermore, the tight-binding result for the $K$-valley is not exactly a mirror version of the result for the $K'$-valley. The latter observation, combined with the observations from the previous subsection, indicate that these differences probably come from trigonal warping.

\begin{figure}[tb]  
  \begin{center}
    \includegraphics{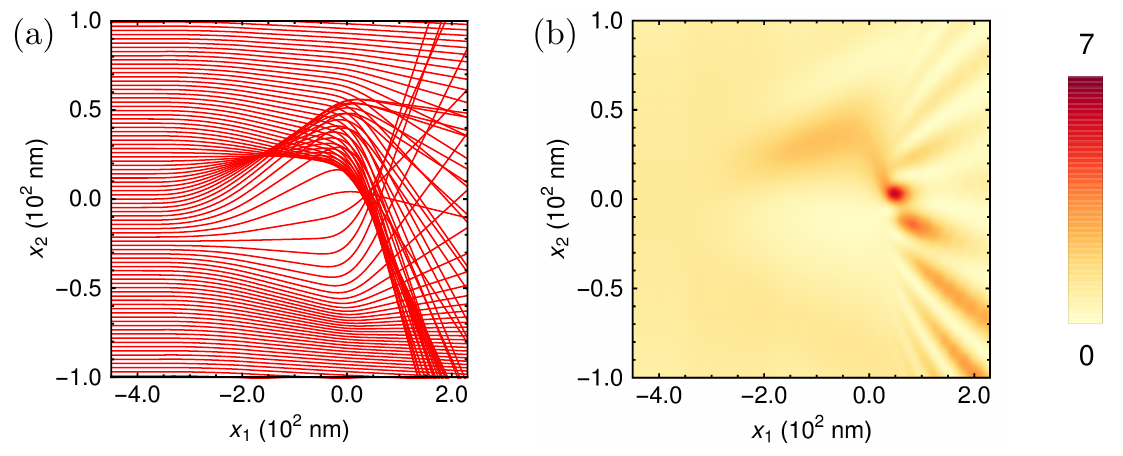}
    \caption{(a) Trajectories for an electron in the $K'$ valley, computed using the modified equations of motion~(\ref{eq:EOM-L1}). (b) Result of a tight-binding calculation for a zigzag sample with a width of $4000~a_{CC} \approx 568$~nm. To produce these figures, we used the same parameters as in figure~\ref{fig:densityMassL35_5Kwant}.
    }
    \label{fig:densityMassL35_5TrajKwant}
  \end{center}
  \end{figure}

Instead of only considering the result of the tight-binding calculation in a region near the cusp, we can also plot it in a much larger region. The intensity, shown in figure~\ref{fig:densityMassL35_5TrajKwant}(b), shows that the displacement of the focus is not the only effect of the semiclassical phase modulator.
We observe that some focusing already takes place in the region before the Gaussian potential well. In particular, we see that, for electrons in the $K'$-valley, the intensity $\lVert \Psi \rVert^2$ is larger above the $x_1$-axis than below it. This is unexpected from the point of view of the WKB approximation~(\ref{eq:PsiRegular-graphene}): the Jacobian~(\ref{eq:JacobianXEikonal}) is the same for all trajectories and there is no interference between different trajectories since the Lagrangian manifold has a single leaf. Hence, the WKB approximation cannot entirely capture the behavior of the observed intensity in this region. We remark that this is acceptible from a semiclassical point of view, since the dimensionless semiclassical parameter $h = 0.09$ is rather large in our problem.

Because the dimensionless semiclassical parameter is fairly large, we could try to explain the observed effect by modifying the equations of motion. Instead of computing the equations of motions using the principal symbol $L_0$, we can also incorporate the subprincipal symbol $L_1^W$ into the equations of motion.
This leads to different equations of motion, which were constructed in Refs.~\cite{Littlejohn91,Xiao10}, see also Refs.~\cite{Kuratsuji85,Kuratsuji87,Karasev90}. 
With the notations that we have used throughout this paper, these modified equations of motion are given by
\begin{equation}  \label{eq:EOM-L1}
\begin{aligned}
  \frac{\text{d} x_j}{\text{d} t} &= \phantom{-}\frac{\partial L_0}{\partial p_j} + h \frac{\partial L_{1,A}^W}{\partial p_j} + h \sum_{k} (\Omega_{pp})_{jk} \frac{\partial L_0}{\partial x_k} - h \sum_{k} (\Omega_{px})_{jk} \frac{\partial L_0}{\partial p_k} , \\
  \frac{\text{d} p_j}{\text{d} t} &= -\frac{\partial L_0}{\partial x_j} - h \frac{\partial L_{1,A}^W}{\partial x_j} - h \sum_{k} (\Omega_{xp})_{jk} \frac{\partial L_0}{\partial x_k} + h \sum_{k} (\Omega_{xx})_{jk} \frac{\partial L_0}{\partial p_k} .
\end{aligned}
\end{equation}
The quantities $\Omega_{ab}$ in this equation are the Berry curvatures~\cite{Berry84}. In our case, these curvatures are two-dimensional matrices, whose elements are purely real. 
For the components of $\Omega_{xp}$, we have,
\begin{equation}  \label{eq:Berry-curvatures}
  (\Omega_{xp})_{jk} = i \bigg( \frac{\partial \chi_0^\dagger}{\partial x_j} \frac{\partial \chi_0}{\partial p_k} 
    - \frac{\partial \chi_0^\dagger}{\partial p_k} \frac{\partial \chi_0}{\partial x_j} \bigg)  .
\end{equation}  
The components of the other Berry curvatures are given by analogous equations.
Comparing these equations of motion with the derivations in section~\ref{sec:adiabatic}, we observe that only $L_{1,A}^W$ enters the equations of motion directly.
The Berry part $L_{1,B}^W$ does not enter the equations of motion directly. This makes sense, since this quantity is not gauge-invariant, see equation~(\ref{eq:L1-gauge-transformation-effect}). Because we do not want the equations of motion to depend on our choice of gauge, $L_{1,B}^W$ cannot directly enter the equations of motion.
Instead, the Berry contribution enters the equations of motion through the Berry curvatures. A simple calculation shows that these are gauge independent, which shows that the modified equations of motion~(\ref{eq:EOM-L1}) are invariant under the gauge transformation~(\ref{eq:chi0-gauge-transformation}).
In the literature, these modified equations of motion have been used to explain the experimentally observed bending of trajectories in graphene on hexagonal boron nitride~\cite{Gorbachev14}.
Furthermore, effects on the transport properties~\cite{Spivak16} and electromagnetic properties~\cite{Pellegrino15} were predicted for Weyl semimetals.

When we numerically integrate the modified equations of motion~(\ref{eq:EOM-L1}) for an electron in the $K'$-valley, we obtain the trajectories shown in figure~\ref{fig:densityMassL35_5TrajKwant}(a).
We observe that these modified trajectories capture the additional focusing very well.
Before the Gaussian potential well, the density of trajectories is increased above the $x_1$-axis and reduced below the $x_1$-axis, in excellent agreement with the observed intensity $\lVert \Psi_{K'} \rVert^2$, shown in figure~\ref{fig:densityMassL35_5TrajKwant}(b).
After the Gaussian potential well, the density of trajectories is larger below the $x_1$-axis, which is also in agreement with the observed intensity. 
Our computations show that these effects mainly come from the terms with $L_{1,A}^W$ and that the Berry curvature only plays a minor role in our problem.

For our scattering setup, the modified trajectories for electrons in the $K$-valley can be obtained from those in the $K'$-valley by a reflection in the $x_1$-axis. 
This makes sense when we remember that the wavefunctions in the two valleys are related by equation~(\ref{eq:symm-between-valleys}). Thus, the two valleys deflect the electrons in opposite directions. 
In the modified equations of motion~(\ref{eq:EOM-L1}), changing the valley under consideration corresponds to changing the sign of $\alpha$. Since both $L_{1,A}^W$ and the Berry curvatures are proportional to $\alpha$, this implies that the terms that are proportional to $h$ change sign. This subsequently leads to a reflection in the $x_1$-axis.
The modified trajectories for electrons in the $K$-valley also show very good agreement with the observed intensity $\lVert \Psi_K \rVert^2$.

\section{Conclusion and outlook}  \label{sec:conclusion}

In this paper, we have given a detailed description of electronic optics in graphene. In particular, we have developed a complete semiclassical theory of focusing by above-barrier scattering for charge carriers in graphene.
We have constructed the wavefunction in regular points and in the vicinity of singular points in the presence of arbitrary smooth potentials and masses.
In particular, we have shown that the semiclassical phase can have a large influence on the position of the focus and on its intensity.
We have compared our semiclassical results with the results of tight-binding calculations and have found very good agreement.

Although the specific combination~(\ref{eq:BPh-pot-mass-combi}) of potential and mass that we considered in section~\ref{subsec:numerics-large-h-semiclassical-phase} may seem somewhat artificial, it provides an important proof of concept.
It shows that the presence of a coordinate-dependent mass can have a large influence on the focus, even in a very symmetric and idealized setup. In experiments, a coordinate-dependent mass naturally arises when one considers graphene on a substrate such as hexagonal boron nitride~\cite{Sachs11,Bokdam14,Slotman15,Wallbank13,Diez14,Chizhova14,Yankowitz14}. In this case, the potential and mass are likely to be much less symmetric than in our example.
For graphene on hexagonal boron nitride one can, nevertheless, model the mass using a combination of sines~\cite{Sachs11}. 
The shape of this function bears quite some resemblance to the shape of the function considered in section~\ref{subsec:numerics-large-h-semiclassical-phase}.
Furthermore, the scales in this model are similar to the ones that we used, since the amplitude of the mass is approximately 30~meV~\cite{Sachs11,Bokdam14} and its periodicity about 15~nm~\cite{Slotman15}, depending on the relative orientation of the graphene and the substrate.
We therefore consider it likely that a large semiclassical phase can also arise in experiments by a similar mechanism. In particular, our results may be important for precise image reconstruction in a Dirac fermion microscope~\cite{Boggild17}.

However, a realistic description of graphene on hexagonal boron nitride also requires the inclusion of a vector potential~\cite{Wallbank13,Slotman15}. This gauge field arises due to the modification of the hopping parameters by the induced strain~\cite{Kane97,Sasaki05,Morozov06,Vozmediano10,Katsnelson13}.
An important next step would therefore be to include this vector potential into our semiclassical description. We believe that this would not be very difficult, as it does not require any conceptually new steps, only the modification of various formulas. Furthermore, some semiclassical results for this case were already obtained in Refs.~\cite{Carmier08,Stegmann16}.
In particular, it was shown~\cite{Carmier08} that including the vector potential into the description leads to the modification of the semiclassical phase. 
This modification is very important in problems in which the semiclassical quantization condition enters~\cite{Shytov08,Young09,Carmier08,Fuchs10}, but will likely also manifest itself in scattering problems.

We have also found that trigonal warping affects the position of the focus, even at very low energies of around 200~meV. For electrons propagating along the zigzag direction, the position of the focus obtained from tight-binding calculations lies slightly before or after the position predicted for the Dirac Hamiltonian, depending on the valley. For electrons propagating along the armchair direction, the numerically observed position coincides with the position predicted for Dirac electrons. 
Based on a similar result that was obtained for focusing in graphene Veselago lenses~\cite{Reijnders17b}, we expect that the position of the focus will deviate from the Dirac prediction for propagation along all orientations except for the armchair orientation. The size of this deviation will be maximal for the zigzag orientation and will smoothly decrease as we rotate towards the armchair direction.
This result shows that one also has to take trigonal warping into account when one wants to perform precise image reconstruction for the Dirac fermion microscope~\cite{Boggild17}. Since the higher-order corrections to the Dirac Hamiltonian are typically larger for other Dirac materials~\cite{Qi11,Moore10,Hasan10,Hasan11,Bansil16}, we expect these effects to be even stronger in such materials.

Besides obtaining specific results for graphene, we also gave an extensive and careful review of the semiclassical methods that we used.
We can identify three key steps in our construction of the semiclassical approximation. The first step consists of the reduction of the matrix equation to an effective scalar equation for a specific scalar eigenmode. 
In this reduction, we mainly work with symbols rather than with the actual operators. Since these symbols are functions on phase space, they are much easier to manipulate than the operators themselves.
The second step is the construction of an asymptotic solution in both regular and singular points. 
In this step, we first lift our problem from the configuration space to the phase space, meaning that we solve Hamilton's equations rather than the Hamilton-Jacobi equation. Subsequently, we make extensive use of the fact that we can introduce eikonal coordinates on our Lagrangian manifold. 
This allows us to perform various simplifications and to make use of a new representation of the canonical operator in the vicinity of singular points. The third and final step is to simplify this integral representation using the uniform approximation. 
For each of these steps, we discussed both its general principle as well as its specific application to graphene. We therefore believe that our paper will not only be useful for graphene researchers, but also for those that would like to learn more about semiclassical techniques.

Since our discussion of the semiclassical techniques was partially general, 
many of the formulas that we discussed in this paper can be directly applied to other problems in which one can introduce eikonal coordinates.
An example of such a problem is focusing in bilayer graphene~\cite{Katsnelson13}, a material for which the electron energy is proportional to $|p|^2$. Since the effective Hamiltonian $L_0$ for this system has the form $F(x,|p|)$, we can also introduce eikonal coordinates in this problem~\cite{Dobrokhotov15}.
To this end, we should construct a function $C(x)$, cf. equation~(\ref{eq:L0Eikonal}), such that the solutions of the Hamiltonian system for $\mathcal{L}_0 = C(x)|p| = 1$ coincide with those of the Hamiltonian system for $L_0 = E$, except for a reparametrization of time. This reparametrization is given by a function $R$, cf. equation~(\ref{eq:timereparametrization}), which can be expressed in terms of $F$ and $C$, see Ref.~\cite{Dobrokhotov15}. 
Once we have determined the functions $C$ and $R$, we can immediately express the wavefunction in the vicinity of singular points using equation~(\ref{eq:PsiSingular}). The other two functions in this expression, the eigenvector $\chi_0$ and the semiclassical phase $\Phi_{sc}$, which is derived from $L_1$, can be determined by performing the adiabatic reduction of the matrix Hamiltonian to an effective scalar Hamiltonian. Finally, we can apply the uniform approximation to the resulting integral expression without any essential modification.
Therefore, the formulas discussed in this paper can also be used to obtain a solution for the scattering problem for bilayer graphene.

We also want to sketch three other possible directions for future research.
A first suggestion concerns the construction of an asymptotic solution in the presence of trigonal warping, the second-order term in the expansion of graphene's tight-binding Hamiltonian.
In section~\ref{subsec:numerics-large-h-comparison}, we concluded that this term is probably responsible for the differences between our semiclassical results and the results of tight-binding calculations. 
In the presence of trigonal warping, the classical Hamiltonian depends on both $|p|$ and $\phi_p$, instead of only on $|p|$. 
Therefore, we can no longer introduce eikonal coordinates into our problem, which means that we can no longer use the new representation that we discussed in section~\ref{subsec:maslovsingular} to construct an asymptotic solution.
We could still construct a semiclassical approximation using the conventional representation, which we briefly discussed in section~\ref{subsec:maslovapproaches}, together with its disadvantages. However, there seems to be a more straightforward way, which relies on two observations. First, the velocity vector $X_t$ remains nonvanishing when we include the trigonal warping term.
Second, in the presence of trigonal warping we can still use the coordinates $t$ and $\phi$ to parametrize our Lagrangian manifold, even though $\langle X_t, X_\phi \rangle$ no longer vanishes.
Together, these observations imply that we can construct an asymptotic solution using the new representation of the canonical operator discussed in Ref.~\cite{Dobrokhotov17}, which is an extension of the new representation from Refs.~\cite{Dobrokhotov14,Dobrokhotov14b} that we discussed in section~\ref{sec:maslov}. We believe that graphene with trigonal warping would provide an interesting case-study for the application of this extended new representation.
Comparing this new asymptotic solution to the results of tight-binding calculations, we would gain additional insight into the importance of trigonal warping in low-energy scattering in graphene.

A second future direction concerns the construction of a semiclassical approximation based on the trajectories that one obtains from the modified equations of motion~(\ref{eq:EOM-L1}). In this paper, we constructed a semiclassical approximation based on the trajectories that we obtained from the classical Hamiltonian $L_0$, see equation~(\ref{eq:Hamilton}).
However, we saw in section~\ref{subsec:numerics-large-h-semiclassical-phase} that this approximation is not fully able to explain the results of tight-binding calculations when we consider our semiclassical phase modulator. In particular, it could not explain the intensity modulations that occur before the Gaussian potential well. On the other hand, we could qualitatively explain the intensity observed in tight-binding calculations using the trajectories that we obtained from the modified equations of motion~(\ref{eq:EOM-L1}). Therefore, we believe that a semiclassical approximation based on these trajectories would be able to accurately describe the observed intensity quantitatively. A starting point for the construction of such a solution can be found in Ref.~\cite{Littlejohn91}.

A third and final possible research direction concerns the continuation of the uniform approximation across the boundary of a singular chart. In section~\ref{subsec:numerics-large-h}, we obtained a continuous result by making an ad hoc replacement: we replaced the absolute value of the Jacobian by minus its value. However, we did not give a rigorous proof of this continuation. In other words, we did not establish that this continuation gives rise to the correct Maslov index. A mathematical proof of this statement would have to take the additional extrema of the action into account and the way in which they affect the second derivative $\tau_{\phi\phi}$.

In conclusion, our paper shows that the semiclassical approximation is a very useful tool to describe electronic optics in graphene. 
Specifically, it allows us to construct the wavefunction in regular points as well as in the vicinity of singular points.
This construction especially highlights the influence of the semiclassical phase on focusing.

\section*{Acknowledgments}

We are grateful to Timur Tudorovskiy, Andrey Zabolotskii, Vladimir {Na\-zai\-kin\-skii} and Misha Titov for helpful discussions.

D.~S.~Minenkov and S.~Yu. Dobrokhotov appreciate financial support by
State Project AAAA-A17-117021310377-1 and Grants RFBR No. 14-01-00521, 16-31-00442.
K.~J.~A. Reijnders and M.~I.~Katsnelson acknowledge financial support from the European Research Council (ERC) Advanced Grant No. 338957
FEMTO/NANO and from the Netherlands Organisation for Scientific Research (NWO) via the Spinoza Prize.

\section*{Competing interests}

The authors declare no competing interests.

\appendix

\section{Properties of the subprincipal symbol $L_1$} \label{app:subprincipal-symbol}

In this appendix, we take a closer look at the subprincipal symbol $L_1$. Assuming that the matrix Hamiltonian $\hat{H}$ is self-adjoint and therefore satisfies equations~(\ref{eq:self-adjoint-relation-a1}) and~(\ref{eq:self-adjoint-Weyl-a1}), we show by explicit calculations that $L_1$ satisfies equations~(\ref{eq:ImL1}) and~(\ref{eq:L1-Weyl-relation-L1}).

We start by constructing the imaginary part of $L_1$. Using equation~(\ref{eq:L1}) and the identity $ \text{Im} \, L_1 = \tfrac{1}{2 i} (L_1 - L_1^*)$, we first obtain
\begin{multline} \label{eq:Im-L1}
  2 \text{Im} \, L_1 = -i \chi_0^\dagger (H_1 - H_1^\dagger) \chi_0 
    - \chi_0^\dagger \left\langle \frac{\partial H_0}{\partial p} , \frac{\partial \chi_0}{\partial x} \right\rangle
    + \chi_0^\dagger \left\langle \frac{\partial L_0}{\partial x} , \frac{\partial \chi_0}{\partial p} \right\rangle \\
    - \left\langle \frac{\partial \chi_0^\dagger}{\partial x} , \frac{\partial H_0}{\partial p} \right\rangle \chi_0
    + \left\langle \frac{\partial \chi_0^\dagger}{\partial p} , \frac{\partial L_0}{\partial x} \right\rangle \chi_0 .
\end{multline}
Subsequently, we consider equation~(\ref{eq:L0}) and apply the operator $\langle \partial/\partial x , \partial/\partial p \rangle$ to both of its sides. Multiplying the result by $\chi_0^\dagger$, using equation~(\ref{eq:L0}) and rearranging the six remaining terms, we obtain an expression for $\langle \partial/\partial x , \partial/\partial p \rangle L_0$. Adding this expression to equation~(\ref{eq:Im-L1}), we arrive at
\begin{equation}
\begin{aligned} \label{eq:Im-L1-plus-L0-deriv}
  2 \text{Im} \, L_1 + \left\langle \frac{\partial}{\partial x} , \frac{\partial}{\partial p} \right\rangle L_0 &= 
    -i \chi_0^\dagger (H_1 - H_1^\dagger) \chi_0 + \chi_0^\dagger \left( \left\langle \frac{\partial}{\partial x} , \frac{\partial}{\partial p} \right\rangle H_0 \right) \chi_0 \\
    & \hspace*{1.0cm} + \chi_0^\dagger \left\langle \frac{\partial H_0}{\partial x} , \frac{\partial \chi_0}{\partial p} \right\rangle
    - \chi_0^\dagger \left\langle \frac{\partial L_0}{\partial p} , \frac{\partial \chi_0}{\partial x} \right\rangle \\
    & \hspace*{1.0cm} - \left\langle \frac{\partial \chi_0^\dagger}{\partial x} , \frac{\partial H_0}{\partial p} \right\rangle \chi_0
    + \left\langle \frac{\partial \chi_0^\dagger}{\partial p} , \frac{\partial L_0}{\partial x} \right\rangle \chi_0 \\
\end{aligned}
\end{equation}
The first two terms on the right-hand side of this equation vanish because $\hat{H}$ is self-adjoint and therefore satisfies equation~(\ref{eq:self-adjoint-relation-a1}). Before we show that the remaining four terms vanish as well, and that $L_1$ therefore satisfies equation~(\ref{eq:ImL1}), we turn our attention to the real part of $L_1$.

We construct the real part of $L_1$ using equation~(\ref{eq:L1}) and the relation $ \text{Re} \, L_1 = \tfrac{1}{2} (L_1 + L_1^*)$. When we subsequently subtract $L_1^W$, we arrive at
\begin{multline} \label{eq:Re-L1-plus-L1-Weyl}
  2i ( \text{Re} \, L_1 - L_1^W) = 2 i \text{Re} ( \chi_0^\dagger H_1 \chi_0 ) 
    - \left\langle \frac{\partial \chi_0^\dagger}{\partial x} , \frac{\partial H_0}{\partial p} \right\rangle \chi_0 
    + \left\langle \frac{\partial \chi_0^\dagger}{\partial p} , \frac{\partial L_0}{\partial x} \right\rangle \chi_0 \\
    - 2i \chi_0^\dagger H_1^W \chi_0 
    + \chi_0^\dagger \left\langle \frac{\partial H_0}{\partial x} , \frac{\partial \chi_0}{\partial p} \right\rangle
    - \chi_0^\dagger \left\langle \frac{\partial L_0}{\partial p} , \frac{\partial \chi_0}{\partial x} \right\rangle .
\end{multline}
Since the matrix Hamiltonian $\hat{H}$ is self-adjoint, the subprincipal symbol $H_1$ satisfies equation~(\ref{eq:self-adjoint-Weyl-a1}) and we have
\begin{equation}
  \text{Re} ( \chi_0^\dagger H_1 \chi_0 ) = \frac{1}{2} \big(\chi_0^\dagger H_1 \chi_0 + \chi_0^\dagger H_1^\dagger \chi_0\big) = \chi_0^\dagger H_1^W \chi_0 .
\end{equation}
Therefore, the first and the fourth term on the right-hand side of equation~(\ref{eq:Re-L1-plus-L1-Weyl}) cancel and we are left with four terms. Comparing these terms with the four last terms of equation~(\ref{eq:Im-L1-plus-L0-deriv}), we see that they coincide. 

We therefore define
\begin{multline}  \label{eq:diff-four-terms}
  \Upsilon = - \left\langle \frac{\partial \chi_0^\dagger}{\partial x} , \frac{\partial H_0}{\partial p} \right\rangle \chi_0 
    + \left\langle \frac{\partial \chi_0^\dagger}{\partial p} , \frac{\partial L_0}{\partial x} \right\rangle \chi_0 
    + \chi_0^\dagger \left\langle \frac{\partial H_0}{\partial x} , \frac{\partial \chi_0}{\partial p} \right\rangle \\
    - \chi_0^\dagger \left\langle \frac{\partial L_0}{\partial p} , \frac{\partial \chi_0}{\partial x} \right\rangle .
\end{multline}
In the remainder of this appendix, we show that $\Upsilon$ vanishes, which, by our previous considerations, implies that $L_1$ satisfies both equations~(\ref{eq:ImL1}) and~(\ref{eq:L1-Weyl-relation-L1}). Taking the derivative with respect to $p$ on both sides of equation~(\ref{eq:L0}), we obtain an expression for $(\partial H_0/\partial p) \chi_0$. Inserting it into equation~(\ref{eq:diff-four-terms}), we have
\begin{multline} \label{eq:diff-four-terms-v2}
  \Upsilon = \left\langle \frac{\partial \chi_0^\dagger}{\partial x} , H_0 \frac{\partial \chi_0}{\partial p} \right\rangle 
    - \left\langle \frac{\partial \chi_0^\dagger}{\partial x} , \frac{\partial L_0}{\partial p} \right\rangle \chi_0
    - L_0 \left\langle \frac{\partial \chi_0^\dagger}{\partial x} , \frac{\partial \chi_0}{\partial p} \right\rangle \\
    + \left\langle \frac{\partial \chi_0^\dagger}{\partial p} , \frac{\partial L_0}{\partial x} \right\rangle \chi_0 
    + \chi_0^\dagger \left\langle \frac{\partial H_0}{\partial x} , \frac{\partial \chi_0}{\partial p} \right\rangle 
    - \chi_0^\dagger \left\langle \frac{\partial L_0}{\partial p} , \frac{\partial \chi_0}{\partial x} \right\rangle .
\end{multline}
Taking the derivative with respect to $p$ on both sides of the equality $\chi_0^\dagger \chi_0 = 1$, we see that
\begin{equation} \label{eq:normalization-derivative}
  \frac{\partial \chi_0^\dagger}{\partial x} \chi_0 + \chi_0^\dagger \frac{\partial \chi_0}{\partial x} = 0 .
\end{equation}
Taking the inner product of this result with $\partial L_0/\partial p$, we see that the second and sixth terms in equation~(\ref{eq:diff-four-terms-v2}) cancel. Naturally, we can replace the derivatives with respect to $x$ in equation~(\ref{eq:normalization-derivative}) by derivatives with respect to $p$. We can subsequently use this expression to shift the derivative in the fourth term of equation~(\ref{eq:diff-four-terms-v2}) from $\chi_0^\dagger$ to $\chi_0$, at the expense of a minus sign. Rearranging the remaining terms in equation~(\ref{eq:diff-four-terms-v2}), we obtain
\begin{equation}
  \Upsilon = \left\langle \frac{\partial}{\partial x} \left( \chi_0^\dagger H_0^\dagger - L_0^\dagger \chi_0^\dagger \right) , \frac{\partial \chi_0}{\partial p} \right\rangle = 0, 
\end{equation}
where the last equality holds by virtue of equation~(\ref{eq:L0}). We have therefore explicitly shown that the subprincipal symbol $L_1$ satisfies equations~(\ref{eq:ImL1}) and~(\ref{eq:L1-Weyl-relation-L1}).

\section{Introducing eikonal coordinates using operator decomposition} \label{app:operatorReduction}

In section~\ref{subsec:eikonal}, we introduced eikonal coordinates on the Lagrangian manifold by a reparametrization of time. We subsequently discussed how this affects the construction of the asymptotic solution in section~\ref{subsec:maslovapproaches}, in particular in equation~(\ref{eq:precanonicalTimeChange}). In the latter section, we also briefly discussed an alternative method to introduce eikonal coordinates. In this method, we use an operator decomposition to introduce a new Hamiltonian, which automatically leads to eikonal coordinates on the Lagrangian manifold. In this appendix, we take a closer look at this alternative method and show that it gives the same results as the method discussed in the main text.
Meanwhile, we also discuss the generalization of several related notions to wider classes of Hamiltonians.

Let us first consider how we can realize the transition from an effective Hamiltonian $L_0$ to an effective Hamiltonian $\mathcal{L}_0 = C(x)|p|$ within a more general setting. This transition can be performed~\cite{Dobrokhotov15} when the original Hamiltonian $L_0$ only depends on the length of the momentum vector and not on its direction, i.e. $L_0 = L_0(x,|p|)$. When the derivative $\partial L_0/\partial |p|$ is nonvanishing, one can solve the equation $L_0(x,|p|)=E$ for $|p|$. Writing the solution as $|p|=1/C(x)$, we obtain $C(x)|p|=1$. We can subsequently consider $\mathcal{L}_0=C(x)|p|$ as our new Hamiltonian and compute its trajectories for the effective energy 1. Note that the function $C(x)$ depends on the original energy $E$.
The computations in section~\ref{subsec:eikonal} provide a specific example of this general procedure, discussed in more detail in Ref.~\cite{Dobrokhotov15}.

With these definitions, we can write down the (seemingly trivial) equality
\begin{equation} \label{eq:symbol-relation-Hams}
  L_0(x,p) - E = B(x,p)^2 \big( \mathcal{L}_0(x,p) - 1 \big),
\end{equation}
where the function $B(x,p)$ is defined by
\begin{equation} \label{eq:B-def}
  B(x,p)^2 = \frac{L_0(x,p) - E}{\mathcal{L}_0(x,p) - 1} .
\end{equation}
For the moment, let us assume that the right-hand side of this equation is positive, and define $B(x,p)$ to be its positive root. We come back to this assumption later on.
Until now, we have looked at the quantities $B$, $L_0$ and $\mathcal{L}_0$ in equation~(\ref{eq:symbol-relation-Hams}) as simple functions. However, we can also interpret them as principal symbols of operators. This line of thought leads to the following operator decomposition:
\begin{equation} \label{eq:operator-relation-Hams}
  L_0(x,\hat{p}) + h L_1(x,\hat{p}) - E = 
    \hat{B}^\dagger \big( \mathcal{L}_0(x,\hat{p}) + h \mathcal{L}_1(x,\hat{p}) - 1 \big) \hat{B} + \mathcal{O}(h^2) ,
\end{equation}
in which the operators are related to their symbols by standard quantization, see equation~(\ref{eq:standard-quantization}). In particular, the symbol of the operator $B$ is given by $B(x,p)$, see equation~(\ref{eq:B-def}). Furthermore, the symbols $L_0$ and $L_1$ are defined by equations~(\ref{eq:L0}) and~(\ref{eq:L1}), respectively, and the symbol $\mathcal{L}_0$ was defined above.

We can then determine the symbol $\mathcal{L}_1$ by passing to symbols in equation~(\ref{eq:operator-relation-Hams}). 
Using equations~(\ref{eq:standard-symbol-product}) and~(\ref{eq:adjoint-symbol-relation}) to compute the symbol of the product of operators on the right-hand side, we obtain a large number of terms.
Collecting all terms of order $h^0$ and using that $B(x,p)$ is real, we naturally recover equation~(\ref{eq:symbol-relation-Hams}). Collecting the terms of order $h$, and separating real and imaginary terms, we find that
\begin{align}
  \text{Re} \, L_1 &= 
    B^2 \text{Re} \, \mathcal{L}_1 ,  \label{eq:symbol-relation-L1-real} \\
  \text{Im} \, L_1 & = 
    B^2 \text{Im} \, \mathcal{L}_1 
      - \left( \left\langle \frac{\partial }{\partial p} , \frac{\partial }{\partial x} \right\rangle B \right) (\mathcal{L}_0 - 1) B 
      - \left\langle \frac{\partial B}{\partial p} , \frac{\partial \mathcal{L}_0}{\partial x} \right\rangle B  \nonumber
      \\ & \hspace*{3.2cm}
      - B \left\langle \frac{\partial \mathcal{L}_0}{\partial p} , \frac{\partial B}{\partial x} \right\rangle
      - (\mathcal{L}_0 - 1) \left\langle \frac{\partial B}{\partial p}, \frac{\partial B}{\partial x} \right\rangle . \label{eq:symbol-relation-L1-imag}
\end{align}
In section~\ref{subsec:separation}, we showed that the effective Hamiltonian $\hat{L}$ is symmetric. Therefore, the imaginary part of $L_1$ is related to $L_0$ by equation~(\ref{eq:ImL1}). Because of the symmetric placement of the operator $\hat{B}$ and its adjoint, the imaginary part of $\mathcal{L}_1$ is related to $\mathcal{L}_0$ by the same equation. In fact, one can define a self-adjoint operator $\hat{\mathcal{L}}$ by adding higher-order expansion coefficients on both sides of equation~(\ref{eq:operator-relation-Hams}). 

For completeness, let us explicitly derive the relation between $\mathcal{L}_1$ and $\mathcal{L}_0$. Rewriting equation~(\ref{eq:symbol-relation-L1-imag}), we have
\begin{multline}
  \text{Im} \, \mathcal{L}_1 = 
    \frac{1}{B^{2}} \text{Im} \, L_1
      + \frac{1}{B} \left( \left\langle \frac{\partial }{\partial p} , \frac{\partial }{\partial x} \right\rangle B \right) (\mathcal{L}_0 - 1)
      + \frac{1}{B} \left\langle \frac{\partial B}{\partial p} , \frac{\partial \mathcal{L}_0}{\partial x} \right\rangle   \\
      + \frac{1}{B} \left\langle \frac{\partial \mathcal{L}_0}{\partial p} , \frac{\partial B}{\partial x} \right\rangle
      + \frac{1}{B^2} (\mathcal{L}_0 - 1) \left\langle \frac{\partial B}{\partial p}, \frac{\partial B}{\partial x} \right\rangle
\end{multline}
Subsequently, we use that $\text{Im}\,L_1$ is related to $L_0$ by equation~(\ref{eq:ImL1}), which is in turn related to $\mathcal{L}_0$ by equation~(\ref{eq:symbol-relation-Hams}). Computing the derivatives, we find that many terms cancel and finally obtain
\begin{equation} \label{eq:ImL1-eikonal}
  \text{Im} \, \mathcal{L}_1(x,p) = -\frac{1}{2} \left\langle \frac{\partial}{\partial x} , \frac{\partial}{\partial p} \right\rangle \mathcal{L}_0(x,p) ,
\end{equation}
as we anticipated.

Because of the relation~(\ref{eq:operator-relation-Hams}), we can study the eigenvalue equation
\begin{equation} \label{eq:eigenvalue-eq-eikonal}
  \big( \mathcal{L}_0(x,\hat{p}) + h \mathcal{L}_1(x,\hat{p}) \big) \tilde{\psi} = \tilde{\psi}
\end{equation}
instead of $\hat{L} \psi = E \psi$. Using the methods that we discussed in the main text, we can construct an asymptotic solution for equation~(\ref{eq:eigenvalue-eq-eikonal}). Of course, the Lagrangian manifold that is induced by this equation coincides with the Lagrangian manifold $\Lambda^2$ that we discussed in section~\ref{sec:lagman}. However, it is now automatically parametrized by eikonal coordinates, since its Hamiltonian has the form $\mathcal{L}_0 = C(x)|p|$, see section~\ref{subsec:eikonal}.
Once we have constructed an asymptotic solution for $\tilde{\psi}$, we obtain an asymptotic solution for $\psi$ by
\begin{equation} \label{eq:wavefunction-Ham-eikonal-relation}
  \psi(x) = (\hat{B})^{-1} \, \tilde{\psi}(x) .
\end{equation}
Since our asymptotic solution is given in the form of the canonical operator, we can compute the action of the $(\hat{B})^{-1}$ on $\tilde{\psi}$ using the commutation formula~(\ref{eq:commutation-relation-canop-gen}). Therefore, we obtain the leading-order term of the asymptotic solution for $\psi(x)$ by multiplying the amplitude of the asymptotic solution for $\tilde{\psi}(x)$ by $1/B(x,p)$. In the remainder of this appendix, we show that this leads to the same result for $\psi$ as reparametrizing the time along the trajectories and subsequently using equation~(\ref{eq:precanonicalTimeChange}).

To this end, we consider the symbol $B(x,p)$ along the solutions of the Hamiltonian system, i.e. on the Lagrangian manifold. Looking at equation~(\ref{eq:B-def}), we observe that both the numerator and the denominator of $B^2$ vanish on $\Lambda^2$, since points on this manifold satisfy both $L_0=E$ and $\mathcal{L}_0=1$. We therefore have to define their ratio in the sense of a limit and set
\begin{equation} \label{eq:R-general}
  R(x) = \lim_{|p| \to 1/C(x)} \, \frac{L_0(x,|p|)-E}{C(x) |p| - 1} = \frac{1}{C(x)} \left. \frac{\partial L_0}{\partial |p|}\right|_{|p|=1/C(x)} .
\end{equation}
Note that we have used the same letter for this quantity and for the variable that reparametrizes the time in equation~(\ref{eq:timereparametrization}). This is no coincidence, since these two quantities actually coincide: one can show~\cite{Dobrokhotov15} that for a general scalar Hamiltonian $L_0$ one obtains eikonal coordinates on the Lagrangian manifold by reparametrizing the time with the factor $R$ defined in equation~(\ref{eq:R-general}). Computing this quantity for the effective Hamiltonian~(\ref{eq:L0Dirac}) that arises from the Dirac equation, one naturally recovers the result~(\ref{eq:Rtransition}).
At this point, we briefly come back to the assumption that $B^2$ is positive, which we made below equation~(\ref{eq:B-def}). We now see that, along the trajectories of the Hamiltonian system, it translates to the requirement that the derivative of $L_0$ is positive. Since this condition is satisfied for electrons in graphene, the method is consistent for our example. When one considers holes in graphene, the derivative is negative. In this case, one should consider the effective Hamiltonian $\mathcal{L}_0 = -C(x)|p|=-1$ instead, which leads to a positive value for $R(x)$. 

Let us now compute the wavefunction $\psi$ using equation~(\ref{eq:wavefunction-Ham-eikonal-relation}). The commutation formula~(\ref{eq:commutation-relation-canop-gen}) states that we obtain it by multiplying the amplitude in the asymptotic solution for $\tilde{\psi}$ by the inverse of the principal symbol of $B$ along the trajectories. In this way, we introduce a factor $1/\sqrt{R(\mathcal{X})}$, which is exactly the same factor as was introduced previously by equation~(\ref{eq:precanonicalTimeChange}).
However, this does not prove the equality of the two solutions yet, since we also have to take a closer look at the semiclassical phase, which is governed by the subprincipal symbol $\mathcal{L}_1$. We have
\begin{equation}
  \Phi_{sc} = -\int_0^\tau \text{Re} \, \mathcal{L}_1 \, \text{d} \tau
    = -\int_0^\tau \frac{1}{B^{2}} \text{Re} \, L_1 \, \text{d} \tau
    = -\int_0^\tau \frac{1}{R} L_1^W \, \text{d} \tau
\end{equation}
This is exactly the same as we obtained in equation~(\ref{eq:BerryPhaseEikonal}). Thus, the semiclassical phases of the two asymptotic solutions also coincide. 

We therefore conclude that the asymptotic solution $\psi$ defined by equation~(\ref{eq:wavefunction-Ham-eikonal-relation}) coincides with our previously obtained asymptotic solution. Hence, introducing eikonal coordinates by applying the operator equality~(\ref{eq:operator-relation-Hams}) leads to the same final result as introducing eikonal coordinates on the Lagrangian manifold.

\section{Conventional representation of the precanonical operator} \label{app:conventionalCanonicalOp}

In section~\ref{subsec:leadingorder}, we discussed how to obtain the leading-order approximation to the wavefunction in the vicinity of fold and cusp points. In that section, we started from the new representation of the precanonical operator corresponding to singular charts, discussed in section~\ref{subsec:maslovsingular}. In this appendix, we show how the same result can be obtained using the conventional form of the precanonical operator corresponding to singular charts, discussed in section~\ref{subsec:maslovapproaches}. 
These calculations turn out to be more involved than those for the new representation, where one can use the iteration method~\cite{Dobrokhotov14}. Instead, for the conventional representation, we manually calculate the derivatives up to fourth order after a change of coordinates.
These calculations also provide some additional insight into the connection between the two different representations.
We remark that the general equivalence of the new representation and the conventional representation was proven in Ref.~\cite{Dobrokhotov14}.

The conventional representation of the precanonical operator corresponding to singular charts is given by equation~(\ref{eq:precanonical-singular-conventional}). We consider this precanonical operator on the Lagrangian manifold parametrized by eikonal coordinates, meaning that we replace $t$ by $\tau$ in equation~(\ref{eq:precanonical-singular-conventional}). One can subsequently obtain the precanonical operator for the original problem using equation~(\ref{eq:precanonicalTimeChange}).
In order to simplify our precanonical operator near a singular point, we first perform a rotation of the coordinate system. Specifically, we consider the velocity vector (in eikonal coordinates) at the singular point, i.e. $\mathcal{X}_\tau^* = (\mathcal{X}^*_{1,\tau},\mathcal{X}^*_{2,\tau})$, where the star indicates that these quantities are to be evaluated at the singular point. We subsequently rotate our coordinates in such a way that, in the new coordinate system, the components of this vector are given by $\mathcal{X}^*_{1,\tau} > 0$ and $\mathcal{X}_{2,\tau}^* = 0$.
Looking at the examples in section~\ref{sec:examples}, we observe that this new coordinate system is often very natural. For instance, when we consider scattering by a Gaussian potential well, we observe that we do not have to perform a rotation for these conditions to be satisfied at the cusp point. For the fold points, we need to perform a rotation over a certain angle, which increases when we move further away from the cusp point.

The rotation that we perform affects the precanonical operator. Suppose that we are able to construct the precanonical operator in our new coordinate system. When we label the points in this new coordinate system with a prime, we can denote this precanonical operator by $(K_{\Lambda^2(\tau,\phi)}^{\Omega_i} A_0)(x')$. This precanonical operator is related to the precanonical operator in the original coordinate system by the relation~\cite{Maslov81}
\begin{equation} \label{eq:precanonicalCoordChange}
  (K_{\Lambda^2(\tau,\phi)}^{\Omega_i} A_0)(x) = \left(\left|\det\frac{\partial x'(x)}{\partial x}\right|\right)^{1/2} (K_{\Lambda^2(\tau,\phi)}^{\Omega_i} A_0)(x'(x)) .
\end{equation}
It is apparent from equation~(\ref{eq:precanonicalX}) that this relation holds for precanonical operators corresponding to regular charts. However, it can also be extended to those corresponding to singular charts~\cite{Maslov81,Maslov65}. Since we consider a rotation, the determinant in equation~(\ref{eq:precanonicalCoordChange}) is one in our case. In the remainder of this appendix, we work in our rotated coordinate system, unless explicitly noted otherwise. To simplify our notation, we henceforth omit the primes on these coordinates.

To show that the new representation coincides with the conventional representation, we would like to change the integration variable in equation~(\ref{eq:precanonical-singular-conventional}) from $p_2$ to $\phi$. In order to perform such a change of variables, we have to express $\tau$ as a function of $x_1$ and $\phi$, instead of as a function of $x_1$ and $p_2$.
Let us therefore take a closer look at the equation
\begin{equation}
  x_1 = \mathcal{X}_1(\tau, \phi) .
\end{equation}
Since $\mathcal{X}^*_{1,\tau} > 0$, this equation has a smooth solution $\tau = \tau(x_1,\phi)$ in the vicinity of a singular point. Taking the total derivative of $x_1 = \mathcal{X}(\tau(x_1,\phi), \phi)$ with respect to $\phi$, we establish that
\begin{equation} \label{eq:derivtauphi}
  0 = \frac{\partial \mathcal{X}_1}{\partial \tau} \frac{\partial \tau}{\partial \phi} + \frac{\partial \mathcal{X}_1}{\partial \phi} , \quad \text{hence} \quad
  \frac{\partial \tau}{\partial \phi} = -\frac{\mathcal{X}_{1,\phi}}{\mathcal{X}_{1,\tau}} .
\end{equation}
Subsequently, we compute the total derivative $\text{d}p_2/\text{d}\phi$ which we need to perform the change of variables. Taking the total derivative of $p_2=\mathcal{P}_2(\tau(x_1,\phi),\phi)$ with respect to $\phi$, we obtain
\begin{equation} \label{eq:change-vars-p2-phi}
  \frac{\text{d} p_2}{\text{d} \phi} = \frac{\partial \mathcal{P}_2}{\partial \tau} \frac{\partial \tau}{\partial \phi} + \frac{\partial \mathcal{P}_2}{\partial \phi} 
    = - \mathcal{P}_{2,\tau} \frac{\mathcal{X}_{1,\phi}}{\mathcal{X}_{1,\tau}} + \mathcal{P}_{2,\phi}
    = \frac{\overline{\mathcal{J}}}{\mathcal{X}_{1,\tau}} ,
\end{equation}
where we have defined
\begin{equation} \label{eq:Jacobian-fourier-p1-eikonal}
  \overline{\mathcal{J}}(\tau,\phi)=\det \frac{\partial(\mathcal{X}_1, \mathcal{P}_2)}{\partial(\tau,\phi)} = \mathcal{X}_{1,\tau} \mathcal{P}_{2,\phi} - \mathcal{P}_{2,\tau} \mathcal{X}_{1,\phi} ,
\end{equation}
in analogy with equation~(\ref{eq:Jacobian-fourier-p1}).

To be able to use the conventional representation~(\ref{eq:precanonical-singular-conventional}) in the vicinity of a singular point, we have to show that the Jacobian~$\overline{\mathcal{J}}$ is nonzero at the singular point. We now do this explicitly, and at the same time establish some convenient properties of our rotated coordinate system. 
First, we observe that the vectors $\mathcal{X}_\tau$ and $\mathcal{P}$ are parallel by the Hamiltonian system~(\ref{eq:Ham-Dirac-2}). Hence, $\mathcal{P}^*_1>0$ and $\mathcal{P}^*_2=0$. 
In section~\ref{subsec:classification}, we showed that $\langle \mathcal{P}^*, \mathcal{P}_\phi^* \rangle = 0$ and that at the same time both the vector $\mathcal{P}_\phi^*$ and the vector $\mathcal{P}_\phi^*$ are nonzero. Since inner products are invariant under a coordinate rotation, these equalities still hold in our new coordinate system. Taking into account that $\mathcal{P}^*_1>0$ and $\mathcal{P}^*_2=0$, the relation $\langle \mathcal{P}^*, \mathcal{P}_\phi^* \rangle = 0$ implies that $\mathcal{P}_{1,\phi}^*=0$. However, since $\mathcal{P}_\phi^* \neq 0$, the second component of this vector cannot vanish, i.e., $\mathcal{P}_{2,\phi}^*\neq0$.
Since $\mathcal{X}_\phi$ vanishes at a singular point, we subsequently find that the Jacobian~(\ref{eq:Jacobian-fourier-p1-eikonal}) at such a point is given by $\overline{\mathcal{J}}^*=\mathcal{X}_{1,\tau}^* \mathcal{P}_{2,\phi}^* \neq 0$. Thus, the Jacobian~$\overline{\mathcal{J}}$ is nonzero at the singular point.

Our next step is to perform the change of variables from $p_2$ to $\phi$ in the conventional representation~(\ref{eq:precanonical-singular-conventional}) of the precanonical operator. Using equation~(\ref{eq:change-vars-p2-phi}), we obtain
\begin{multline} \label{eq:precanonical-singular-conventional-coord-change}
  (K_{\Lambda^2(\tau,\phi)}^{\Omega_i^s} A_0)(x) =  \frac{e^{i\pi/4}}{\sqrt{2\pi h}} e^{-\frac{i \pi}{2} \mu_{\Omega_i^s} } \\
  \times \int_{-\infty}^\infty \text{d} \phi
  \left. \left| \frac{\overline{\mathcal{J}}(\tau,\phi)}{\mathcal{X}_{1,\tau}(\tau,\phi)} \right|
  \frac{A_0(\tau,\phi)}{| \overline{\mathcal{J}}(\tau,\phi) |^{1/2} } 
  e^{\frac{i}{h} \big(S(\tau,\phi) + \mathcal{P}_2(\tau,\phi) [ x_2 - \mathcal{X}_2(\tau,\phi) ] \big)}
  \right|_{\tau=\tau(x_1,\phi)} .
\end{multline}
We now want to show that this representation gives rise to the same leading-order approximations as equation~(\ref{eq:precanonicalXsingular}). To this end, we have to show that, at singular points, both the amplitudes of the integrals and the expansion coefficients of the phase functions coincide. 

Let us start by showing that the amplitudes coincide. First of all, we note that the Maslov indices in both representations are the same, see section~\ref{subsec:maslovindex} and Ref.~\cite{Dobrokhotov14}. Second, we observe that $A_0(\tau,\phi)$ is a function on the Lagrangian manifold, which implies that it is the same in both representations. Finally, we consider the determinants. We have
\begin{equation} \label{eq:det-amplitude-conventional}
  \left| \frac{\overline{\mathcal{J}}^*}{\mathcal{X}_{1,\tau}^*} \right| \frac{1}{| \overline{\mathcal{J}}^* |^{1/2} } 
    = \frac{ \sqrt{\mathcal{X}_{1,\tau}^* |\mathcal{P}_{2,\phi}^*|}}{\mathcal{X}_{1,\tau}^*} = \sqrt{|\mathcal{P}^*| |\mathcal{P}_{\phi}^*|} ,
\end{equation}
where the first equality holds because of the properties that we proved in the previous paragraphs. The last equality is somewhat more intricate. Since we have $\mathcal{P}_1^*>0$ and $\mathcal{P}_2^*=0$, we also have $|\mathcal{P}^*|=\mathcal{P}_1^*$. Combining this observation with Hamilton's equation $\mathcal{X}_{1,\tau} = C \mathcal{P}_1/|\mathcal{P}|$ and the fact that $C|\mathcal{P}|=1$, we obtain $|\mathcal{P}^*|=\mathcal{P}_1^*=1/\mathcal{X}_{1,\tau}^*$. Furthermore, since $\mathcal{P}_{1,\phi}^*=0$, we have $|\mathcal{P}_{\phi}^*| = |\mathcal{P}_{2,\phi}^*|$, which then implies the last equality in equation~(\ref{eq:det-amplitude-conventional}). Hence, taking into account that $\langle \mathcal{P}^*, \mathcal{P}_\phi^* \rangle = 0$, we observe that also the determinants of both representations coincide. Thus, we have shown that the amplitudes of both representations coincide at a singular point.

Our final step is to show that the expansion coefficients of the phase functions of both representations coincide. To this end, we have to show that we obtain the coefficients~(\ref{eq:phi-expansion-coeffs}) when we perform a Taylor expansion of the phase function 
\begin{equation} \label{eq:phase-function-conventional}
  \Phi_{cv} = S(\tau(x_1,\phi),\phi) + \mathcal{P}_2(\tau(x_1,\phi),\phi) \, [ \, x_2 - \mathcal{X}_2(\tau(x_1,\phi),\phi) \, ]
\end{equation}
with respect to $\phi$ and $x_1$ around a singular point. This derivation is fairly cumbersome and is performed step by step in the next paragraphs.

Let us consider a singular point $(\tau^*, \phi^*)$ on the Lagrangian manifold. The point $x$ in configuration space that corresponds to this point is defined by $x=\mathcal{X}^*=\mathcal{X}(\tau^*, \phi^*)$. When we consider the inverse function $\tau(x_1,\phi)$, we therefore have $\tau^* = \tau(\mathcal{X}_1^*,\phi^*)$.
We start by calculating the simplest coefficient, $a_0$. It can be found by evaluating $\Phi_{cv}$ at the point $\phi^*$, setting $x=\mathcal{X}^*$. Hence,
\begin{equation}
  a_0 = S^* = \tau^* ,
\end{equation}
where the last equality is implied by equation~(\ref{eq:actionTau}).

Taking the derivative of equation~(\ref{eq:phase-function-conventional}) with respect to $x_1$, we find that
\begin{equation}
  \frac{\partial \Phi_{cv}}{\partial x_1} = \frac{\partial S}{\partial \tau} \frac{\partial \tau}{\partial x_1} + \frac{\partial \mathcal{P}_2}{\partial \tau} \frac{\partial \tau}{\partial x_1} (x_2 - \mathcal{X}_2) - \mathcal{P}_2 \frac{\partial \mathcal{X}_2}{\partial \tau} \frac{\partial \tau}{\partial x_1} .
\end{equation}
Specializing to the singular point, taking into account that $\mathcal{P}_2^* = 0$, we subsequently find that
\begin{equation}
  b_{0,1} = \frac{1}{\mathcal{X}_{1,\tau}^*} = \mathcal{P}_1^* .
\end{equation}
Since $\mathcal{P}_2^* = 0$, we immediately observe that $b_{0,2} = 0$. Therefore, $b_0 = \mathcal{P}^*$, in accordance with equation~(\ref{eq:phi-expansion-coeffs}).

To compute the higher-order coefficients, we have to take derivatives of $\Phi_{cv}$ with respect to $\phi$. It is important to note that the (total) derivative with respect to $\phi$ consists of two parts, i.e.
\begin{equation} \label{eq:total-deriv-phi}
  \frac{\text{d}}{\text{d}\phi} = \frac{\partial}{\partial\phi} + \frac{\partial \tau}{\partial \phi} \frac{\partial}{\partial \tau} 
    = \frac{\partial}{\partial\phi} -\frac{\mathcal{X}_{1,\phi}}{\mathcal{X}_{1,\tau}} \frac{\partial}{\partial \tau},
\end{equation}
where the last equality follows from equation~(\ref{eq:derivtauphi}). Taking into account that $\partial S/\partial \phi$ vanishes, we have
\begin{equation}
  \frac{\text{d} \Phi_{cv}}{\text{d}\phi} = - \frac{\mathcal{X}_{1,\phi}}{\mathcal{X}_{1,\tau}} \frac{\partial S}{\partial \tau} 
    - \mathcal{P}_2 \mathcal{X}_{2,\phi} + \frac{\mathcal{X}_{1,\phi}}{\mathcal{X}_{1,\tau}} \mathcal{P}_2 \mathcal{X}_{2,\tau}
    + \left( \mathcal{P}_{2,\phi} - \frac{\mathcal{X}_{1,\phi}}{\mathcal{X}_{1,\tau}} \mathcal{P}_{2,\tau} \right) (x_2 - \mathcal{X}_2) .
\end{equation}
By virtue of the first equality in equation~(\ref{eq:eikonalCoordinates}), this expression becomes
\begin{equation}
  \frac{\text{d} \Phi_{cv}}{\text{d}\phi} = - \frac{\mathcal{X}_{1,\phi}}{\mathcal{X}_{1,\tau}} \mathcal{P}_1 \mathcal{X}_{1,\tau}
    - \mathcal{P}_2 \mathcal{X}_{2,\phi} + \left( \mathcal{P}_{2,\phi} - \frac{\mathcal{X}_{1,\phi}}{\mathcal{X}_{1,\tau}} \mathcal{P}_{2,\tau} \right) (x_2 - \mathcal{X}_2) .
\end{equation}
The first two terms of this expression cancel because of the second equality in equation~(\ref{eq:eikonalCoordinates}), while the third term can be rewritten using our definition~(\ref{eq:Jacobian-fourier-p1-eikonal}). We therefore obtain
\begin{equation} \label{eq:Phicd-deriv}
  \frac{\text{d} \Phi_{cv}}{\text{d}\phi} = \frac{\overline{\mathcal{J}}}{\mathcal{X}_{1,\tau}} (x_2 - \mathcal{X}_2) .
\end{equation}
Note that this equality also holds when we are not at the singular point. Specializing to a singular point, we immediately find that $a_1 = 0$. Furthermore, we find that $b_{1,2} = \mathcal{P}^*_{2,\phi}$. To obtain $b_{1,1}$, we take the derivative of equation~(\ref{eq:Phicd-deriv}) with respect to $x_1$ and subsequently specialize to the singular point. This gives
\begin{equation}
  b_{1,1} = -\frac{\overline{\mathcal{J}}^*}{\mathcal{X}_{1,\tau}^*} \frac{\mathcal{X}_{2,\tau}^*}{\mathcal{X}_{1,\tau}^*} = 0 .
\end{equation}
Since $\mathcal{P}^*_{1,\phi}$ also vanishes, we conclude that $b_1 = \mathcal{P}_\phi^*$, in agreement with equation~(\ref{eq:phi-expansion-coeffs}).

To compute $a_2$ and $b_2$, we compute the derivative of equation~(\ref{eq:Phicd-deriv}). This gives
\begin{equation} \label{eq:Phicd-second-deriv}
  \frac{\text{d}^2 \Phi_{cv}}{\text{d}\phi^2} = \frac{\text{d}}{\text{d} \phi} \left( \frac{\overline{\mathcal{J}}}{\mathcal{X}_{1,\tau}} \right) (x_2 - \mathcal{X}_2) 
    - \frac{\overline{\mathcal{J}} \mathcal{J}}{\mathcal{X}_{1,\tau}^2} ,
\end{equation}
where $\mathcal{J} = \mathcal{X}_{1,\tau} \mathcal{X}_{2,\phi} - \mathcal{X}_{1,\phi} \mathcal{X}_{2,\tau}$ is the regular Jacobian and we have used equation~(\ref{eq:total-deriv-phi}). Since the Jacobian $\mathcal{J}$ vanishes at the singular point, equation~(\ref{eq:Phicd-second-deriv}) shows that $a_2=0$.
Furthermore, we observe that the coefficient $b_{2,2}$ is given by the total derivative of $\overline{\mathcal{J}}/\mathcal{X}_{1,\tau}$ with respect to $\phi$, evaluated at the singular point. Looking at the prescription~(\ref{eq:total-deriv-phi}) for the total derivative and realizing that $\mathcal{X}_{1,\phi}^*=0$, we conclude that it is sufficient to take the partial derivative with respect to $\phi$ in this expression. We therefore obtain
\begin{equation}
  b_{2,2} = \frac{1}{\mathcal{X}_{1,\tau}^*}(\mathcal{X}_{1,\tau\phi}^*\mathcal{P}_{2,\phi}^* + \mathcal{X}_{1,\tau}^*\mathcal{P}_{2,\phi\phi}^* 
  - \mathcal{P}_{2,\tau}^* \mathcal{X}_{1,\phi\phi}^*) - \frac{\overline{\mathcal{J}}^*}{(\mathcal{X}_{1,\tau}^*)^2} \mathcal{X}_{1,\tau\phi}^* .
\end{equation}
We first prove that $\mathcal{X}_{1,\tau\phi}^*$ vanishes. By the Hamiltonian system, $\mathcal{X}_{1,\tau} = C^2 \mathcal{P}_1$, whence $\mathcal{X}_{1,\tau\phi} = C^2 \mathcal{P}_{1,\phi} + 2 C \langle \partial C/\partial \mathcal{X}, \mathcal{X}_\phi \rangle \mathcal{P}_1$. Since both $\mathcal{P}_{1,\phi}^*$ and $\mathcal{X}_\phi^*$ vanish, we conclude that $\mathcal{X}_{1,\tau\phi}^*$ vanishes. Since we are only interested in the coefficient $b_2$ for cusp points, we can also set $\mathcal{X}_{\phi\phi}^*$ to zero. Therefore, we conclude that $b_{2,2} = \mathcal{P}_{2,\phi\phi}^*$ for a cusp point. 
To find the coefficient $b_{2,1}$, we have to take the derivative of equation~(\ref{eq:Phicd-second-deriv}) with respect to $x_1$ and subsequently specialize to the singular point. Since terms proportional to $(x_2 - \mathcal{X}_2)$ vanish when we take this final step, we omit them from the very beginning. We therefore obtain
\begin{equation} \label{eq:b21-pre}
  b_{2,1} = - \left[ \frac{\text{d}}{\text{d} \phi} \left( \frac{\overline{\mathcal{J}}}{\mathcal{X}_{1,\tau}} \right) \right]^* \frac{\mathcal{X}_{2,\tau}^*}{\mathcal{X}_{1,\tau}^*} 
    - \bigg[ \frac{\partial}{\partial \tau} \bigg( \frac{\overline{\mathcal{J}}}{\mathcal{X}_{1,\tau}^2} \bigg) \bigg]^* \frac{\mathcal{J}^*}{\mathcal{X}_{1,\tau}^*}
    - \frac{\overline{\mathcal{J}}^*}{(\mathcal{X}_{1,\tau}^*)^3} \mathcal{J}_\tau^* .
\end{equation}
It is clear that both the first and the second term in this equation vanish, since $\mathcal{X}_{2,\tau}^* = \mathcal{J}^* = 0$. In section~\ref{subsec:classification}, we derived that $\mathcal{J}_\tau^* = (C^4)^* \widetilde{\mathcal{J}}^*$, see equation~(\ref{eq:JtauCaustic}). This equation remains valid in our rotated coordinate system. Since we also have $1/\mathcal{X}_{1,\tau}^*=|\mathcal{P}^*|$, equation~(\ref{eq:b21-pre}) becomes 
\begin{equation}
  b_{2,1} = |\mathcal{P}|^3 \mathcal{X}_{1,\tau}^* \mathcal{P}_{2,\phi}^* (C^*)^4 |\mathcal{P}^*| |\mathcal{P}_\phi^*| = (\mathcal{P}_{2,\phi}^*)^2 / \mathcal{P}_1^* ,
\end{equation}
where the last equality follows from the equalities $\mathcal{P}_{1,\phi}^*=0$ and $C|\mathcal{P}|=1$. 
Taking the second derivative of the relation $C|\mathcal{P}|=1$ with respect to $\phi$, we find that $(\mathcal{P}_{2,\phi}^*)^2 = \mathcal{P}_1^* \mathcal{P}_{1,\phi\phi}^*$ for a cusp point. Therefore, $b_{2,1} = \mathcal{P}_{1,\phi\phi}^*$ and we find that $b_2 = \mathcal{P}_{\phi\phi}^*$ for a cusp point, in accordance with equation~(\ref{eq:phi-expansion-coeffs}).

Subsequently, we compute the coefficient $a_3$. Since we are not interested in the coefficient $b_3$, we can immediately specialize to the singular point and set $x=\mathcal{X}^*$ in these calculations. Taking the derivative of equation~(\ref{eq:Phicd-second-deriv}) with respect to $\phi$ and retaining only the nonzero terms, we obtain
\begin{equation}
\begin{aligned}
  a_3 &= \bigg[ - \frac{\text{d}}{\text{d} \phi} \left( \frac{\overline{\mathcal{J}}}{\mathcal{X}_{1,\tau}} \right) \frac{\text{d} \mathcal{X}_2}{\text{d} \phi} 
           - \frac{\overline{\mathcal{J}}}{\mathcal{X}_{1,\tau}^2} \frac{\text{d} \mathcal{J}}{\text{d} \phi} \bigg]^* \\
      &= \bigg[ - \frac{\text{d}}{\text{d} \phi} \left( \frac{\overline{\mathcal{J}}}{\mathcal{X}_{1,\tau}} \right) \bigg]^* \mathcal{X}_{2,\phi}^* 
           - \frac{\mathcal{X}_{1,\tau}^* \mathcal{P}_{2,\phi}^*}{(\mathcal{X}_{1,\tau}^*)^2}
           \mathcal{X}_{1,\tau}^*\mathcal{X}_{2,\phi\phi}^* 
      = -\mathcal{P}_{2,\phi}^* \mathcal{X}_{2,\phi\phi}^*
\end{aligned}
\end{equation}
Since $\mathcal{P}_{1,\phi}^* = 0$, we conclude that $a_3 = -\langle \mathcal{P}_{\phi}^* , \mathcal{X}_{\phi\phi}^* \rangle$. Since the vectors $\mathcal{P}_{\phi}^*$ and $\mathcal{X}_{\phi\phi}^*$ are parallel, see section~\ref{subsec:leadingorder}, we can also write $|a_3| = | \mathcal{P}_{\phi}^* || \mathcal{X}_{\phi\phi}^* |$.

Finally, we compute the coefficient $a_4$ for a cusp point. Since $\mathcal{X}_{\phi\phi}^*$ is zero at such a point, the derivative $(\text{d} \mathcal{J}/\text{d} \phi)^*$ also vanishes at a cusp point. Therefore, when we take the second (total) derivative of equation~(\ref{eq:Phicd-second-deriv}) with respect to $\phi$ and specialize to the singular point, the only nonzero term is
\begin{equation}
  a_4 = - \frac{\overline{\mathcal{J}}^*}{(\mathcal{X}_{1,\tau}^*)^2} \bigg[ \frac{\text{d}^2 \mathcal{J}}{\text{d} \phi^2} \bigg]^* 
      = - \frac{\mathcal{X}_{1,\tau}^* \mathcal{P}_{2,\phi}^*}{(\mathcal{X}_{1,\tau}^*)^2} \mathcal{X}_{1,\tau}^* \mathcal{X}_{2,\phi\phi\phi}^* = -\mathcal{P}_{2,\phi}^* \mathcal{X}_{2,\phi\phi\phi}^* .
\end{equation}
Hence, $a_4 = -\langle \mathcal{P}_{\phi}^* , \mathcal{X}_{\phi\phi\phi}^* \rangle$ for a cusp point, in agreement with equation~(\ref{eq:phi-expansion-coeffs}). We can also write $|a_4| = | \mathcal{P}_{\phi}^* | | \mathcal{X}_{\phi\phi\phi}^* |$, since the vectors $\mathcal{P}_{\phi}^*$ and $\mathcal{X}_{\phi\phi\phi}^*$ are parallel, see section~\ref{subsec:leadingorder}.

In conclusion, we have shown that the expansion coefficients of the phase function~(\ref{eq:phase-function-conventional}) coincide with the expansion coefficients~(\ref{eq:phi-expansion-coeffs}) of the phase function of the new representation. Although we used a special coordinate system in this section, all of the coefficients take the form of an inner product. Since inner products are invariant under coordinate rotations, these coefficients have the same form in the original coordinate system.
Combining all the results from this appendix, we find that, in the vicinity of singular points, the conventional representation~(\ref{eq:precanonical-singular-conventional}) gives rise to the same leading-order appproximations as the new representation~(\ref{eq:precanonicalXsingular}) of the precanonical operator corresponding to singular charts.


\end{document}